\documentclass[a4paper,12pt]{article}
\usepackage{amssymb}
\usepackage{amsmath}
\usepackage{epsfig}

\usepackage{graphics}

\usepackage{latexsym}
\usepackage{rotating}
\usepackage{titlesec}

\titleclass{\subsubsubsection}{straight}[\subsection]

\newcounter{subsubsubsection}[subsubsection]
\renewcommand\thesubsubsubsection{\thesubsubsection.\arabic{subsubsubsection}}
\titleformat{\subsubsubsection}
  {\normalfont\normalsize\bfseries}{\thesubsubsubsection}{1em}{}
\titlespacing*{\subsubsubsection}
{0pt}{3.25ex plus 1ex minus .2ex}{1.5ex plus .2ex}

\makeatletter
\renewcommand\paragraph{\@startsection{paragraph}{5}{\z@}%
  {3.25ex \@plus1ex \@minus.2ex}%
  {-1em}%
  {\normalfont\normalsize\bfseries}}
\renewcommand\subparagraph{\@startsection{subparagraph}{6}{\parindent}%
  {3.25ex \@plus1ex \@minus .2ex}%
  {-1em}%
  {\normalfont\normalsize\bfseries}}
\def\toclevel@subsubsubsection{4}
\def\toclevel@paragraph{5}
\def\toclevel@paragraph{6}
\def\l@subsubsubsection{\@dottedtocline{4}{7em}{4em}}
\def\l@paragraph{\@dottedtocline{5}{10em}{5em}}
\def\l@subparagraph{\@dottedtocline{6}{14em}{6em}}
\makeatother

\title{Nonlocal Modified KdV Equations and Their Soliton Solutions}

\author{Metin G\"{u}rses \thanks{gurses@fen.bilkent.edu.tr}\\
{\small Department of Mathematics, Faculty of Science}\\
{\small Bilkent University, 06800 Ankara - Turkey}\\
Asl{\i} Pekcan \thanks{Email:aslipekcan@hacettepe.edu.tr} \\
{\small Department of Mathematics, Faculty of Science} \\
{\small Hacettepe University, 06800 Ankara - Turkey}
}

\date{\nonumber}
\begin{document}
\maketitle
\date{\nonumber}
\newtheorem{thm}{Theorem}[section]
\newtheorem{Le}{Lemma}[section]
\newtheorem{defi}{Definition}[section]
\newtheorem{ex}{Example}[section]
\newtheorem{pro}{Proposition}[section]
\baselineskip 17pt

\numberwithin{equation}{section}

\begin{abstract}
We study the nonlocal modified Korteweg-de Vries (mKdV) equations obtained from
AKNS scheme by Ablowitz-Musslimani type nonlocal reductions. We first find soliton solutions of
the coupled mKdV system by using the Hirota direct method. Then by using the Ablowitz-Musslimani reduction formulas, we
find one-, two-, and three-soliton solutions of local and nonlocal complex mKdV and mKdV equations. The soliton solutions of these equations are of two types.
We give one-soliton solutions of both types and present only first type of two- and three-soliton solutions.
We illustrate our soliton solutions by plotting their graphs  for particular values of the parameters.

\noindent
\textbf{Keywords:} Ablowitz-Musslimani reduction, Nonlocal mKdV equations, Hirota bilinear form, Soliton solutions.
\end{abstract}

\newpage

\tableofcontents{}

\newpage

\section{Introduction}

When the Lax pair is a cubic polynomial of the spectral parameter we obtain  coupled modified Korteweg-de Vries system of equations from the AKNS formalism \cite{AKNS}. These equations are given by
\begin{eqnarray}
aq_t&=&-\frac{1}{4}q_{xxx}+\frac{3}{2}rqq_x\label{mKdV1}\\
ar_t&=&-\frac{1}{4}r_{xxx}+\frac{3}{2}rqr_x,\label{mKdV2}
\end{eqnarray}
where $q(t,x)$ and $r(t,x)$ are in general complex dynamical variables, $a$ is a complex constant. We call the above system of coupled equations as nonlinear modified Korteweg-de Vries system (mKdV system). We have two different local (standard) reductions of this system:
\begin{eqnarray}
&a)\, r(t,x)= k \bar{q}(t,x), \label{non0}\\
&b)\, r(t,x)=kq(t,x), \label{non00}
\end{eqnarray}
where $k$ is a real constant and $\bar{q}$ is the complex conjugate of the function $q$. When we apply the first reduction (\ref{non0}) to the
equations (\ref{mKdV1}) and (\ref{mKdV2}) we obtain complex modified Korteweg-de Vries (cmKdV) equation
\begin{equation}\label{mKdV3}\displaystyle
aq_t=-\frac{1}{4}q_{xxx}+\frac{3}{2}k\bar{q}qq_x,
\end{equation}
provided that $\bar{a}=a$. The second reduction (\ref{non00}) gives
the usual mKdV equation
\begin{equation}\label{usualmKdV}\displaystyle
aq_t=-\frac{1}{4}q_{xxx}+\frac{3}{2}kq^2q_x,
\end{equation}
with no condition on $a$.

\noindent In \cite{AbMu1}-\cite{AbMu3} Ablowitz and Musslimani introduced an integrable nonlocal reduction of the mKdV system  (\ref{mKdV1}) and (\ref{mKdV2}). The first one is given by
\begin{equation}
r(t,x)=k \bar{q}(\varepsilon_{1} t, \varepsilon_{2} x), \label{non}
\end{equation}
where $(\varepsilon_{1})^2=(\varepsilon_{2})^2=1$. Under this
condition the mKdV system (\ref{mKdV1}) and (\ref{mKdV2}) reduce to
\begin{equation}
aq_{t}(t,x)=-\frac{1}{4} q_{xxx}(t,x) +\frac{3}{2}k \bar{q}(\varepsilon_{1} t, \varepsilon_{2} x)q(t,x)q_x(t,x), \label{mKdV4}
\end{equation}
provided that $\bar{a}=\varepsilon_{1}\varepsilon_2 a$. The case for $(\varepsilon_{1}, \varepsilon_{2})=(1,1)$ yields the local equation (\ref{mKdV3}). There are three different nonlocal reductions where $(\varepsilon_{1}, \varepsilon_{2})=\{(-1,1),(1,-1),(-1,-1)\}$. Hence for these values of $\varepsilon_{1}$ and  $\varepsilon_{2}$ and for different signs of $k$ (sign($k$)=$\pm 1$), we have six different nonlocal integrable cmKdV equations obtained by Ablowitz-Musslimani reduction (\ref{non}). They are respectively the time reflection symmetric ($T$-symmetric), the space reflection symmetric ($S$-symmetric), and the space-time reflection symmetric ($ST$-symmetric) nonlocal cmKdV equations given below in part $A$.

\vspace{0.5cm}
\noindent
\textbf{A.} $r(t,x)=k \bar{q}(\varepsilon_{1} t, \varepsilon_{2} x)$ (Nonlocal cmKdV equations)\\

\noindent 1. $T$-symmetric cmKdV equation:
\begin{equation}\label{mKdV41}\displaystyle
aq_t(t,x)=-\frac{1}{4}q_{xxx}(t,x)+\frac{3}{2}k\bar{q}(-t,x)q(t,x)q_x(t,x), ~~~\bar{a}=-a.
\end{equation}

\vspace{0.3cm}

\noindent 2. $S$-symmetric cmKdV equation:
\begin{equation}\label{mKdV42}
aq_t(t,x)=-\frac{1}{4}q_{xxx}(t,x)+\frac{3}{2}k\bar{q}(t,-x)q(t,x)q_x(t,x),~~~\bar{a}=-a.
\end{equation}

\vspace{0.3cm}
\noindent
3. $ST$-symmetric cmKdV equation:
\begin{equation}\label{mKdV43}
aq_t(t,x)=-\frac{1}{4}q_{xxx}(t,x)+\frac{3}{2}k\bar{q}(-t,-x)q(t,x)q_x(t,x),~~~\bar{a}=a.
\end{equation}

\vspace{0.5cm}
\noindent
Another nonlocal reduction of the mKdV system is given by
\begin{equation}\label{nonsecond}
r(t,x)=kq(\varepsilon_1t,\varepsilon_2x),
\end{equation}
yielding the equation
\begin{equation}\displaystyle
aq_t(t,x)=-\frac{1}{4}q_{xxx}(t,x)+\frac{3}{2}kq(t,x)q(\varepsilon_1t,\varepsilon_2x)q_x(t,x)
\end{equation}
provided that $\varepsilon_1\varepsilon_2=1$. Therefore we have only one possibility $(\varepsilon_1,\varepsilon_2)=(-1,-1)$ to have a nonlocal equation, without any additional condition on the parameter $a$. The  $ST$-symmetric nonlocal
mKdV equation obtained here is given below in part $B$.

\vspace{0.5cm}
\noindent \textbf{B.} $r(t,x)=k q(\varepsilon_{1} t, \varepsilon_{2} x)$ (Nonlocal mKdV equation)\\

\noindent 1. Real $ST$-symmetric MKdV equation
\begin{equation}\label{mKdV44}
aq_t(t,x)=-\frac{1}{4}q_{xxx}(t,x)+\frac{3}{2}kq(-t,-x)q(t,x)q_x(t,x).
\end{equation}
\bigskip

\noindent The nonlocal cmKdV and mKdV equations have the focusing and defocusing cases when $k<0$ and $k>0$, respectively. All the above equations are integrable.

\noindent Recently, Ablowitz and Musslimani proposed other nonlocal nonlinear integrable equations such as reverse space-time and reverse time nonlocal nonlinear Schr\"{o}dinger (NLS) equation, sine-Gordon equation, $(1+1)$ and $(2+1)$ dimensional three-wave interaction, Davey-Stewartson equation, and derivative NLS equation arising from symmetry reductions of general AKNS scattering problem \cite{AbMu1}-\cite{AbMu3}. They discussed Lax pairs, an infinite number of conservation laws, inverse scattering transforms and found one-soliton solutions of these equations. Specifically, they studied complex and real $ST$-symmetric  nonlocal mKdV equations. In \cite{ma}, Ma, Shen, and Zhu showed that $ST$-symmetric nonlocal complex mKdV equation is gauge equivalent to a spin-like model. They constructed Darboux transformations for nonlocal cmKdV, different type of exact solutions including dark-soliton, W-type soliton, M-type soliton, and periodic solutions are obtained. Ji and Zhu obtained soliton, kink, anti-kink, complexiton, breather, rogue-wave solutions, and nonlocalized solutions with singularities of real $ST$-symmetric nonlocal mKdV equation through Darboux transformation and inverse scattering transform \cite{JZ1}, \cite{JZ2}. In \cite{Yang}, the authors showed that many nonlocal integrable equations like Davey-Stewartson equation, $T$-symmetric NLS equation, nonlocal derivative NLS equation, and $ST$-symmetric
cmKdV equation can be converted to local integrable equations by simple variable transformations. They used these transformations to obtain solutions of the nonlocal equations from the solutions of the local equations and to derive new nonlocal integrable equations like complex and real $ST$- and $T$-symmetric NLS equations and nonlocal complex short pulse equations. Some of our solutions coincide with the solutions given in \cite{AbMu3}-\cite{Yang}.  There is an increasing interest in obtaining the nonlocal reductions of systems of integrable equations and analyzing their solutions and properties \cite{fok}-\cite{Vincent}.

\noindent In a previous paper \cite{GurPek}, we have studied the soliton solutions of the NLS system and nonlocal NLS equations.
Main purpose of this paper is to find soliton solutions of the nonlocal mKdV equations of all types. For this purpose
we first find soliton solutions of the coupled mKdV system (\ref{mKdV1}) and (\ref{mKdV2}) by using
the Hirota direct method. Then by using the Ablowitz-Musslimani reductions (\ref{non}) and (\ref{nonsecond}),
we obtain soliton solutions of the nonlocal  complex mKdV (including $T$-, $S$-, and $ST$-symmetric equations) and nonlocal $ST$-symmetric mKdV equations. We
show that there are two different types of one-soliton solution of the reduced mKdV system. We give  the corresponding two- and three-soliton solutions of the
first type. We also present the graphs of some solutions for certain values of the parameters. They include one-, two-, and three-soliton waves, breather and kink-type of waves.

\noindent The lay out of the paper is as follows. In Section 2 we apply Hirota method to the coupled mKdV system (\ref{mKdV1}) and (\ref{mKdV2}) and find soliton solutions. In Section 3 we obtain soliton solutions of the two different local mKdV equations by using local reductions on soliton solutions of the coupled mKdV system. In Section 4 we find soliton solutions of $T$-symmetric, $S$-symmetric, and two different $ST$-symmetric
mKdV equations and we give some examples for one-, two-, and three-soliton solutions together with their graphs.

\section{Hirota Method for the Coupled MKdV System}

\noindent Let $\displaystyle q=\frac{F}{f}$  and $\displaystyle r=\frac{G}{f}$. Equation (\ref{mKdV1})
becomes
\begin{align}
&4aF_tf^3-4aFf_tf^2+F_{xxx}f^3-3F_{xx}f_xf^2+6F_xf_x^2f-3F_xf_{xx}f^2\nonumber\\
&-6Ff_x^3+6Ff_xf_{xx}f-Ff_{xxx}f^2-6GFF_xf+6GF^2f_x=0,
\end{align}
which is equivalent to
\begin{equation*}
f^2(4aD_t+D_x^3)F\cdot f-3(D_x^2f\cdot f+2GF)(D_x F\cdot f)=0.
\end{equation*}
Similarly, the equation (\ref{mKdV2}) can be written as
\begin{equation*}
f^2(4aD_t+D_x^3)G\cdot f-3(D_x^2f\cdot f+2GF)(D_x G\cdot f)=0.
\end{equation*}
Hence the Hirota bilinear form of the mKdV system is
\begin{align}
& P_1(D)\{F\cdot f\}=(4aD_t+D_x^3-3 \alpha D_{x})\{F\cdot f\}=0\label{mkdvH1}\\
& P_2(D)\{G\cdot f\}=(4aD_t+D_x^3-3 \alpha D_{x})\{G\cdot f\}=0\label{mkdvH2}\\
& P_3(D)\{f\cdot f\}=(D_x^2-\alpha) \{f\cdot f\}=-2GF,\label{mkdvH3}
\end{align}
where $\alpha$ is an arbitrary constant. Note that for mKdV system we obtain similar solutions as in the nonlinear Schr\"{o}dinger
system case \cite{GurPek}.

\subsection{One-Soliton Solution of the MKdV System}
To find one-soliton solution we use the following expansions for the functions $F$, $G$, and $f$,
\begin{equation}\label{expansionONE}
F=\varepsilon F_1, \quad G=\varepsilon G_1, \quad f=1+\varepsilon^2 f_2,
\end{equation}
where
\begin{equation}\label{MKdVF1G1}
F_1=e^{\theta_1}, \quad G_1=e^{\theta_2}, \quad \theta_i=k_ix+\omega_i t+\delta_i,\, i= 1, 2.
\end{equation}
We insert these expansions into (\ref{mkdvH1})-(\ref{mkdvH3}). The coefficient of $\varepsilon^0$ gives
\begin{equation}
(D_x^2-\alpha)\{1\cdot 1\}=0
\end{equation}
yielding that $\alpha=0$. Hence the coefficients of $\varepsilon^n$, $1\leq n \leq 4$ becomes
\begin{align}
\varepsilon^1:&\, 4aF_{1,t}+F_{1,xxx}=0,\label{d1}\\
&\, 4aG_{1,t}+G_{1,xxx}=0,\label{d2}\\
\varepsilon^2:&\, f_{2,xx}+F_1G_1=0,\label{d3}\\
\varepsilon^3:&\, 4a(F_{1,t}f_2-F_1f_{2,t})+F_{1,xxx}f_2-3F_{1.xx}f_{2,x}+3F_{1,x}f_{2,xx}-F_1f_{2,xxx}=0,\\
&\, 4a(G_{1,t}f_2-G_1f_{2,t})+G_{1,xxx}f_2-3G_{1,xx}f_{2,x}+3G_{1,x}f_{2,xx}-G_1f_{2,xxx}=0,\\
\varepsilon^4:&\, f_2f_{2,xx}-f_{2,x}^2=0.
\end{align}
The equation (\ref{d1}) and (\ref{d2}) give the dispersion relations
\begin{equation}\label{dispersionONE}\displaystyle
\omega_i=-\frac{k_i^3}{4a}, \quad i=1, 2.
\end{equation}
From the coefficient of $\varepsilon^2$ we obtain the function $f_2$ as
\begin{equation}\label{f_2ONE}
\displaystyle f_2=-\frac{ e^{(k_1+k_2)x+(\omega_1+\omega_2)t+\delta_1+\delta_2} }{(k_1+k_2)^2}.
\end{equation}
The coefficients of $\varepsilon^3$ and $\varepsilon^4$ vanish directly by the dispersion relation (\ref{dispersionONE}) and (\ref{f_2ONE}). Take $\varepsilon=1$. Hence a pair of solutions of the mKdV system (\ref{mKdV1}) and (\ref{mKdV2})
is given by $(q(t,x),r(t,x))$ where
\begin{equation}\label{mKdVsystemonesol}
\displaystyle q(t,x)=\frac{e^{\theta_1}}{1+Ae^{\theta_1+\theta_2}}, \quad \quad r(t,x)=\frac{e^{\theta_2}}{1+Ae^{\theta_1+\theta_2}},
\end{equation}
with $\displaystyle \theta_i=k_ix-\frac{k_i^3}{4a}t+\delta_i$, $i=1, 2$, and $\displaystyle A=-\frac{1}{(k_1+k_2)^2}$. Here $k_{1}$, $k_{2}$, $\delta_{1},$ and $\delta_{2}$ are arbitrary complex numbers.

\subsection{Two-Soliton Solution of the MKdV System}

For two-soliton solution, we take
\begin{equation}\label{expansionmKdV2}
F=\varepsilon F_1+\varepsilon^3 F_3, \quad G=\varepsilon G_1+\varepsilon^3 G_3, \quad f=1+\varepsilon^2 f_2+\varepsilon^4 f_4,
\end{equation}
where
\begin{equation}\label{funcmKdV2}
F_1=e^{\theta_1}+e^{\theta_2}, \quad G_1=e^{\eta_1}+e^{\eta_2},
\end{equation}
with $\theta_i=k_i x+\omega_i t+\delta_i$, $\eta_i=\ell_ix+m_it+\alpha_i$ for $i=1, 2$. When we insert above expansions into (\ref{mkdvH1})-(\ref{mkdvH3}),
we get the coefficients of $\varepsilon^n$, $1\leq n \leq 8$ as
\begin{align}
\varepsilon:&\, 4aF_{1,t}+F_{1,xxx}=0,\label{twod1} \\
&\, 4aG_{1,t}+G_{1,xxx}=0,\label{twod2} \\
\varepsilon^2:&\, f_{2,xx}+G_1F_1=0,\label{twod3}\\
\varepsilon^3:&\, 4a(F_{1,t}f_2-F_1f_{2,t})+F_{1,xxx}f_2-3F_{1,xx}f_{2,x}+3F_{1,x}f_{2,xx}-F_1f_{2,xxx}\nonumber\\
&\, +4aF_{3,t}+F_{3,xxx}=0\label{twod4},\\
&\, 4a(G_{1,t}f_2-G_1f_{2,t})+G_{1,xxx}f_2-3G_{1,xx}f_{2,x}+3G_{1,x}f_{2,xx}-G_1f_{2,xxx}\nonumber\\
&\, +4aG_{3,t}+G_{3,xxx}=0,\label{twod5}\\
\varepsilon^4:&\, f_2f_{2,xx}-f_{2,x}^2+f_{4,xx}+G_1F_3+G_3F_1=0,\label{twod6}\\
\varepsilon^5:&\,  4a(F_{3,t}f_2-F_3f_{2,t})+F_{3,xxx}f_2-3F_{3,xx}f_{2,x}+3F_{3,x}f_{2,xx}-F_3f_{2,xxx}\nonumber\\
&\,+4a(F_{1,t}f_4-F_1f_{4,t})+F_{1,xxx}f_4-3F_{1,xx}f_{4,x}+3F_{1,x}f_{4,xx}-F_1f_{4,xxx}=0,\label{twod7}\\
&\, 4a(G_{3,t}f_2-G_3f_{2,t})+G_{3,xxx}f_2-3G_{3,xx}f_{2,x}+3G_{3,x}f_{2,xx}-G_3f_{2,xxx}\nonumber\\
&\,+4a(G_{1,t}f_4-G_1f_{4,t})+G_{1,xxx}f_4-3G_{1,xx}f_{4,x}+3G_{1,x}f_{4,xx}-G_1f_{4,xxx}=0,\label{twod8}
\end{align}
\begin{align}
\varepsilon^6:&\, f_{2,xx}f_4-2f_{2,x}f_{4,x}+f_2f_{4,xx}+G_3F_3=0,\label{twod9}\\
\varepsilon^7:&\, 4a(F_{3,t}f_4-F_3f_{4,t})+F_{3,xxx}f_4-3F_{3,xx}f_{4,x}+3F_{3,x}f_{4,xx}-F_3f_{4,xxx}=0,\label{twod10}\\
&\, 4a(G_{3,t}f_4-G_3f_{4,t})+G_{3,xxx}f_4-3G_{3,xx}f_{4,x}+3G_{3,x}f_{4,xx}-G_3f_{4,xxx}=0,\label{twod11}\\
\varepsilon^8:&\, f_4f_{4,xx}-f_{4,x}^2=0.\label{twod12}
\end{align}
The equations (\ref{twod1}) and (\ref{twod2}) give the dispersion relations
\begin{equation}\label{dispersiontwo}\displaystyle
\omega_i=-\frac{k_i^3}{4a}, \quad m_i=-\frac{\ell_i^3}{4a}, \quad i=1, 2.
\end{equation}
From the coefficient of $\varepsilon^2$ we obtain the function $f_2$,
\begin{equation}\displaystyle
f_2= e^{\theta_1+\eta_1+\alpha_{11}}+e^{\theta_1+\eta_2+\alpha_{12}}+e^{\theta_2+\eta_1+\alpha_{21}}+e^{\theta_2+\eta_2+\alpha_{22}}=\sum_{1\leq i,j\leq 2}e^{\theta_i+\eta_j+\alpha_{ij}},
\end{equation}
where
\begin{equation}
\displaystyle e^{\alpha_{ij}}=-\frac{1}{(k_i+\ell_j)^2},\, 1\leq i,j\leq 2.
\end{equation}
The equations (\ref{twod4}) and (\ref{twod5}) give the functions $F_3$ and $G_3$,
\begin{equation}\label{F_3G_3}
F_3=\gamma_1e^{\theta_1+\theta_2+\eta_1}+\gamma_2e^{\theta_1+\theta_2+\eta_2},\quad G_3=\beta_1e^{\theta_1+\eta_1+\eta_2}+\beta_2e^{\theta_2+\eta_1+\eta_2},
\end{equation}
where
\begin{equation}\label{gamma_ibeta_i}
\displaystyle \gamma_i=-\frac{(k_1-k_2)^2}{(k_1+\ell_i)^2(k_2+\ell_i)^2}, \quad \beta_i=-\frac{(\ell_1-\ell_2)^2}{(\ell_1+k_i)^2(\ell_2+k_i)^2},\, i=1, 2.
\end{equation}

\noindent Finally, the equation (\ref{twod6}) yields the function $f_4$ as
\begin{equation}
f_4=Me^{\theta_1+\theta_2+\eta_1+\eta_2},
\end{equation}
where
\begin{equation}\label{M}
\displaystyle M=\frac{(k_1-k_2)^2(l_1-l_2)^2}{(k_1+l_1)^2(k_1+l_2)^2(k_2+l_1)^2(k_2+l_2)^2}.
\end{equation}
Other equations (\ref{twod7})-(\ref{twod12}) vanish directly by the dispersion relations (\ref{dispersiontwo}) and the functions
$f_2, f_4, F_3$, and $G_3$.

\noindent Let us also take $\varepsilon=1$. Then two-soliton solution of the
mKdV system (\ref{mKdV1}) and (\ref{mKdV2}) is given with the pair $(q(t,x),r(t,x))$,
\begin{align}
\displaystyle& q(t,x)=\frac{e^{\theta_1}+e^{\theta_2}+\gamma_1e^{\theta_1+\theta_2+\eta_1}+\gamma_2e^{\theta_1+\theta_2+\eta_2}}
{1+e^{\theta_1+\eta_1+\alpha_{11}}+e^{\theta_1+\eta_2+\alpha_{12}}+e^{\theta_2+\eta_1+\alpha_{21}}+e^{\theta_2+\eta_2+\alpha_{22}}
+Me^{\theta_1+\theta_2+\eta_1+\eta_2}},\label{mKdVtwosolq(t,x)}\\
&r(t,x)=\frac{e^{\eta_1}+e^{\eta_2}+\beta_1e^{\theta_1+\eta_1+\eta_2}+\beta_2e^{\theta_2+\eta_1+\eta_2}}
{1+e^{\theta_1+\eta_1+\alpha_{11}}+e^{\theta_1+\eta_2+\alpha_{12}}+e^{\theta_2+\eta_1+\alpha_{21}}+e^{\theta_2+\eta_2+\alpha_{22}}
+Me^{\theta_1+\theta_2+\eta_1+\eta_2}},\label{mKdVtwosolr(t,x)}
\end{align}
with $\displaystyle \theta_i=k_ix-\frac{k_i^3}{4a}t+\delta_i$, $\displaystyle \eta_i=\ell_ix-\frac{\ell_i^3}{4a}t+\alpha_i$ for $i=1, 2$. Here $k_{i}$, $\ell_{i}, \delta_{i}$, and $\alpha_{i}$, $i=1, 2$ are arbitrary complex numbers.

\subsection{Three-Soliton Solution of the MKdV System}

To find three-soliton solution, we take
\begin{equation}\label{expansionMKdV3}
f=1+\varepsilon^2 f_2+\varepsilon^4 f_4+\varepsilon^6 f_6,\quad G=\varepsilon G_1+\varepsilon^3 G_3+\varepsilon^5 G_5,\quad F=\varepsilon F_1+\varepsilon^3 F_3+\varepsilon^5 F_5,
\end{equation}
and
\begin{equation}\label{threeF1G1MKdV}
F_1=e^{\theta_1}+e^{\theta_2}+e^{\theta_3}, \quad G_1=e^{\eta_1}+e^{\eta_2}+e^{\eta_3},
\end{equation}
where $\theta_i=k_i x+\omega_i t+\delta_i$, $\eta_i=\ell_ix+m_it+\alpha_i$ for $i=1, 2, 3$. Inserting (\ref{expansionMKdV3}) into (\ref{mkdvH1})-(\ref{mkdvH3})
give us the coefficients of $\varepsilon^n$, $1\leq n\leq 12$ as
\begin{align}
\varepsilon:&\, 4aF_{1,t}+F_{1,xxx}=0,\label{threed1}\\
&\, 4aG_{1,t}+G_{1,xxx}=0,\label{threed2}\\
\varepsilon^2:&\, f_{2,xx}+G_1F_1=0,\label{threed3}\\
\varepsilon^3:&\,4a(F_{1,t}f_2-F_1f_{2,t})+F_{1,xxx}f_2-3F_{1,xx}f_{2,x}+3F_{1,x}f_{2,xx}-F_1f_{2,xxx}\nonumber\\
&+4aF_{3,t}+F_{3,xxx}=0,\label{threed4}\\
&\, 4a(G_{1,t}f_2-G_1f_{2,t})+G_{1,xxx}f_2-3G_{1,xx}f_{2,x}+3G_{1,x}f_{2,xx}-G_1f_{2,xxx}\nonumber\\
&+4aG_{3,t}+G_{3,xxx}=0,\label{threed5}\\
\varepsilon^4:&\, f_{4,xx}+f_2f_{2,xx}-f_{2,x}^2+G_1F_3+G_3F_1=0,\label{threed6}\\
\varepsilon^5:&\,4aF_{5,t}+F_{5,xxx}+4a(F_{3,t}f_2-F_3f_{2,t})+F_{3,xxx}f_2-3F_{3,xx}f_{2,x}+3F_{3,x}f_{2,xx}-F_3f_{2,xxx}\nonumber\\
&+4a(F_{1,t}f_4-F_1f_{4,t})+F_{1,xxx}f_4-3F_{1,xx}f_{4,x}+3F_{1,x}f_{4,xx}-F_1f_{4,xxx}=0,\label{threed7}\\
&\, 4aG_{5,t}+G_{5,xxx}+4a(G_{3,t}f_2-G_3f_{2,t})+G_{3,xxx}f_2-3G_{3,xx}f_{2,x}+3G_{3,x}f_{2,xx}-G_3f_{2,xxx}\nonumber\\
&+4a(G_{1,t}f_4-G_1f_{4,t})+G_{1,xxx}f_4-3G_{1,xx}f_{4,x}+3G_{1,x}f_{4,xx}-G_1f_{4,xxx}=0,\label{threed8}\\
\varepsilon^6:&\, f_{2,xx}f_4-2f_{2,x}f_{4,x}+f_2f_{4,xx}+f_{6,xx}+G_5F_1+G_1F_5+G_3F_3=0,\label{threed9}\\
\varepsilon^7:&\,4a(F_{3,t}f_4-F_3f_{4,t})+F_{3,xxx}f_4-3F_{3,xx}f_{4,x}+3F_{3,x}f_{4,xx}-F_3f_{4,xxx}\nonumber\\
&+4a(F_{5,t}f_2-F_5f_{2,t})+F_{5,xxx}f_2-3F_{5,xx}f_{2,x}+3F_{5,x}f_{2,xx}-F_5f_{2,xxx}\nonumber\\
&+4a(F_{1,t}f_6-F_1f_{6,t})+F_{1,xxx}f_6-3F_{1,xx}f_{6,x}+3F_{1,x}f_{6,xx}-F_1f_{6,xxx}=0,\label{threed10}\\
&\,4a(G_{3,t}f_4-G_3f_{4,t})+G_{3,xxx}f_4-3G_{3,xx}f_{4,x}+3G_{3,x}f_{4,xx}-G_3f_{4,xxx}\nonumber\\
&+4a(G_{5,t}f_2-G_5f_{2,t})+G_{5,xxx}f_2-3G_{5,xx}f_{2,x}+3G_{5,x}f_{2,xx}-G_5f_{2,xxx}\nonumber\\
&+4a(G_{1,t}f_6-G_1f_{6,t})+G_{1,xxx}f_6-3G_{1,xx}f_{6,x}+3G_{1,x}f_{6,xx}-G_1f_{6,xxx}=0,\label{threed11}\\
\varepsilon^8:&\, f_{2,xx}f_6-2f_{2,x}f_{6,x}+f_2f_{6,xx}+f_4f_{4,xx}-f_{4,x}^2+G_3F_5+G_5F_3=0,\label{threed12}\\
\varepsilon^9:&\, 4a(F_{3,t}f_6-F_3f_{6,t})+F_{3,xxx}f_6-3F_{3,xx}f_{6,x}+3F_{3,x}f_{6,xx}-F_3f_{6,xxx}\nonumber\\
&+4a(F_{5,t}f_4-F_5f_{4,t})+F_{5,xxx}f_4-3F_{5,xx}f_{4,x}+3F_{5,x}f_{4,xx}-F_5f_{4,xxx}=0, \label{threed13}\\
&\, 4a(G_{3,t}f_6-G_3f_{6,t})+G_{3,xxx}f_6-3G_{3,xx}f_{6,x}+3G_{3,x}f_{6,xx}-G_3f_{6,xxx}\nonumber\\
&+4a(G_{5,t}f_4-G_5f_{4,t})+G_{5,xxx}f_4-3G_{5,xx}f_{4,x}+3G_{5,x}f_{4,xx}-G_5f_{4,xxx}=0, \label{threed14}
\end{align}
\begin{align}
\varepsilon^{10}:&\, f_{4,xx}f_6-2f_{4,x}f_{6,x}+f_4f_{6,xx}+G_5F_5=0,\label{threed15}\\
\varepsilon^{11}:&\, 4a(F_{5,t}f_6-F_5f_{6,t})+F_{5,xxx}f_6-3F_{5,xx}f_{6,x}+3F_{5,x}f_{6,xx}-F_5f_{6,xxx}=0,\label{threed16}\\
&\,  4a(G_{5,t}f_6-G_5f_{6,t})+G_{5,xxx}f_6-3G_{5,xx}f_{6,x}+3G_{5,x}f_{6,xx}-G_5f_{6,xxx}=0,\label{threed17}\\
\varepsilon^{12}:&\, f_6f_{6,xx}-f_{6,x}^2=0.\label{threed18}
\end{align}
This part is also very similar to the part related to three-soliton solution of the coupled NLS system given in \cite{GurPek}. From (\ref{threed1}) and
(\ref{threed2}) we get the dispersion relations
\begin{equation}\displaystyle
\omega_i=-\frac{k_i^3}{4a},\quad m_i=-\frac{\ell_i^3}{4a}, i=1, 2, 3.
\end{equation}

\noindent The coefficient of $\varepsilon^2$ gives the function $f_2$ as
\begin{equation}\displaystyle
f_2=\sum_{1\leq i, j\leq 3} e^{\theta_i+\eta_j+\alpha_{ij}}, \quad e^{\alpha_{ij}}=-\frac{1}{(k_i+\ell_j)^2}, 1\leq i, j\leq 3.
\end{equation}
From the equations (\ref{threed4}) and (\ref{threed5}), we obtain the functions $F_3$ and $G_3$
\begin{align}\displaystyle
&F_3=\sum_{\substack{1\leq i,j,s\leq 3 \\ i<j}} A_{ijs}e^{\theta_i+\theta_j+\eta_s}, \quad A_{ijs}=-\frac{(k_i-k_j)^2}{(k_i+\ell_s)^2(k_j+\ell_s)^2}, 1\leq i,j,s\leq 3, i<j,\\
&G_3=\sum_{\substack{1\leq i,j,s\leq 3 \\ i<j}} B_{ijs}e^{\eta_i+\eta_j+\theta_s}, \quad B_{ijs}=-\frac{(\ell_i-\ell_j)^2}{(\ell_i+k_s)^2(\ell_j+k_s)^2}, 1\leq i,j,s\leq 3, i<j.
\end{align}
We get the function $f_4$ from the coefficient of $\varepsilon^4$ as
\begin{equation}
f_4=\sum_{\substack{1\leq i<j\leq 3 \\ 1\leq p<r \leq3}} M_{ijpr}e^{\theta_i+\theta_j+\eta_p+\eta_r},
\end{equation}
where
\begin{equation}\displaystyle
M_{ijpr}=\frac{(k_i-k_j)^2(l_p-l_r)^2}{(k_i+l_p)^2(k_i+l_r)^2(k_j+l_p)^2(k_j+l_r)^2},
\end{equation}
for $1\leq i< j\leq 3$, $1\leq p< r\leq 3$.
From (\ref{threed7}) and (\ref{threed8}) we obtain the functions $F_5$ and $G_5$,
\begin{align}
&F_5=V_{12}e^{\theta_1+\theta_2+\theta_3+\eta_1+\eta_2}+V_{13}e^{\theta_1+\theta_2+\theta_3+\eta_1+\eta_3}+V_{23}e^{\theta_1+\theta_2+\theta_3+\eta_2+\eta_3},\\
&G_5=W_{12}e^{\theta_1+\theta_2+\eta_1+\eta_2+\eta_3}+W_{13}e^{\theta_1+\theta_2+\eta_1+\eta_2+\eta_3}+W_{23}e^{\theta_2+\theta_3+\eta_1+\eta_2+\eta_3},
\end{align}
where
\begin{eqnarray}\displaystyle
&&V_{ij}=\frac{S_{ij}}{4a(\omega_1+\omega_2+\omega_3+m_i+m_j)+(k_1+k_2+k_3+\ell_i+\ell_j)^3},\\
 &&W_{ij}=\frac{Q_{ij}}{4a(\omega_i+\omega_j+m_1+m_2+m_3)+(k_i+k_j+\ell_1+\ell_2+\ell_3)^3},
\end{eqnarray}
for $1\leq i<j\leq 3$. Here $S_{ij}$ and $Q_{ij}$ are given in Appendix.
The equation (\ref{threed9}) gives the function $f_6$
\begin{equation}\displaystyle
f_6=He^{\theta_1+\theta_2+\theta_3+\eta_1+\eta_2+\eta_3},
\end{equation}
where the coefficient $H$ is represented in Appendix. The rest of the equations (\ref{threed10})-(\ref{threed18})
are satisfied directly. Let us also take $\varepsilon=1$. Hence three-soliton solution of the coupled mKdV system (\ref{mKdV1}) and (\ref{mKdV2}) is given with the pair $(q(t,x),r(t,x))$ where
\begin{align}\displaystyle
q(t,x)&=\frac{e^{\theta_1}+e^{\theta_2}+e^{\theta_3}+\sum_{\substack{1\leq i,j,s\leq 3 \\ i<j}} A_{ijs}e^{\theta_i+\theta_j+\eta_s}
+\sum_{\substack{1\leq i,j\leq 3 \\ i<j}}V_{ij}e^{\theta_1+\theta_2+\theta_3+\eta_i+\eta_j}}{1+\sum_{1\leq i,j\leq3}e^{\theta_i+\eta_j+\alpha_{ij}}+\sum_{\substack{1\leq i<j\leq 3 \\ 1\leq p<r \leq3}} M_{ijpr}e^{\theta_i+\theta_j+\eta_p+\eta_r}+He^{\theta_1+\theta_2+\theta_3+\eta_1+\eta_2+\eta_3}}, \label{mKdV3solq(t,x)}\\
r(t,x)&=\frac{e^{\eta_1}+e^{\eta_2}+e^{\eta_3}+\sum_{\substack{1\leq i,j,s\leq 3 \\ i<j}} B_{ijs}e^{\eta_i+\eta_j+\theta_s}+
\sum_{\substack{1\leq i,j\leq 3 \\ i<j}}W_{ij}e^{\theta_i+\theta_j+\eta_1+\eta_2+\eta_3}}{1+\sum_{1\leq i,j\leq3}e^{\theta_i+\eta_j+\alpha_{ij}}+\sum_{\substack{1\leq i<j\leq 3 \\ 1\leq p<r \leq3}} M_{ijpr}e^{\theta_i+\theta_j+\eta_p+\eta_r}+He^{\theta_1+\theta_2+\theta_3+\eta_1+\eta_2+\eta_3}}.\label{mKdV3solr(t,x)}
\end{align}

Having obtained the one-, two-, and three-soliton solutions of the mKdV system we now ready to obtain such
soliton solutions of the local and nonlocal reductions of the mKdV system.

\section{Local Reductions of the MKdV System}

In this section we will consider the local reductions; the reduction (\ref{non0}) of the mKdV system (\ref{mKdV1}) and (\ref{mKdV2}) yielding the
equation (\ref{mKdV3}) with the condition
\begin{equation}\label{standardcond1}
\bar{a}=a
\end{equation}
satisfied, and the reduction (\ref{non00}) giving the usual mKdV equation (\ref{usualmKdV})
with no condition on the constant $a$ is provided.

\subsection{One-Soliton Solution for the Local CMKdV Equation:\\$(r=k\bar{q}(t,x))$}

Firstly, we obtain the conditions on the parameters of one-soliton solution of the mKdV system to satisfy the equality (\ref{non0}) i.e.,
\begin{equation}\label{stan1one}
\displaystyle \frac{e^{k_2x-\frac{k_2^3}{4a}t+\delta_2}}{1+Ae^{(k_1+k_2)x-\frac{(k_1^3+k_2^3)}{4a}t+\delta_1+\delta_2  }}
=k\frac{e^{\bar{k}_1x-\frac{\bar{k}_1^3}{4\bar{a}}t+\bar{\delta}_1}}
{1+\bar{A}e^{(\bar{k}_1+\bar{k}_2)x-\frac{(\bar{k}_1^3+\bar{k}_2^3)}{4\bar{a}}t+\bar{\delta}_1+\bar{\delta}_2}}.
\end{equation}

\noindent Here we will present two types of solutions for the parameters satisfying the above equation.

 \subsubsection{Type 1.}

Let us choose the parameters satisfying the following conditions:
\begin{align}\label{stanoneconds}
\displaystyle &i)\, k_2=\bar{k}_1,\quad  ii)\,\frac{k_2^3}{4a}=\frac{\bar{k}_1^3}{4\bar{a}},\quad iii)\, e^{\delta_2}=ke^{\bar{\delta}_1},\quad
iv)\, \bar{A}=A,\nonumber\\
 &v)\,k_1+k_2=\bar{k}_1+\bar{k}_2,\quad
 vi)\,\frac{(k_1^3+k_2^3)}{4a}=\frac{(\bar{k}_1^3+\bar{k}_2^3)}{4\bar{a}},\quad vii)\, e^{\delta_1+\delta_2}=e^{\bar{\delta}_1+\bar{\delta}_2}.
\end{align}
Clearly, the conditions $ii)$, $iv)$, $v)$, and $vi)$ are satisfied directly by $i)$ and (\ref{standardcond1}). Now use the condition $iii)$ in $vii)$. Note also that since $k$ is a real constant we have $\bar{k}=k$. Hence we have
\begin{equation*}
e^{\delta_1+\delta_2}=ke^{\delta_1}e^{\bar{\delta}_1}\quad \mathrm{and} \quad e^{\bar{\delta}_1+\bar{\delta}_2}=ke^{\bar{\delta}_1}e^{\delta_1},
\end{equation*}
yielding the equality $e^{\delta_1+\delta_2}=e^{\bar{\delta}_1+\bar{\delta}_2}$.

\noindent Therefore in this type the parameters of
one-soliton solution of the equation (\ref{mKdV3}) must have the following properties:
\begin{equation}
1)\,\bar{a}=a,\quad 2)\,k_2=\bar{k}_1,\quad 3)\, e^{\delta_2}=ke^{\bar{\delta}_1}.
\end{equation}
In general we have four arbitrary parameters left in the solution. If we let $k_1=\alpha+i\beta$, $\alpha, \beta \in \mathbb{R}$, then one-soliton solution of (\ref{mKdV3}) becomes
\begin{equation}\displaystyle
q(t,x)=\frac{e^{(\alpha+i\beta)x-\frac{(\alpha^3-3\alpha\beta^2)+i(3\alpha^2\beta-\beta^3)}{4a}t+\delta_1}}
{1-\frac{k}{4\alpha^2}e^{2\alpha x+\frac{(3\alpha\beta^2-\alpha^3)}{2a}t+\delta_1+\bar{\delta}_1}},
\end{equation}
so
\begin{equation}\label{localone1}\displaystyle
|q(t,x)|^2=\frac{e^{2\alpha x-\frac{(\alpha^3-3\alpha\beta^2)}{2a}t+\delta_1+\bar{\delta}_1}}{(1-\frac{k}{4\alpha^2}e^{2\alpha x+\frac{(3\alpha\beta^2-\alpha^3)}{2a}t+\delta_1+\bar{\delta}_1})^2}.
\end{equation}
For $k< 0$, the solution (\ref{localone1}) can be written as
\begin{equation}\displaystyle
|q(t,x)|^2=-\frac{\alpha^2}{k}\mathrm{sech}^2(\alpha x+\frac{(3\alpha \beta^2-\alpha^3)}{4a}t+\frac{\delta_1+\bar{\delta}_1}{2}+\delta),
\end{equation}
where $\delta=\frac{1}{2}\ln(-\frac{k}{4\alpha^2})$. The above solution is non-singular. Consider the following example.

\noindent \textbf{Example 1.}\, For the parameters $(k_1, k_2, e^{\delta_1}, e^{\delta_2}, k, a)=(1+\frac{i}{2}, 1-\frac{i}{2}, 1+i, -1+i, -1, 2)$, the solution becomes
\begin{equation}\displaystyle
q(t,x)=\frac{(1+i)e^{(1+\frac{1}{2}i)x-(\frac{1}{32}+\frac{11}{64}i)t}}{1+\frac{1}{2}e^{2x-\frac{1}{16}t}},
\end{equation}
so
\begin{equation}\label{standardone}\displaystyle
|q(t,x)|^2=\frac{2e^{2x-\frac{1}{16}t}}{(1+\frac{1}{2}e^{2x-\frac{1}{16}t})^2}=\mathrm{sech}^2(x-\frac{1}{8}t+\delta),
\end{equation}
where $\delta=-\frac{\ln 2}{2}$. The graph of the solution (\ref{standardone}) is given in Figure 1.
\begin{center}
\begin{figure}[h]
\centering
\begin{minipage}[t]{1\linewidth}
\centering
\includegraphics[angle=0,scale=.24]{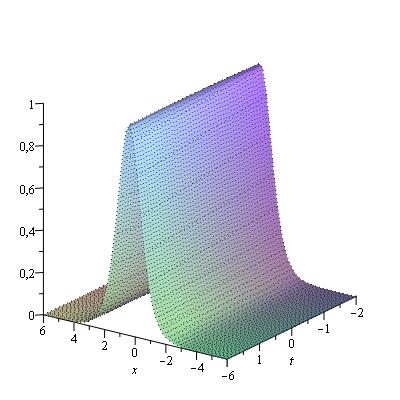}
\caption{One-soliton solution for (\ref{localone1}) with the parameters $k_1=\alpha+i\beta=1+\frac{i}{2}$, $e^{\delta_1}=1+i$, $k=-1$, $a=2$.}
\end{minipage}\hfill
\end{figure}
\end{center}

\subsubsection{Type 2.}

Consider the cross multiplication in (\ref{stan1one}). We have
\begin{align}
e^{k_2x-\frac{k_2^3}{4a}t+\delta_2}&+\bar{A}e^{(\bar{k}_1+\bar{k}_2+k_2)x-\Big[\frac{\bar{k}_1^3+\bar{k}_2^3}{4\bar{a}}+\frac{k_2^3}{4a}  \Big]t+\bar{\delta}_1+\bar{\delta}_2+\delta_2}\nonumber\\
&=ke^{\bar{k}_1x-\frac{\bar{k}_1^3}{4\bar{a}}t+\bar{\delta}_1}
+Ake^{(k_1+k_2+\bar{k}_1)x-\Big[\frac{k_1^3+k_2^3}{4a}+\frac{\bar{k}_1^3}{4\bar{a}}\Big]t+\delta_1+\delta_2+\bar{\delta}_1}.
\end{align}
Now take the parameters satisfying the following conditions so that the above equality holds:
\begin{align}
\displaystyle
&i)\, k_2=k_1+k_2+\bar{k}_1,\quad ii)\, \frac{k_2^3}{4a}=\frac{k_1^3+k_2^3}{4a}+\frac{\bar{k}_1^3}{4\bar{a}},
\quad iii)\,e^{\delta_2}=Ake^{\delta_1+\delta_2+\bar{\delta}_1}\nonumber\\
&iv)\, \bar{k}_1=\bar{k}_1+\bar{k}_2+k_2,\quad v)\, \frac{\bar{k}_1^3+\bar{k}_2^3}{4\bar{a}}+\frac{k_2^3}{4a}=\frac{\bar{k}_1^3}{4\bar{a}},
\quad vi)\, \bar{A}e^{\bar{\delta}_1+\bar{\delta}_2+\delta_2}=ke^{\bar{\delta}_1}.
\end{align}
These conditions hold when
\begin{equation}\label{cond1}
1)\,\bar{a}=a,\quad 2)\,k_1=-\bar{k}_1,\quad 3)\, k_2=-\bar{k}_2,\quad 4)\, Ake^{\delta_1+\bar{\delta}_1}=1,\quad 5)\, Ae^{\delta_2+\bar{\delta}_2}=k.
\end{equation}
Although there are more conditions in this type, the number of free parameters in the solution is also four. Here let $k_1=\alpha i$, $k_2=\beta i$,
$e^{\delta_1}=a_1+ib_1$, and $e^{\delta_2}=a_2+ib_2$ for $\alpha, \beta, a_j, b_j\in \mathbb{R}$, $j=1, 2$. Therefore one-soliton solution of (\ref{mKdV3})
becomes
\begin{equation}\displaystyle
q(t,x)=\frac{e^{i\alpha x+i\frac{\alpha^3}{4a}t}(a_1+ib_1)}{1+\frac{1}{(\alpha+\beta)^2}e^{i(\alpha+\beta)x+i\frac{(\alpha^3+\beta^3)}{4a}t}(a_1+ib_1)(a_2+ib_2)},
\end{equation}
hence the corresponding function $|q(t,x)|^2$ is
\begin{align}\label{localone2}\displaystyle
&|q(t,x)|^2\nonumber\\
&=\frac{a_1^2+b_1^2}{2+\frac{2}{(\alpha+\beta)^2}
[(a_1a_2-b_1b_2)\cos((\alpha+\beta)x+\frac{(\alpha^3+\beta^3)}{4a}t)-(a_1b_2+a_2b_1)\sin((\alpha+\beta)x+\frac{(\alpha^3+\beta^3)}{4a}t)]},
\end{align}
where $a_1^2+b_1^2=\frac{(\alpha+\beta)^2}{k}$ and $a_2^2+b_2^2=k(\alpha+\beta)^2$. Now let $a_1a_2-b_1b_2=B\cos\omega_0$, $a_1b_2+a_2b_1=B\sin\omega_0$ where $B>0$. Clearly, we have
\begin{equation}
B=\sqrt{(a_1^2+b_1^2)(a_2^2+b_2^2)}=(\alpha+\beta)^2
\end{equation}
by the conditions $4)$ and $5)$ in (\ref{cond1}). Therefore the solution (\ref{localone2}) becomes
\begin{equation}\displaystyle
|q(t,x)|^2=\frac{a_1^2+b_1^2}{1+\cos\theta}=\frac{a_1^2+b_1^2}{4}\sec^2(\frac{\theta}{2}),
\end{equation}
where $\theta=(\alpha+\beta)x+\frac{(\alpha^3+\beta^3)}{4a}t+\omega_0$. This is a singular solution for any choice of the parameters.

\noindent Let us give the following example.

\noindent \textbf{Example 2.}\, Choose the parameters as $(k_1, k_2, e^{\delta_1}, e^{\delta_2}, k, a)=(i, i, i, -4i, 4, \frac{1}{2})$. Then
the solution becomes
\begin{equation}\displaystyle
q(t,x)=\frac{ie^{ix+\frac{1}{2}it}}{1+e^{2ix+it}}
\end{equation}
so the function $|q(t,x)|^2$ is
\begin{equation}\label{solex2}\displaystyle
|q(t,x)|^2=\frac{1}{4}\sec^2(\frac{2x+t}{2}).
\end{equation}
The solution is singular and the wave is of periodical breather type. Its graph is given in Figure 2.
\begin{center}
\begin{figure}[h]
\centering
\begin{minipage}[t]{1\linewidth}
\centering
\includegraphics[angle=0,scale=.23]{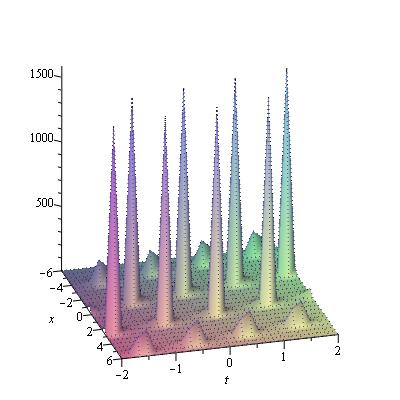}
\caption{A singular periodical breather-type wave for (\ref{localone2}) with the parameters
$k_1=\alpha i=i, k_2=\beta i=i, e^{\delta_1}=a_1+ib_1=i, e^{\delta_2}=a_2+ib_2=-4i, k=4, a=\frac{1}{2}$.}
\end{minipage}\hfill
\end{figure}
\end{center}
\noindent In finding two- and three-soliton solutions we will only use the approach presented in Type 1.

\subsection{Two-Soliton Solution for the Local CMKdV Equation:\\
 $(r=k\bar{q}(t,x))$}

We now use two-soliton solution of the mKdV system given by the pair (\ref{mKdVtwosolq(t,x)})-(\ref{mKdVtwosolr(t,x)}) in the local reduction (\ref{non0})
and obtain the following constraints:
\begin{equation}\label{twostandardmKdVcond}
1)\, \bar{a}=a, \quad 2)\, \ell_i=\bar{k}_i, i=1, 2,\quad 3)\, e^{\alpha_i}=ke^{\bar{\delta}_i}, i=1, 2.
\end{equation}

\noindent \textbf{Example 3.}\, Let us illustrate an example for two-soliton solutions of (\ref{mKdV3}). Choose the parameters as $(k_1,\ell_1,k_2,\ell_2)=(\frac{1}{4},\frac{1}{4},\frac{1}{10},\frac{1}{10})$ with
$(e^{\alpha_j},e^{\delta_j},k,a)=(-1+i,1+i,-1,2)$, $j=1, 2$. In this case we obtain the function $|q(t,x)|^2$ as
\begin{equation}\label{stantwoex2}\displaystyle
|q(t,x)|^2=\frac{2[e^{\frac{1}{4}x-\frac{1}{512}t}+e^{\frac{1}{10}x-\frac{1}{8000}t}+\frac{72}{49}e^{\frac{3}{5}x-\frac{129}{32000}t}
+\frac{450}{49}e^{\frac{9}{20}x-\frac{141}{64000}t}]^2}{[1+8e^{\frac{1}{2}x-\frac{1}{256}t}
+\frac{1600}{49}e^{\frac{7}{20}x-\frac{133}{64000}t}+50e^{\frac{1}{5}x-\frac{1}{4000}t}+\frac{32400}{2401}e^{\frac{7}{10}x-\frac{133}{32000}t}]^2}.
\end{equation}
The graph of (\ref{stantwoex2}) is given in Figure 3.
\begin{center}
\begin{figure}[h]
\centering
\begin{minipage}[t]{1\linewidth}
\centering
\includegraphics[angle=0,scale=.24]{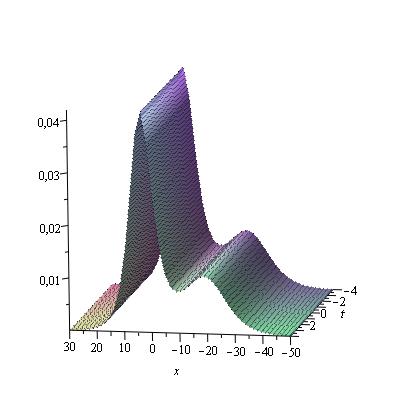}
\caption{Two solitons plotted for (\ref{stantwoex2}) with the parameters $k_1=\ell_1=\frac{1}{4}, k_2=\ell_2=\frac{1}{10},
e^{\alpha_j}=-1+i,e^{\delta_j}=1+i$, $j=1, 2$, $k=-1, a=2$.}
\end{minipage}%
\end{figure}
\end{center}

\subsection{Three-Soliton Solution for the Local CMKdV Equation:\\
 $(r=k\bar{q}(t,x))$}

Similar to previous section we first consider three-soliton solution of the mKdV system with the local reduction (\ref{non0}) and obtain the
following constraints on the parameters:
\begin{equation}\label{threestandardmKdVcond}
1)\, \bar{a}=a, \quad 2)\, \ell_i=\bar{k}_i, i=1, 2, 3, \quad 3)\, e^{\alpha_i}=ke^{\bar{\delta}_i}, i=1, 2, 3.
\end{equation}

\noindent \textbf{Example 4.}\, Now we present an example for three-soliton solution of the local mKdV equation (\ref{mKdV3}).
Take $(k_1, \ell_1, k_2, \ell_2, k_3, \ell_3)=(\frac{1}{4},\frac{1}{4},-\frac{1}{2},-\frac{1}{2},-\frac{1}{6},-\frac{1}{6})$
with $(e^{\alpha_j},e^{\delta_j},k ,a)=(-1+i,1+i,-1,10)$, $j=1, 2, 3$. In this case we have
\begin{equation}\label{solex4}\displaystyle
|q(t,x)|^2=\frac{A}{B},
\end{equation}
where
\begin{align*}
A=&2\Big[e^{\frac{1}{4}x-\frac{1}{2560}t}+e^{-\frac{1}{2}x+\frac{1}{320}t}+e^{-\frac{1}{6}x+\frac{1}{8640}t}+72e^{\frac{3}{1280}t}+18e^{-\frac{3}{4}x+\frac{3}{512}t}\\
&+889e^{-\frac{5}{12}x+\frac{197}{69120}t}+200e^{\frac{1}{3}x-\frac{23}{34560}t}+450e^{-\frac{1}{2}x-\frac{11}{69120}t}+\frac{1}{2}e^{-\frac{7}{6}x+\frac{11}{1728}t}\\
&+\frac{9}{2}e^{-\frac{5}{6}x+\frac{29}{8640}t}+5400e^{-\frac{2}{3}x+\frac{193}{34560}t}+202500e^{-\frac{1}{3}x+\frac{89}{34560}t}
+\frac{2025}{4}e^{-\frac{13}{12}x+\frac{421}{69120}t}\Big]^2
\end{align*}
and
\begin{align*}
B=&\Big[1+9e^{-\frac{2}{3}x+\frac{7}{2160}t}+576e^{\frac{1}{2}x-\frac{19}{69120}t}+64e^{-\frac{1}{4}x+\frac{7}{2560}t}+16200e^{-\frac{1}{6}x+\frac{17}{6912}t}\\
&+18e^{-\frac{1}{3}x+\frac{1}{4320}t}+7200e^{-\frac{7}{12}x+\frac{41}{13824}t}+\frac{9}{4}e^{-\frac{4}{3}x+\frac{7}{1080}t}+90000e^{\frac{1}{6}x-\frac{19}{34560}t}\\
&+2592e^{-\frac{11}{12}x+\frac{413}{69120}t}+1296e^{-\frac{1}{2}x+\frac{7}{1280}t}+2e^{-x+\frac{1}{160}t}+8e^{\frac{1}{2}x-\frac{1}{1280}t}
+903474e^{-\frac{5}{6}x+\frac{197}{34560}t}\Big]^2.
\end{align*}
The graph of the above solution is given in Figure 4.
\begin{center}
\begin{figure}[h]
\centering
\begin{minipage}[t]{1\linewidth}
\centering
\includegraphics[angle=0,scale=.24]{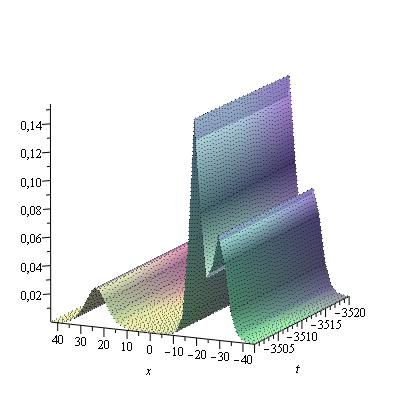}
\caption{Three solitons plotted for (\ref{solex4}) with the parameters $k_1=\ell_1=\frac{1}{4}, k_2=\ell_2=-\frac{1}{2},
k_3=\ell_3=-\frac{1}{6}, e^{\alpha_j}=-1+i, e^{\delta_j}=1+i, j=1, 2, 3$, $k=-1, a=10$.}
\end{minipage}%
\end{figure}
\end{center}

\subsection{One-Soliton Solution for the Local MKdV Equation:\\
 $(r=kq(t,x))$}

In Section 3.1, we gave two types of solutions to find one-soliton solution for local cmKdV equation (\ref{mKdV3}), but for local mKdV equation (\ref{usualmKdV}), Type 2 gives $k_1=k_2=0$ yielding trivial solution. Hence we will only present Type 1 here. When we consider the relation $r(t,x)=kq(t,x)$ for one-soliton solution of the mKdV system (\ref{mKdV1}) and (\ref{mKdV2}) we obtain the
following constraints:
\begin{equation}
1) \, k_1=k_2,\quad 2)\, e^{\delta_2}=ke^{\delta_1}.
\end{equation}
\noindent Let $k_1=k_2=\alpha+i\beta$ and $e^{\delta_1}=a_1+ib_1$, $\alpha, \beta, a_1, b_1\in \mathbb{R}$. Then one-soliton solution of (\ref{usualmKdV}) becomes
\begin{equation}\displaystyle
q(t,x)=\frac{e^{(\alpha+i\beta)x-\frac{(\alpha^3-3\alpha\beta^2)+i(3\alpha^2\beta-\beta^3)}{4a}t}(a_1+ib_1)}
{1-\frac{k}{4(\alpha^2+\beta^2)^2}e^{2(\alpha+i\beta)x-\frac{(\alpha^3-3\alpha\beta^2)+i(3\alpha^2\beta-\beta^3)}{2a}t}(a_1+ib_1)^2(\alpha-i\beta)^2}.
\end{equation}
Therefore we obtain the function
\begin{equation}\label{localone3}\displaystyle
|q(t,x)|^2=\frac{Y}{W},
\end{equation}
where
\begin{align}\displaystyle
Y=&e^{2\alpha x-\frac{(\alpha^3-3\alpha\beta^2)}{2a}t}(a_1^2+b_1^2)\nonumber\\
W=&1-\frac{k}{2(\alpha^2+\beta^2)^2}\Big[A_1\cos\Big(2\beta x-\frac{(3\alpha^2\beta-\beta^3)}{2a}t\Big)-B_1\sin\Big(2\beta x-\frac{(3\alpha^2\beta-\beta^3)}{2a}t\Big)\Big]\nonumber\\
&+\frac{k^2}{16(\alpha^2+\beta^2)^4}e^{4\alpha x-\frac{\alpha^3-3\alpha\beta^2}{a}t}(A_1^2+B_1^2),
\end{align}
for
\begin{equation*}
A_1=(a_1\alpha+b_1\beta)^2-(a_1\beta-b_1\alpha)^2,\quad B_1=2a_1b_1(\alpha^2-\beta^2)-2\alpha\beta(a_1^2-b_1^2).
\end{equation*}
Let $A_1=B\cos\omega_0$, $B_1=B\sin\omega_0$ where $B>0$. Hence
\begin{equation}
A_1^2+B_1^2=B^2=(a_1^2+b_1^2)^2(\alpha^2+\beta^2)^2,
\end{equation}
so $B=(a_1^2+b_1^2)(\alpha^2+\beta^2)$. Then the denominator $W$ of the solution (\ref{localone3}) becomes
\begin{equation}\displaystyle
W=1-\gamma_1\cos\theta+\frac{\gamma_1^2}{4}e^{\phi}=\frac{\gamma_1^2}{4}\Big[\frac{4}{\gamma_1^2}(1-\gamma_1\cos\theta)+e^{\phi}\Big],
\end{equation}
where $\theta=2\beta x-\frac{(3\alpha^2\beta-\beta^3)}{2a}t+\omega_0$, $\phi=4\alpha x-\frac{(\alpha^3-3\alpha\beta^2)}{a}t$,
and $\gamma_1=\frac{kB}{2(\alpha^2+\beta^2)^2}=\frac{k(a_1^2+b_1^2)}{2(\alpha^2+\beta^2)}$. Hence we conclude that if $|\gamma_1|\leq 1$ the solution (\ref{localone3}) is non-singular.

\noindent Consider the following example.

\noindent \textbf{Example 5.}\, Let us find a particular solution for (\ref{usualmKdV}). Choose $(k_1, k_2, e^{\delta_1}, e^{\delta_2}, k, a)=(\frac{1}{4},\frac{1}{4}, i, i, 1, 2)$. Then we get one-soliton solution of (\ref{usualmKdV}) as
\begin{equation}\displaystyle
q(t,x)=\frac{ie^{\frac{1}{4}x-\frac{1}{512}t}}{1+8e^{\frac{1}{2}x-\frac{1}{256}t}},
\end{equation}
so
\begin{equation}\label{STAN2oneex1}\displaystyle
|q(t,x)|^2=\frac{e^{\frac{1}{2}x-\frac{1}{256}t}}{(1+8e^{\frac{1}{2}x-\frac{1}{256}t})^2}=\frac{1}{32}\mathrm{sech}^2(\frac{1}{4}x-\frac{1}{512}t+\delta),
\end{equation}
where $\delta=\frac{3\ln 2}{2}$. The graph of (\ref{STAN2oneex1}) is given in Figure 5.
\begin{center}
\begin{figure}[h]
\centering
\begin{minipage}[t]{1\linewidth}
\centering
\includegraphics[angle=0,scale=.21]{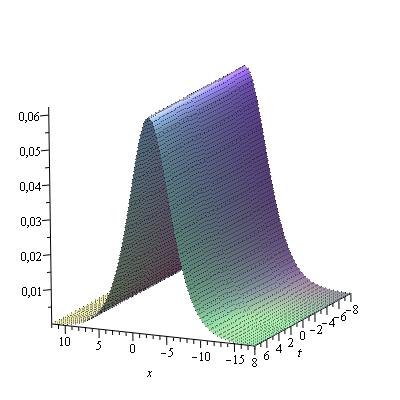}
\caption{One soliton for (\ref{localone3}) with the parameters $k_1=\alpha+i\beta=\frac{1}{4}$, $e^{\delta_1}=i, k=1, a=2$.}
\end{minipage}%
\end{figure}
\end{center}
\subsection{Two-Soliton Solution for the Local MKdV Equation:\\
 $(r=kq(t,x))$}

The conditions on the parameters to satisfy the relation (\ref{non00}) by two-soliton solution of the mKdV system are found as
\begin{equation}
1)\, k_i=\ell_i,\quad 2)\, e^{\alpha_i}=ke^{\delta_i}, i=1, 2.
\end{equation}

\noindent \textbf{Example 6.}\, Consider this particular example. Choose $(k_1,\ell_1,k_2,\ell_2)=(\frac{1}{4},\frac{1}{4},-\frac{1}{8},-\frac{1}{8})$ with
$(e^{\alpha_j},e^{\delta_j},k,a)=(i,i,1,10)$, $j=1, 2$. Hence we obtain two-soliton solution of (\ref{usualmKdV}) given by
\begin{equation}\label{solex6}\displaystyle
|q(t,x)|^2=\frac{\Big[e^{\frac{1}{4}x-\frac{1}{2560}t}+e^{-\frac{1}{8}x+\frac{1}{20480}t}+36e^{\frac{3}{8}x-\frac{3}{4096}t}+144e^{-\frac{3}{10240}t}\Big]^2}
{\Big[1+4e^{\frac{1}{2}x-\frac{1}{1280}t}+128e^{\frac{1}{8}x-\frac{7}{20480}t}+16e^{-\frac{1}{4}x+\frac{1}{10240}t}+5184e^{\frac{1}{4}x-\frac{7}{10240}t}\Big]^2}.
\end{equation}
The graph of the above function is given in Figure 6.
\begin{center}
\begin{figure}[h]
\centering
\begin{minipage}[t]{1\linewidth}
\centering
\includegraphics[angle=0,scale=0.23]{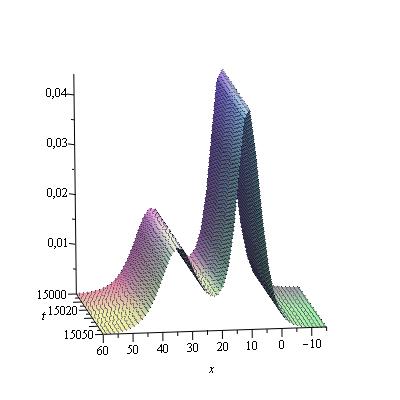}
\caption{Two solitons plotted for (\ref{solex6}) with the parameters $k_1=\ell_1=\frac{1}{4}, k_2=\ell_2=-\frac{1}{8}, e^{\alpha_j}=e^{\delta_j}=i, j=1, 2, k=1, a=10$.}
\end{minipage}%
\end{figure}
\end{center}

\subsection{Three-Soliton Solution for the Local MKdV Equation:\\
 $(r=kq(t,x))$}

Similar to the previous section we find the conditions on the parameters to satisfy the relation (\ref{non00}) by three-soliton solution of the mKdV system as
\begin{equation}
1)\, k_i=\ell_i,\quad 2)\, e^{\alpha_i}=ke^{\delta_i}, i=1, 2, 3.
\end{equation}

\noindent \textbf{Example 7.}\, Let us give an example of three-soliton solution of the local mKdV equation (\ref{usualmKdV}). Take
$(k_1,\ell_1,k_2,\ell_2,k_3,\ell_3)=(\frac{1}{4},\frac{1}{4},-\frac{1}{2},-\frac{1}{2},-\frac{1}{6},-\frac{1}{6})$ with $(e^{\alpha_j},e^{\delta_j},k,a)=(i,i,1,10)$, $j=1, 2, 3$. Then three-soliton solution of (\ref{usualmKdV}) is given by
\begin{equation}\label{solex7}\displaystyle
|q(t,x)|^2=\frac{A}{B},
\end{equation}
where
\begin{align*}
A=&\Big[e^{\frac{1}{4}x-\frac{1}{2560}t}+e^{-\frac{1}{2}x+\frac{1}{320}t}+e^{-\frac{1}{6}x+\frac{1}{8640}t}+36e^{\frac{3}{1280}t}+9e^{-\frac{3}{4}x+\frac{3}{512}t}\\
&+\frac{889}{2}e^{-\frac{5}{12}x+\frac{197}{69120}t}+100e^{\frac{1}{3}x-\frac{23}{34560}t}+225e^{-\frac{1}{12}x-\frac{11}{69120}t}
+\frac{1}{4}e^{-\frac{7}{6}x+\frac{11}{1728}t}\\
&+\frac{9}{4}e^{-\frac{5}{6}x+\frac{29}{8640}t}+1350e^{-\frac{2}{3}x+\frac{193}{34560}t}+50625e^{-\frac{1}{3}x+\frac{89}{34560}t}
+\frac{2025}{16}e^{-\frac{13}{12}x+\frac{421}{69120}t}\Big]^2
\end{align*}
and
\begin{align*}
B=&\Big[1+32e^{-\frac{1}{4}x+\frac{7}{2560}t}+4050e^{-\frac{1}{6}x+\frac{17}{6912}t}+288e^{\frac{1}{12}x-\frac{19}{69120}t}+324e^{-\frac{1}{2}x+\frac{7}{1280}t}\\
&+\frac{9}{2}e^{-\frac{2}{3}x+\frac{7}{2160}t}+648e^{-\frac{11}{12}x+\frac{413}{69120}t}+1800e^{-\frac{7}{12}x+\frac{41}{13824}t}+9e^{-\frac{1}{3}x+\frac{1}{4320}t}\\
&+\frac{9}{16}e^{-\frac{4}{3}x+\frac{7}{1080}t}+4e^{\frac{1}{2}x-\frac{1}{1280}t}+\frac{451737}{4}e^{-\frac{5}{6}x+\frac{197}{34560}t}
+22500e^{\frac{1}{6}x-\frac{19}{34560}t}+e^{-x+\frac{1}{160}t}\Big]^2.
\end{align*}
The graph of the above solution is given in Figure 7.
\begin{center}
\begin{figure}[h]
\centering
\begin{minipage}[t]{1\linewidth}
\centering
\includegraphics[angle=0,scale=.28]{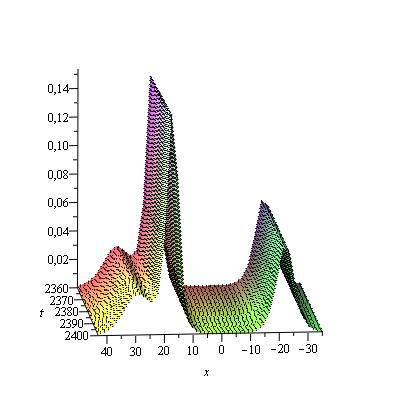}
\caption{Three solitons plotted for (\ref{solex7}) with the parameters $k_1=\ell_1=\frac{1}{4}, k_2=\ell_2=-\frac{1}{2}, k_3=\ell_3=-\frac{1}{6},
e^{\alpha_j}=e^{\delta_j}=i, j=1, 2, 3, k=1, a=10$.}
\end{minipage}%
\end{figure}
\end{center}

\section{Nonlocal Reductions of the MKdV System}

In this section we will first use the reduction (\ref{non}) given by Ablowitz and Musslimani \cite{AbMu1}-\cite{AbMu3} and obtain
soliton solutions for three different nonlocal cmKdV equations (\ref{mKdV41})-(\ref{mKdV43}) with the condition
\begin{equation}\label{nonlocalmKdVcond1}
\bar{a}=\varepsilon_{1}\varepsilon_2 a
\end{equation}
satisfied. Secondly, we will deal with the reduction (\ref{nonsecond}) and obtain one- and two-soliton solutions of the real nonlocal mKdV equation (\ref{mKdV44}).

\subsection{One-Soliton Solution for the Nonlocal CMKdV Equations:\\
 $(r=k\bar{q}(\varepsilon_1t,\varepsilon_2x))$}

Similar to local cmKdV equation, we will present Type 1 and Type 2 to find one-soliton solution for the nonlocal cmKdV equations. One can also use the same approach in Type 2 to obtain two- and three-soliton solutions for nonlocal cmKdV equation.

\subsubsection{Type 1.}

\noindent Firstly, we find the conditions on the parameters of one-soliton solution of the mKdV system to satisfy the equality (\ref{non}).
We must have
\begin{equation}\label{r=kbarq}
\displaystyle \frac{e^{k_2x-\frac{k_2^3}{4a}t+\delta_2}}{1+Ae^{(k_1+k_2)x-\frac{(k_1^3+k_2^3)}{4a}t+\delta_1+\delta_2  }}
=k\frac{e^{\bar{k}_1\varepsilon_2x-\frac{\bar{k}_1^3}{4\bar{a}}\varepsilon_1t+\bar{\delta}_1}}
{1+\bar{A}e^{(\bar{k}_1+\bar{k}_2)\varepsilon_2x-\frac{(\bar{k}_1^3+\bar{k}_2^3)}{4\bar{a}}\varepsilon_1t+\bar{\delta}_1+\bar{\delta}_2}}.
\end{equation}
This equation can be satisfied for some different choices of the parameters.
One of the case includes the following equalities that must be satisfied by the parameters:
\begin{align}\label{7conds}
\displaystyle &i)\, k_2=\varepsilon_2\bar{k}_1,\quad  ii)\, \frac{k_2^3}{4a}=\frac{\bar{k}_1^3}{4\bar{a}}\varepsilon_1,\quad iii)\, e^{\delta_2}=ke^{\bar{\delta}_1},\quad
iv)\, \bar{A}=A,\nonumber\\
 &v)\,(k_1+k_2)=(\bar{k}_1+\bar{k}_2)\varepsilon_2,\quad
 vi)\,\frac{(k_1^3+k_2^3)}{4a}=\frac{(\bar{k}_1^3+\bar{k}_2^3)}{4\bar{a}}\varepsilon_1,\quad vii)\, e^{\delta_1+\delta_2}=e^{\bar{\delta}_1+\bar{\delta}_2}.
\end{align}

\noindent If we use the conditions (\ref{nonlocalmKdVcond1}) and $i)$ on the left hand side of the equality $ii)$, it is clear that this equality is satisfied directly
since
\begin{equation*}\displaystyle
\frac{k_2^3}{4a}=\frac{\varepsilon_2\bar{k}_1^3}{4\varepsilon_1\varepsilon_2\bar{a}}=\frac{\bar{k}_1^3}{4\bar{a}}\varepsilon_1.
\end{equation*}

\noindent With the condition given in $i)$ it is obvious that $iv)$ is satisfied directly since
\begin{equation*}\displaystyle
-\frac{1}{(k_1+k_2)^2}=-\frac{1}{(\bar{k}_2\varepsilon_2+\bar{k}_1\varepsilon_2)^2}=-\frac{1}{(\bar{k}_1+\bar{k}_2)^2}.
\end{equation*}

\noindent The condition $v)$ is satisfied since
\begin{equation*}
(k_1+k_2)=(\bar{k}_2\varepsilon_2+\bar{k}_1\varepsilon_2)=(\bar{k}_1+\bar{k}_2)\varepsilon_2
\end{equation*}
by the condition $k_2=\varepsilon_2 \bar{k}_1$ or equivalently $k_1=\varepsilon_2\bar{k}_2$. By the same manner $vi)$
is already true since
\begin{equation*}
\displaystyle \frac{(k_1^3+k_2^3)}{4a}=\frac{(\bar{k}_2^3+\bar{k}_1^3)}{4\varepsilon_1\varepsilon_2\bar{a}}=\frac{(\bar{k}_1^3+\bar{k}_2^3)}{4\bar{a}}\varepsilon_1.
\end{equation*}

\noindent Finally, consider the relation $e^{\delta_2}=ke^{\bar{\delta}_1}$ or $e^{\bar{\delta}_2}=ke^{\delta_1}$ given in $vii)$. Since $k$ is a real constant we have $\bar{k}=k$. Therefore we have
\begin{equation*}
e^{\delta_1+\delta_2}=ke^{\delta_1}e^{\bar{\delta}_1}\quad \mathrm{and} \quad e^{\bar{\delta}_1+\bar{\delta}_2}=ke^{\bar{\delta}_1}e^{\delta_1},
\end{equation*}
yielding the equality $e^{\delta_1+\delta_2}=e^{\bar{\delta}_1+\bar{\delta}_2}$.

\noindent Thus the parameters of
one-soliton solution of the equation (\ref{mKdV4}) must have the following properties:
\begin{equation}\label{1nonlocalmKdVcond}
1)\,\bar{a}=\varepsilon_1\varepsilon_2a,\quad 2)\,k_2=\varepsilon_2\bar{k}_1,\quad 3)\, e^{\delta_2}=ke^{\bar{\delta}_1}.
\end{equation}

\noindent The case $(\varepsilon_1,\varepsilon_2)=(1,1)$ gives local equation. For particular choice of the parameters let us check the solutions of the nonlocal reductions of the mKdV system for $(\varepsilon_1,\varepsilon_2)=\{(-1,1), (1,-1), (-1, -1)\}$.\\

\noindent \subsubsubsection{Case a. ($T$-symmetric): $(\varepsilon_1,\varepsilon_2)=(-1,1)$, $r=k\bar{q}(-t,x)$}

\noindent This case gives $\bar{a}=-a$ and $k_2=\bar{k}_1$, and
\begin{equation}\label{nonlocalmKdVcasea}
\displaystyle aq_t(t,x)=-\frac{1}{4}q_{xxx}(t,x)+\frac{3}{2}k \bar{q}(-t,x)q(t,x)q_x(t,x),
\end{equation}
with $e^{\delta_2}=ke^{\bar{\delta}_1}$. Since $\bar{a}=-a$, $a$ is pure imaginary say $a=ib$, for nonzero $b\in \mathbb{R}$. Let $k_1=\alpha+i\beta$ so $k_2=\alpha-i\beta$ for $\alpha, \beta \in \mathbb{R}$, $\alpha\neq 0$. Then the solution of (\ref{nonlocalmKdVcasea}) becomes
\begin{equation}\displaystyle
q(t,x)=\frac{e^{(\alpha+i\beta)x+\frac{i\alpha^3-3\alpha^2\beta-3i\alpha\beta^2+\beta^3}{4b}t+\delta_1} }{1-\frac{k}{4\alpha^2}e^{2\alpha x+i\frac{\alpha^3-3\alpha\beta^2}{2b}t+\delta_1+\bar{\delta}_1 }   }.
\end{equation}

\noindent This solution is also given in \cite{Yang}. To obtain a real-valued solution we find the function $|q(t,x)|^2=q(t,x)\bar{q}(t,x)$,
\begin{equation}\label{nonlocalonegena}\displaystyle
|q(t,x)|^2=\frac{e^{2\alpha x+\frac{(\beta^3-3\alpha^2\beta)}{2b}t+\delta_1+\bar{\delta}_1 }}{\Big[\frac{k}{4\alpha^2}e^{2\alpha x+\delta_1+\bar{\delta}_1}
-\cos(\frac{(\alpha^3-3\alpha\beta^2)}{2b}t) \Big]^2+\sin^2(\frac{(\alpha^3-3\alpha\beta^2)}{2b}t) }.
\end{equation}

\noindent  When $\alpha^3-3\alpha\beta^2\neq 0$ and $t=\frac{2nb\pi}{\alpha^3-3\alpha\beta^2}$,
$\frac{k}{4\alpha^2}e^{2\alpha x+\delta_1+\bar{\delta}_1}-(-1)^n=0$ where $n$ is an integer, for both focusing and defocusing cases,
the solution is singular. When $\alpha^3-3\alpha\beta^2=0$ the solution for focusing case is non-singular.

\noindent Now for particular choices of the parameters satisfying the conditions (\ref{1nonlocalmKdVcond}) we will give some solutions of
the equation (\ref{nonlocalmKdVcasea}) and present the graphs of the solutions.\\

\noindent\textbf{Example 8.}\, For the parameters $(k_1, k_2, e^{\delta_1}, e^{\delta_2}, k, a)=(2\sqrt{3}+2i, 2\sqrt{3}-2i, 1+i, -1+i, -1, i)$ we obtain the non-singular solution of (\ref{nonlocalmKdVcasea}) as
\begin{equation}\displaystyle
q(t,x)=\frac{24(1+i)e^{(2\sqrt{3}+2i)x-16t}}{24+e^{4\sqrt{3}x}},
\end{equation}
so the function $|q(t,x)|^2$ is
\begin{equation}\label{solex8}\displaystyle
|q(t,x)|^2=\frac{1152e^{4\sqrt{3}x-32t}}{(24+e^{4\sqrt{3}x})^2}.
\end{equation}
The graph of (\ref{solex8}) is given in
Figure 8.\\

\noindent\textbf{Example 9.}\, For the parameters $(k_1, k_2, e^{\delta_1}, e^{\delta_2}, k, a)=(\frac{1}{4}, \frac{1}{4}, 1+i, -1+i, -1, 10i)$, the solution $q(t,x)$ becomes
\begin{equation}\displaystyle
q(t,x)=\frac{(1+i)e^{\frac{1}{4}x+\frac{1}{2560}it}}{1+8e^{\frac{1}{2}x+\frac{1}{1280}it}},
\end{equation}
so the function $|q(t,x)|^2$ is
\begin{equation}\label{solex9}\displaystyle
|q(t,x)|^2=\frac{2e^{\frac{1}{2}x}}{[8e^{\frac{1}{2}x}+\cos(\frac{1}{1280}t)]^2+\sin^2(\frac{1}{1280}t)}.
\end{equation}
The solution (\ref{solex9}) has singularity at $(t,x)=(1280(2n+1)\pi,-6\ln 2)$, $n$ is an integer and its graph is given in
Figure 9.
\newpage

\begin{center}
\begin{figure}[h]
\centering
\begin{minipage}[t]{0.4\linewidth}
\centering
\includegraphics[angle=0,scale=.26]{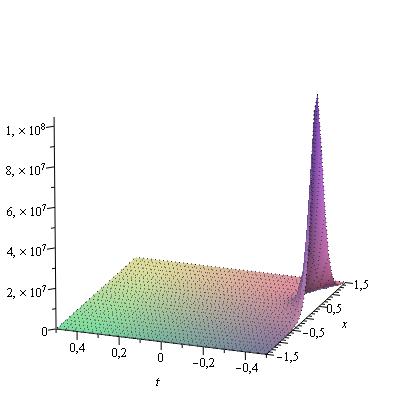}
\caption{A non-singular localized (one-soliton) wave for (\ref{nonlocalonegena})
with the parameters $k_1=\alpha+i\beta=2\sqrt{3}+2i, e^{\delta_1}=1+i, k=-1, a=ib=i$.}
\end{minipage}%
\hfill
\begin{minipage}[t]{0.4\linewidth}
\centering
\includegraphics[angle=0,scale=.26]{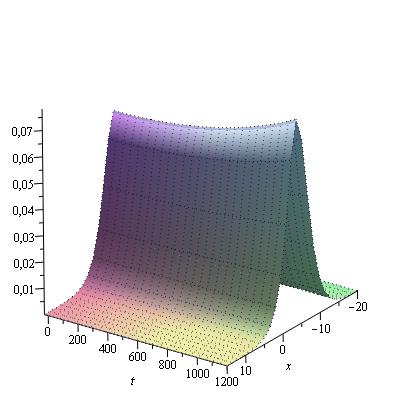}
\caption{A singular (one-soliton) wave for (\ref{nonlocalonegena})
with the parameters $k_1=\alpha+i\beta=\frac{1}{4}, e^{\delta_1}=1+i, k=-1, a=ib=10i$.}
\end{minipage}%
\end{figure}
\end{center}

\subsubsubsection{Case b. ($S$-symmetric): $(\varepsilon_1,\varepsilon_2)=(1,-1)$, $r=k\bar{q}(t,-x)$}

\noindent In this case we have $\bar{a}=-a$, $k_2=-\bar{k}_1$, and
\begin{equation}\label{nonlocalmKdVcaseb}
\displaystyle aq_t(t,x)=-\frac{1}{4}q_{xxx}(t,x)+\frac{3}{2}k \bar{q}(t,-x)q(t,x)q_x(t,x)
\end{equation}
with $e^{\delta_2}=ke^{\bar{\delta}_1}$. Since $\bar{a}=-a$, it is pure imaginary, say $a=ib$ for nonzero $b \in \mathbb{R}$.
Let also $k_1=\alpha+i\beta$ and so $k_2=-\alpha+i\beta$ for $\alpha, \beta \in \mathbb{R}$, $\beta\neq 0$. Then the solution of (\ref{nonlocalmKdVcaseb}) becomes
\begin{equation}\displaystyle
q(t,x)=\frac{e^{(\alpha+i\beta)x+\frac{i\alpha^3-3\alpha^2\beta-3i\alpha\beta^2+\beta^3}{4b}t+\delta_1} }{1+\frac{k}{4\beta^2}e^{2i\beta x+i\frac{\alpha^3-3\alpha\beta^2}{2b}t+\delta_1+\bar{\delta}_1 }   },
\end{equation}
and so the function $|q(t,x)|^2$ is
\begin{equation}\label{nonlocalonegenb}\displaystyle
|q(t,x)|^2=\frac{e^{2\alpha x+\frac{(\beta^3-3\alpha^2\beta)}{2b}t+\delta_1+\bar{\delta}_1}}{\Big[\frac{k}{4\beta^2}e^{\frac{(\beta^3-3\alpha^2\beta)}{2b}t+\delta_1+\bar{\delta}_1}+
\cos(2\beta x)    \Big]^2+\sin^2(2\beta x)}.
\end{equation}
This function is singular at $x=\frac{n\pi}{2\beta}$, $\frac{k}{4\beta^2}e^{\frac{(\beta^3-3\alpha^2\beta)}{2b}t+\delta_1+\bar{\delta}_1}+(-1)^n=0$, where $n$
is an integer for both focusing and defocusing cases.

\noindent Let us give some examples of solutions of (\ref{nonlocalmKdVcaseb}) given for particular choices of the parameters.\\

\noindent\textbf{Example 10.}\, For the parameters $(k_1, k_2, e^{\delta_1}, e^{\delta_2}, k, a)=(\frac{i}{4}, \frac{i}{4}, 1+i, 1-i, 1, \frac{i}{2})$, we get the solution $q(t,x)$ of (\ref{nonlocalmKdVcaseb}) as
\begin{equation}\displaystyle
q(t,x)=\frac{(1+i)e^{\frac{1}{4}ix+\frac{1}{128}t}}{1+8e^{\frac{1}{2}ix+\frac{1}{64}t}}.
\end{equation}
Hence the function $|q(t,x)|^2$ is
\begin{equation}\label{solex10}\displaystyle
|q(t,x)|^2=\frac{2e^{\frac{1}{64}t}}{[8e^{\frac{1}{64}t}+\cos(\frac{1}{2}x) ]^2+\sin^2(\frac{1}{2}x)}.
\end{equation}
This solution has singularity at $(t,x)=(-192\ln 2,2(2n+1)\pi)$ for $n$ integer. The graph of (\ref{solex10})
is given in Figure 10.\\

\noindent\textbf{Example 11.}\, For another set of the parameters $(k_1, k_2, e^{\delta_1}, e^{\delta_2}, k, a)=(1-\frac{i}{4}, -1-\frac{i}{4}, 1+i, -1+i, -1, i)$, the solution $q(t,x)$ becomes
\begin{equation}\displaystyle
q(t,x)=\frac{(1+i)e^{(1-\frac{1}{4}i)x+(\frac{47}{256}+\frac{13}{64}i)t}}{1-8e^{-\frac{1}{2}ix+\frac{47}{128}t}},
\end{equation}
and so we have the function $|q(t,x)|^2$ as
\begin{equation}\label{solex11}\displaystyle
|q(t,x)|^2=\frac{2e^{2x+\frac{47}{128}t}}{[8e^{\frac{47}{128}t}-\cos(\frac{1}{2}x)]^2+\sin^2(\frac{1}{2}x)},
\end{equation}
which is again a singular function. It has singularity at $(t,x)=(-\frac{128}{47}\ln 8, 4n\pi)$. The graph of (\ref{solex11}) is
given in Figure 11.

\begin{center}
\begin{figure}[h]
\centering
\begin{minipage}[t]{0.4\linewidth}
\centering
\includegraphics[angle=0,scale=.24]{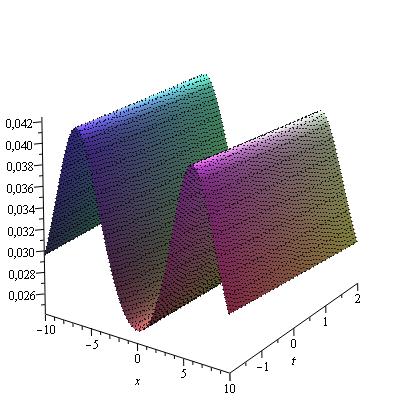}
\caption{A singular periodical (one soliton) wave for (\ref{nonlocalonegenb}) with the parameters $k_1=\alpha+i\beta=\frac{i}{4}, e^{\delta_1}=1+i, k=1, a=ib=\frac{i}{2}$.}
\end{minipage}%
\hfill
\begin{minipage}[t]{0.4\linewidth}
\centering
\includegraphics[angle=0,scale=.24]{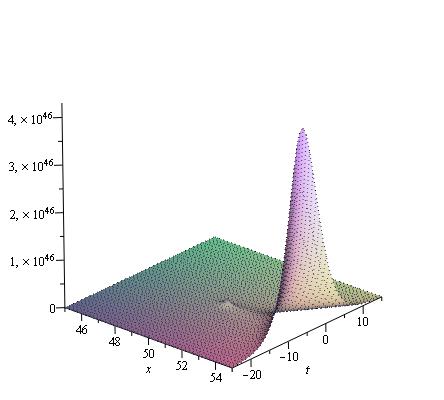}
\caption{A singular localized (one soliton) wave for (\ref{nonlocalonegenb}) with the parameters
 $k_1=\alpha+i\beta=1-\frac{i}{4}, e^{\delta_1}=1+i, k=-1, a=ib=i$.}
\end{minipage}%
\end{figure}
\end{center}

\subsubsubsection{Case c. ($ST$-symmetric): $(\varepsilon_1,\varepsilon_2)=(-1,-1)$, $r=k\bar{q}(-t,-x)$}

\noindent For this case we have $\bar{a}=a$, $k_2=-\bar{k}_1$, and
\begin{equation}\label{nonlocalmKdVcasec}
\displaystyle aq_t(t,x)=-\frac{1}{4}q_{xxx}(t,x)+\frac{3}{2}k \bar{q}(-t,-x)q(t,x)q_x(t,x),
\end{equation}
with $e^{\delta_2}=ke^{\bar{\delta}_1}$. Let $k_1=\alpha+i\beta$ and so $k_2=-\alpha+i\beta$ for $\alpha, \beta \in \mathbb{R}$, $\beta\neq 0$. Then the solution $q(t,x)$ of (\ref{nonlocalmKdVcasec}) becomes
\begin{equation}\label{casecnonlocal}\displaystyle
q(t,x)=\frac{e^{(\alpha+i\beta)x-\frac{\alpha^3+3\alpha^2i\beta-3\alpha\beta^2-i\beta^3}{4a}t+\delta_1}}
{1+\frac{k}{4\beta^2}e^{2i\beta x-i\frac{(6\alpha^2\beta-2\beta^3)}{4a}t+\delta_1+\bar{\delta}_1}}.
\end{equation}
Then we obtain the function $|q(t,x)|^2$ as
\begin{equation}\label{nonlocalonegenc}\displaystyle
|q(t,x)|^2=\frac{e^{2\alpha x+\frac{(3\alpha\beta^2-\alpha^3)}{2a}t+\delta_1+\bar{\delta}_1}}
{\Big[\frac{k}{4\beta^2}e^{\delta_1+\bar{\delta}_1}+\cos(2\beta x+\frac{(\beta^3-3\alpha^2\beta)}{2a}t)\Big]^2+\sin^2(2\beta x+\frac{(\beta^3-3\alpha^2\beta)}{2a}t)}.
\end{equation}
This function is singular on the line $2\beta x+\frac{(\beta^3-3\alpha^2\beta)}{2a}t=n\pi$ if the condition
$\frac{k}{4\beta^2}e^{\delta_1+\bar{\delta}_1}+(-1)^n=0$, where $n$ is an integer, is satisfied, otherwise for both focusing and defocusing cases we have non-singular solutions. More specifically, let $2\alpha x+\frac{(3\alpha\beta^2-\alpha^3)}{2a}t+\delta_1+\bar{\delta}_1=\theta$, $2\beta x+\frac{(\beta^3-3\alpha^2\beta)}{2a}t=\phi$, and $\frac{k}{2\beta^2}e^{\delta_1+\bar{\delta}_1}=\mu$. In this case the solution (\ref{nonlocalonegenc}) can be written as
\begin{equation}\label{casecnewform}\displaystyle
|q(t,x)|^2=\frac{e^{\theta}}{1+\frac{\mu^2}{4}+\mu\cos \phi}=\frac{e^{\theta}}{\mu\Big[(\frac{1}{\mu}+\frac{\mu}{4})+\cos\phi\Big]}.
\end{equation}

\noindent Note that if we take $k=1$, $a=\frac{1}{4}$, $\alpha=2b$, $\beta=2a$, $e^{\delta_1}=-4ai$, in our solution (\ref{casecnonlocal}) then it reduces to one of the solutions given in \cite{ma}. The solution (\ref{casecnewform}) is non-singular for all $\mu$ except $\mu=\pm 2$.

\noindent For particular choice of parameters let us present some examples for the Case c.\\

\noindent\textbf{Example 12.}\, For the parameters $(k_1, k_2, e^{\delta_1}, e^{\delta_2}, k, a)=(\frac{i}{4}, \frac{i}{4}, 1+i, 1-i, 1, 2)$, we have the solution
of (\ref{nonlocalmKdVcasec}) as
\begin{equation}\displaystyle
q(t,x)=\frac{(1+i)e^{\frac{1}{4}ix+\frac{1}{512}it}}{1+8e^{\frac{1}{2}ix+\frac{1}{256}it}}.
\end{equation}
Then we obtain the function $|q(t,x)|^2$,
\begin{equation}\label{solex12}\displaystyle
|q(t,x)|^2=\frac{2}{65+16\cos(\frac{1}{2}x+\frac{1}{256}t)},
\end{equation}
which is non-singular. The graph of (\ref{solex12}) is given in Figure 12.\\

\noindent\textbf{Example 13.}\, For the parameters $(k_1, k_2, e^{\delta_1}, e^{\delta_2}, k, a)=(\frac{4}{10}+\frac{42}{100}i, -\frac{4}{10}+\frac{42}{100}i, 1, 1, 1, -4)$, we have the solution
of (\ref{nonlocalmKdVcasec}),
\begin{equation}\displaystyle
q(t,x)=\frac{e^{(\frac{2}{5}+\frac{21}{50}i)x+(-\frac{923}{100000}+\frac{15939}{2000000}i)t }}
{1+\frac{625}{441}e^{\frac{21}{25}ix+\frac{15939}{1000000}it} }.
\end{equation}
Therefore the function $|q(t,x)|^2$ is found as
\begin{equation}\label{solex13}\displaystyle
|q(t,x)|^2=\frac{194481e^{\frac{4}{5}x-\frac{923}{50000}t}}{2(292553+275625\cos(\frac{21}{25}x+\frac{15939}{1000000}t))}.
\end{equation}
The above function is non-singular. Its graph is given in Figure 13.

\begin{center}
\begin{figure}[h]
\centering
\begin{minipage}[t]{0.4\linewidth}
\centering
\includegraphics[angle=0,scale=.24]{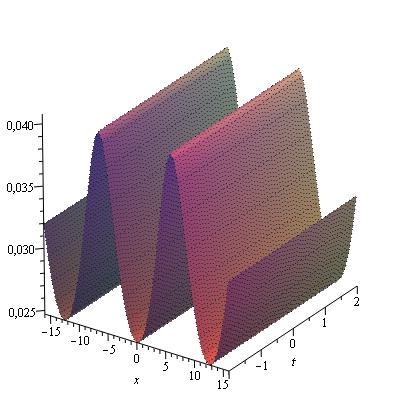}
\caption{A periodical (one soliton) wave for (\ref{nonlocalonegenc}) with the parameters $k_1=\alpha+i\beta=\frac{i}{4},e^{\delta_1}=1+i, k=1,a=2$.}
\end{minipage}%
\hfill
\begin{minipage}[t]{0.3\linewidth}
\centering
\includegraphics[angle=0,scale=.24]{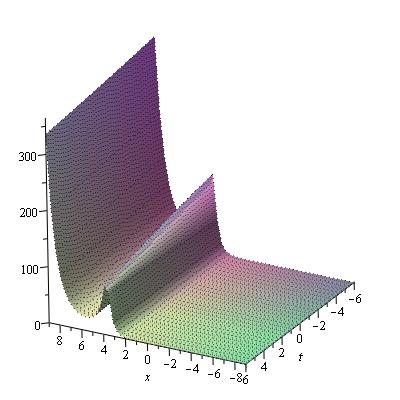}
\caption{A non-singular (one soliton) wave for (\ref{nonlocalonegenc}) with the parameters $k_1=\alpha+i\beta=\frac{4}{10}+\frac{42}{100}i, e^{\delta_1}=1, k=1,a=-4$.}
\end{minipage}%
\end{figure}
\end{center}

\subsubsection{Type 2.}

\noindent Here we again consider one-soliton solution of the mKdV system with the nonlocal reduction (\ref{non}). We use cross multiplication. We have
\begin{align}
e^{k_2x-\frac{k_2^3}{4a}t+\delta_2}&+\bar{A}e^{[(\bar{k}_1+\bar{k}_2)\varepsilon_2+k_2]x
-\Big[\frac{(\bar{k}_1^3+\bar{k}_2^3)\varepsilon_1}{4\bar{a}}+\frac{k_2^3}{4a}\Big]t+\bar{\delta}_1+\bar{\delta}_2+\delta_2}\nonumber\\
&=ke^{\bar{k}_1\varepsilon_2x-\frac{\bar{k}_1^3}{4\bar{a}}\varepsilon_1t+\bar{\delta}_1}
+Ake^{[(k_1+k_2)+\bar{k}_1\varepsilon_2]x-\Big[\frac{(k_1^3+k_2^3)}{4a}+\frac{\bar{k}_1^3}{4\bar{a}}\varepsilon_1  \Big]t+\delta_1+\delta_2+\bar{\delta}_1}.
\end{align}
In addition to the Type 1 conditions, we find the following new conditions in Type 2:
\begin{align*}
&i)\, k_2=k_1+k_2+\bar{k}_1\varepsilon_2,\quad ii)\,\frac{k_2^3}{4a}=\frac{k_1^3+k_2^3}{4a}+\frac{\bar{k}_1^3}{4\bar{a}}\varepsilon_1,
\quad iii)\, e^{\delta_2}=Ake^{\delta_1+\delta_2+\bar{\delta}_1},\\
&iv)\,\bar{k}_1\varepsilon_2+\bar{k}_2\varepsilon_2+k_2=\bar{k}_1\varepsilon_2,
\quad v)\, \frac{\bar{k}_1^3+\bar{k}_2^3}{4\bar{a}}\varepsilon_1+\frac{k_2^3}{4a}=\frac{\bar{k}_1^3}{4\bar{a}}\varepsilon_1,
\quad vi)\,\bar{A}e^{\bar{\delta}_1+\bar{\delta}_2+\delta_2}=ke^{\bar{\delta}_1}.
\end{align*}
Clearly, these conditions are simplified as
\begin{equation}
1)\, \bar{a}=\varepsilon_1\varepsilon_2a,\quad 2)\, k_1=-\bar{k}_1\varepsilon_2,\quad 3)\, k_2=-\bar{k}_2\varepsilon_2,\quad
4)\, Ake^{\delta_1+\bar{\delta}_1}=1,\quad 5)\,Ae^{\delta_2+\bar{\delta}_2}=k.
\end{equation}

\noindent Let us present some particular examples satisfying the above constraints.

\subsubsubsection{Case a. ($T$-symmetric): $(\varepsilon_1,\varepsilon_2)=(-1,1)$, $r=k\bar{q}(-t,x)$}

 Here the parameters satisfy $\bar{a}=-a$, $k_1=-\bar{k}_1$, $k_2=-\bar{k}_2$ with $Ake^{\delta_1+\bar{\delta}_1}=1$, $Ae^{\delta_2+\bar{\delta}_2}=k$. In this case the parameters $a$, $k_1$, and $k_2$ are pure imaginary, say $a=i\alpha$, $k_1=i\beta$, and $k_2=i\gamma$ for $\alpha, \beta,\gamma \in \mathbb{R}$. Therefore we have $e^{\delta_1+\bar{\delta}_1}=\frac{(\beta+\gamma)^2}{k}$ and $e^{\delta_2+\bar{\delta}_2}=k(\beta+\gamma)^2$. Let also $e^{\delta_1}=a_1+ib_1$ and $e^{\delta_2}=a_2+ib_2$ for $a_1,b_1, a_2, b_2 \in \mathbb{R}$. Then one-soliton solution becomes
\begin{equation}\displaystyle
q(t,x)=\frac{e^{i\beta x+\frac{\beta^3}{4\alpha}t}(a_1+ib_1)}{1+\frac{1}{(\beta+\gamma)^2}e^{i(\beta+\gamma)x+\frac{(\beta^3+\gamma^3)}{4\alpha}t}(a_1+ib_1)(a_2+ib_2)}.
\end{equation}
Hence the function $|q(t,x)|^2$ is
\begin{align}\label{nonlocalonetype2a}\displaystyle
&|q(t,x)|^2\nonumber\\
&=\frac{e^{\frac{\beta^3}{2\alpha}t}(a_1^2+b_1^2)}{1+\frac{2}{(\beta+\gamma)^2}
e^{\frac{(\beta^3+\gamma^3)}{4\alpha}t}[(a_1a_2-b_1b_2)\cos((\beta+\gamma)x)-(a_1b_2+a_2b_1)\sin((\beta+\gamma)x)]+e^{\frac{(\beta^3+\gamma^3)}{2\alpha}t}        },
\end{align}
where $a_1^2+b_1^2=\frac{(\beta+\gamma)^2}{k}$ and $a_2^2+b_2^2=k(\beta+\gamma)^2$, $\beta\neq -\gamma$. Let $a_1a_2-b_1b_2=B\cos\omega_0$, $a_1b_2+a_2b_1=B\sin\omega_0$ for $B>0$. Hence
\begin{equation}
B^2=(a_1^2+b_1^2)(a_2^2+b_2^2)=(\beta+\gamma)^4,
\end{equation}
yielding $B=(\beta+\gamma)^2$. In this case, the solution (\ref{nonlocalonetype2a}) can be written as
\begin{equation}\displaystyle
|q(t,x)|^2=\frac{e^{\frac{\beta^3}{2\alpha}t}(a_1^2+b_1^2)}{1+2\cos\theta+e^{\frac{(\beta^3+\gamma^3)}{2\alpha}t}}
=\frac{e^{\frac{(\beta^3-\gamma^3)}{4\alpha}t}(a_1^2+b_1^2)}{2[\cosh(\frac{(\beta^3+\gamma^3)}{4\alpha}t)+\cos\theta]},
\end{equation}
where $\theta=(\beta+\gamma)x+\omega_0$. This solution is singular at $t=0$, $\theta=(2n+1)\pi$ for $n$ integer, and non-singular for $t\neq 0$.

\noindent \textbf{Example 14.}\, Choose $(k_1, k_2, e^{\delta_1}, e^{\delta_2}, k, a)=(i,-\frac{i}{2}, 1, \frac{1}{4},\frac{1}{4}, 2i)$.
Hence the solution becomes
\begin{equation}\displaystyle
q(t,x)=\frac{e^{ix+\frac{1}{8}t}}{1+e^{\frac{1}{2}ix+\frac{7}{64}t}},
\end{equation}
so we obtain the function $|q(t,x)|^2$ as
\begin{equation}\label{solex14}\displaystyle
|q(t,x)|^2=\frac{e^{\frac{9}{64}t}}{2[\cosh(\frac{7}{64}t)+\cos(\frac{1}{2}x)]}.
\end{equation}
The above function has singularity at $(t,x)=(0,2(2n+1)\pi)$, $n$ integer. The graph of this function is given in Figure 14.

\subsubsubsection{Case b. ($S$-symmetric): $(\varepsilon_1,\varepsilon_2)=(1,-1)$, $r=k\bar{q}(t,-x)$}

For this case we have $\bar{a}=-a$, $k_1=\bar{k}_1$, $k_2=\bar{k}_2$ with $Ake^{\delta_1+\bar{\delta}_1}=1$, $Ae^{\delta_2+\bar{\delta}_2}=k$. In this case $a$ is pure imaginary, say $a=i\alpha$, $\alpha\in \mathbb{R}$. The parameters $k_1$ and $k_2$ are real. Let $e^{\delta_1}=a_1+ib_1$ and $e^{\delta_2}=a_2+ib_2$ for $a_1,b_1, a_2, b_2 \in \mathbb{R}$. Hence one-soliton solution becomes
\begin{equation}\displaystyle
q(t,x)=\frac{e^{k_1x+i\frac{k_1^3}{4\alpha}t}(a_1+ib_1)}{1-\frac{1}{(k_1+k_2)^2}e^{(k_1+k_2)x+i\frac{(k_1^3+k_2^3)}{4\alpha}t}(a_1+ib_1)(a_2+ib_2)}.
\end{equation}
Therefore we obtain the function $|q(t,x)|^2$ as
\begin{align}\label{nonlocalonetype2b}\displaystyle
|q(t,x)|^2&=\nonumber\\
&\frac{e^{2k_1x}(a_1^2+b_1^2)}{1-\frac{2e^{(k_1+k_2)x}}{(k_1+k_2)^2}[\cos(\frac{(k_1^3+k_2^3)}{4\alpha}t)(a_1a_2-b_1b_2)
-\sin(\frac{(k_1^3+k_2^3)}{4\alpha}t)(a_1b_2+b_1a_2)]+e^{2(k_1+k_2)x}},
\end{align}
where $a_1^2+b_1^2=-\frac{(k_1+k_2)^2}{k}$ and $a_2^2+b_2^2=-k(k_1+k_2)^2$, $k_1\neq -k_2$. Let $a_1a_2-b_1b_2=B\cos\omega_0$, $a_1b_2+b_1a_2=B\sin\omega_0$ for $B>0$. Therefore $B=(k_1+k_2)^2$. The solution (\ref{nonlocalonetype2b}) can be expressed as
\begin{equation}\displaystyle
|q(t,x)|^2=\frac{e^{(k_1-k_2)x}(a_1^2+b_1^2)}{2[\cosh((k_1+k_2)x)-\cos\theta]},
\end{equation}
where $\theta=\frac{(k_1^3+k_2^3)}{4}t+\omega_0$. This solution has singularity at $x=0$, $\theta=2n\pi$ for $n$ integer, and non-singular for $x\neq 0$.\\

\noindent \textbf{Example 15.}\, Take $(k_1, k_2, e^{\delta_1}, e^{\delta_2}, k, a)=(1, 1, -2, 2, -1, 2i)$.
So the solution $q(t,x)$ becomes
\begin{equation}\displaystyle
q(t,x)=\frac{-2e^{x+\frac{1}{8}it}}{1+e^{2x+\frac{1}{4}it}},
\end{equation}
and the function $|q(t,x)|^2$ is
\begin{equation}\label{solex15}\displaystyle
|q(t,x)|^2=\frac{2}{\cosh(2x)+\cos(\frac{1}{4}t)}.
\end{equation}
The above function has singularity at $(t,x)=(4(2n+1)\pi,0)$, $n$ integer. The graph of this function is given in Figure 15.

\subsubsubsection{Case c. ($ST$-symmetric): $(\varepsilon_1,\varepsilon_2)=(-1,-1)$, $r=k\bar{q}(-t,-x)$}

\noindent Here the parameters satisfy $\bar{a}=a$, $k_1=\bar{k}_1$, $k_2=\bar{k}_2$ with $Ake^{\delta_1+\bar{\delta}_1}=1$, $Ae^{\delta_2+\bar{\delta}_2}=k$. Therefore $a$, $k_1$, and $k_2$ are real. Let also
$e^{\delta_1}=a_1+ib_1$ and $e^{\delta_2}=a_2+ib_2$ for $a_1,b_1, a_2, b_2 \in \mathbb{R}$. Then one-soliton solution becomes
\begin{equation}\displaystyle
q(t,x)=\frac{e^{k_1x-\frac{k_1^3}{4a}t}(a_1+ib_1)}{1-\frac{1}{(k_1+k_2)^2}e^{(k_1+k_2)x-\frac{(k_1^3+k_2^3)}{4a}t}(a_1+ib_1)(a_2+ib_2)}.
\end{equation}
Therefore we have the function $|q(t,x)|^2$ as
\begin{equation}\label{nonlocalonetype2c}\displaystyle
|q(t,x)|^2=\frac{e^{2k_1x-\frac{k_1^3}{2a}t}(a_1^2+b_1^2)}{1-\frac{2(a_1a_2-b_1b_2)}{(k_1+k_2)^2}e^{(k_1+k_2)x-\frac{(k_1^3+k_2^3)}{4a}t}
+e^{2(k_1+k_2)x-\frac{(k_1^3+k_2^3)}{2a}t}},
\end{equation}
where $a_1^2+b_1^2=-\frac{(k_1+k_2)^2}{k}$ and $a_2^2+b_2^2=-k(k_1+k_2)^2$, $k_1\neq -k_2$. If we let $(k_1+k_2)x-\frac{(k_1^3+k_2^3)}{4a}t=\theta$,
$2k_1x-\frac{k_1^3}{2a}t=\phi$, and $\frac{(a_1a_2-b_1b_2)}{(k_1+k_2)^2}=\gamma$, then the solution (\ref{nonlocalonetype2c}) becomes
\begin{equation}\displaystyle
|q(t,x)|^2=\frac{e^{\phi}}{1-2\gamma e^{\theta}+e^{2\theta}}.
\end{equation}
The above function has singularity when the function $f(\theta)=e^{2\theta}-2\gamma e^{\theta}+1$ vanishes. It becomes zero when $e^{\theta}=\gamma\pm\sqrt{\gamma^2-1}$. Hence for $\gamma<1$ the solution is non-singular.

\noindent Note that if we let $k=-1$, $a=\frac{1}{4}$, $k_1=-2\bar{\eta}$, $k_2=-2\eta$,
$e^{\delta_1}=-2(\eta+\bar{\eta})e^{\bar{\theta} i}$, and $e^{\delta_2}=2(\eta+\bar{\eta})e^{i \theta}$, then one-soliton solution that we obtain here turns
to be the same solution given by Ablowitz and Musslimani \cite{AbMu3}.

\noindent \textbf{Example 16.}\, Consider the following set of the parameters: $(k_1, k_2, e^{\delta_1}, e^{\delta_2}, k, a)=(\frac{1}{2},\frac{1}{4}, -\frac{3}{4},\\
\frac{3}{4}, -1, 2)$. We obtain the following non-singular solution as
\begin{equation}\label{solex16}\displaystyle
q(t,x)=\frac{-3e^{\frac{1}{2}x-\frac{1}{64}t}}{4(1+e^{\frac{3}{4}x-\frac{9}{512}t})}.
\end{equation}
The graph of this function is given in Figure 16.
\begin{center}
\begin{figure}[h]
\centering
\begin{minipage}[t]{0.3\linewidth}
\centering
\includegraphics[angle=0,scale=.23]{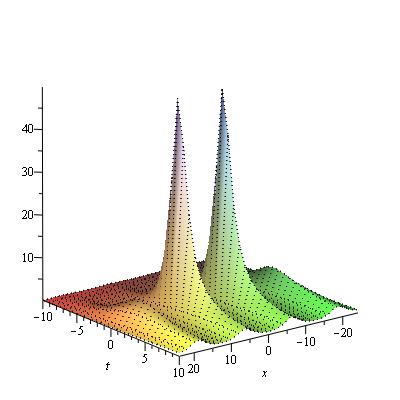}
\caption{A breather-type (one soliton) wave for (\ref{nonlocalonetype2a}) with the parameters $k_1=i\beta=i, k_2=i\gamma=-\frac{i}{2}, e^{\delta_1}=a_1+ib_1=1,
e^{\delta_2}=a_2+ib_2=\frac{1}{4}, k=\frac{1}{4}, a=i\alpha=2i$.}
\end{minipage}%
\hfill
\begin{minipage}[t]{0.3\linewidth}
\centering
\includegraphics[angle=0,scale=.23]{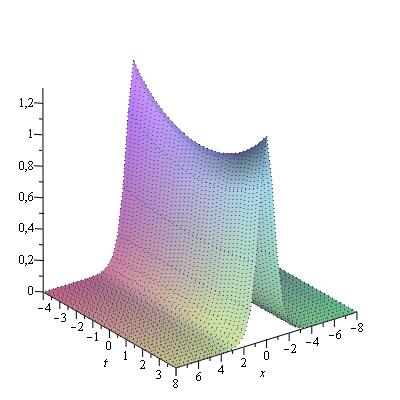}
\caption{A singular localized (one soliton) wave for (\ref{nonlocalonetype2b}) with the parameters $k_1=k_2=1, e^{\delta_1}=a_1+ib_1=-2, e^{\delta_2}=a_2+ib_2=2, k=-1, a=i\alpha=2i.$}
\end{minipage}%
\hfill
\begin{minipage}[t]{0.3\linewidth}
\centering
\includegraphics[angle=0,scale=.23]{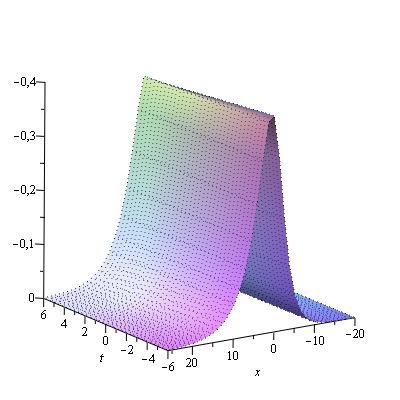}
\caption{One soliton for (\ref{nonlocalonetype2c}) with the parameters $k_1=\frac{1}{2}, k_2=\frac{1}{4}, e^{\delta_1}=a_1+ib_1=-\frac{3}{4},
e^{\delta_2}=a_2+ib_2=\frac{3}{4}, k=-1, a=2$.}
\end{minipage}%
\end{figure}
\end{center}

\subsection{Two-Soliton Solution for the Nonlocal CMKdV Equations:\\
 $(r=k\bar{q}(\varepsilon_1t,\varepsilon_2x))$}

\noindent We first obtain the conditions on the parameters of two-soliton solution of the mKdV system to satisfy (\ref{non}). Here
the function $r(t,x)$ is given in (\ref{mKdVtwosolr(t,x)}) and $k\bar{q}(\varepsilon_1t,\varepsilon_2x)$ is
\begin{equation}
\displaystyle k\bar{q}(\varepsilon_1t,\varepsilon_2x)=k\frac{e^{\bar{\theta}_1}+e^{\bar{\theta}_2}
+\bar{\gamma}_1e^{\bar{\theta}_1+\bar{\theta}_2+\bar{\eta}_1}+\bar{\gamma}_2e^{\bar{\theta}_1+\bar{\theta}_2+\bar{\eta}_2}}
{1+e^{\bar{\theta}_1+\bar{\eta}_1+\bar{\alpha}_{11}}+e^{\bar{\theta}_1+\bar{\eta}_2+\bar{\alpha}_{12}}+e^{\bar{\theta}_2+\bar{\eta}_1+\bar{\alpha}_{21}}
+e^{\bar{\theta}_2+\bar{\eta}_2+\bar{\alpha}_{22}}+\bar{M}e^{\bar{\theta}_1+\bar{\theta}_2+\bar{\eta}_1+\bar{\eta}_2}},
\end{equation}
where
\begin{align*}\displaystyle
& \bar{\theta}_i=\varepsilon_2\bar{k}_ix-\varepsilon_1\frac{\bar{k}_i^3}{4\bar{a}}t+\bar{\delta}_i, \,i=1, 2,\\
& \bar{\eta}_i=\varepsilon_2\bar{\ell}_ix-\varepsilon_1\frac{\bar{\ell}_i^3}{4\bar{a}}t+\bar{\alpha}_i, \,i=1, 2.
\end{align*}
We get the following conditions that must be satisfied:
\begin{align}
&i)\, e^{{\eta}_i}=ke^{\bar{\theta}_i}, i=1, 2,\quad ii)\, e^{\theta_1+\eta_1+\eta_2}=ke^{\bar{\theta}_1+\bar{\theta}_2+\bar{\eta}_1},\quad
iii)\,  e^{\theta_2+\eta_1+\eta_2}=ke^{\bar{\theta}_1+\bar{\theta}_2+\bar{\eta}_2}, \nonumber\\
&iv)\, \beta_i=\bar{\gamma}_i, i=1, 2, \quad v)\, e^{\theta_1+\eta_1}=e^{\bar{\theta}_1+\bar{\eta}_1}, \quad
vi)\, e^{\theta_1+\eta_2}=e^{\bar{\theta}_2+\bar{\eta}_1},\nonumber\\
&vii)\, e^{\theta_2+\eta_1}=e^{\bar{\theta}_1+\bar{\eta}_2},\quad viii)\, e^{\theta_2+\eta_2}=e^{\bar{\theta}_2+\bar{\eta}_2},\quad ix) \, e^{{\alpha}_{ij}}=e^{\bar{\alpha}_{ji}}, i, j=1, 2,\nonumber\\
&x)\, M=\bar{M},\quad xi)\, e^{\theta_1+\theta_2+\eta_1+\eta_2}=e^{\bar{\theta}_1+\bar{\theta}_2+\bar{\eta}_1+\bar{\eta}_2}.\nonumber\\
&
\end{align}
From the condition $i)$ we have $e^{\ell_ix-\frac{\ell_i^3}{4a}t+\alpha_i}=ke^{\varepsilon_2\bar{k}_ix-\varepsilon_1\frac{\bar{k}_i^3}{4\bar{a}}t+\bar{\delta}_i}$, $i=1, 2$ which gives $\ell_i=\varepsilon_2\bar{k}_i$ and $e^{\alpha_i}=ke^{\bar{\delta}_i}$, $i=1, 2$. Since
\begin{equation*}\displaystyle
-\frac{l_i^3}{4a}=-\frac{\varepsilon_2\bar{k}_i^3}{4\bar{a}\varepsilon_1\varepsilon_2}=-\frac{\bar{k}_i^3}{4\bar{a}},
 \end{equation*}
the coefficients of $t$ are equal without any additional condition. The other conditions are also satisfied by the following constraints:
\begin{equation}\label{twononlocalmKdVcond}
1)\, \bar{a}=\varepsilon_1\varepsilon_2a, \quad 2)\, \ell_i=\varepsilon_2\bar{k}_i, i=1, 2,\quad 3)\, e^{\alpha_i}=ke^{\bar{\delta}_i}, i=1, 2.
\end{equation}

\noindent Now by giving particular values to the parameters satisfying the constraints (\ref{twononlocalmKdVcond}), we present examples of two-soliton solutions for three different types of nonlocal cmKdV equations. Note that since the expressions for the real-valued function $|q(t,x)|^2$ are very long, we will only give the functions $q(t,x)$
and illustrate the graphs of $|q(t,x)|^2$.

 \subsubsection{Case a. ($T$-symmetric): $(\varepsilon_1,\varepsilon_2)=(-1,1)$, $r=k\bar{q}(-t,x)$}

\noindent In this case we have $\bar{a}=-a$, $\ell_i=\bar{k}_i$, and
$e^{\alpha_i}=ke^{\bar{\delta}_i}$, $i=1, 2$. We give the examples below for this case.

\noindent\textbf{Example 17.}\, For the parameters $(k_1, \ell_1, k_2, \ell_2)=(\frac{1}{4}, \frac{1}{4}, -\frac{1}{2}, -\frac{1}{2})$ with $(e^{\alpha_j},e^{\delta_j},k,a)=(-1+i, 1+i, -1, 10i)$, $j=1, 2$, then the solution
becomes
\begin{equation}\label{solex17}\displaystyle
q(t,x)=\frac{(1+i)[e^{\frac{1}{4}x+\frac{1}{2560}it}+e^{-\frac{1}{2}x-\frac{1}{320}it}+72e^{-\frac{3}{1280}it}+18e^{-\frac{3}{4}x-\frac{3}{512}it}]}
{1+8e^{\frac{1}{2}x+\frac{1}{1280}it}+64e^{-\frac{1}{4}x-\frac{7}{2560}it}+2e^{-x-\frac{1}{160}it}+1296e^{-\frac{1}{2}x-\frac{7}{1280}it}}
 \end{equation}
 and the graph of the function $|q(t,x)|^2$ is given in Figure 17. It is a singular solution.\\

\noindent\textbf{Example 18.}\, Choose the parameters as  $(k_1, \ell_1, k_2, \ell_2)=(\frac{4}{10}+\frac{5}{10}i, \frac{4}{10}-\frac{5}{10}i, -\frac{42}{100}-\frac{5}{10}i,  -\frac{42}{100}+\frac{5}{10}i)$ with $(e^{\alpha_j},e^{\delta_j},k,a)=(-1+i, 1+i, -1, 2i)$, $j=1, 2$. Then we have the solution $q(t,x)$,
\begin{equation}\label{solex18}\displaystyle
q(t,x)=\frac{A}{B},
\end{equation}
where
\begin{align*}\displaystyle
A=&(1+i)e^{(\frac{2}{5}+\frac{1}{2}i)x-(\frac{23}{1600}+\frac{59}{2000}i)t}+(1+i)e^{-(\frac{21}{50}+\frac{1}{2}i)x+(\frac{349}{20000}+\frac{15057}{500000}i)t}
\\&+\frac{(315517025-192682975i)}{50040008}e^{(\frac{19}{50}-\frac{1}{2}i)x+(\frac{349}{20000}-\frac{14443}{500000}i)t}
\\&+
\frac{(14955601250-10863898750i)}{2758455441}e^{(-\frac{11}{25}+\frac{1}{2}i)x+(-\frac{23}{1600}+\frac{3841}{125000}i)t},
\end{align*}
and
\begin{align*}\displaystyle
B=&1+\frac{25}{8}e^{\frac{4}{5}x-\frac{59}{1000}it}-\frac{(12495000-500000i)}{6255001}e^{(-\frac{1}{50}+i)x+(-\frac{1273}{40000}+\frac{307}{500000}i)t }+\frac{1250}{441}e^{-\frac{21}{25}x+\frac{15057}{250000}it}\\
&-\frac{(12495000+500000i)}{6255001}e^{-(\frac{1}{50}+i)x+(\frac{1273}{40000}+\frac{307}{500000}i)t}
+\frac{273136890625}{11033821764}e^{-\frac{1}{25}x+\frac{307}{250000}it},
\end{align*}
which is a singular function. The graph of the corresponding function $|q(t,x)|^2$ is given in Figure 18.

\begin{center}
\begin{figure}[h]
\centering
\begin{minipage}[t]{0.4\linewidth}
\centering
\includegraphics[angle=0,scale=.23]{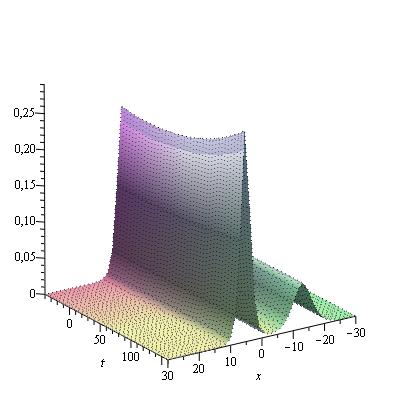}
\caption{A singular localized (two solitons) wave plotted for $|q(t,x)|^2$ corresponding to (\ref{solex17}) with the parameters
$k_1=\ell_1=\frac{1}{4}, k_2=\ell_2=-\frac{1}{2}, e^{\alpha_j}=-1+i, e^{\delta_j}=1+i, j=1, 2, k=-1, a=10i$.}
\end{minipage}%
\hfill
\begin{minipage}[t]{0.4\linewidth}
\centering
\includegraphics[angle=0,scale=.23]{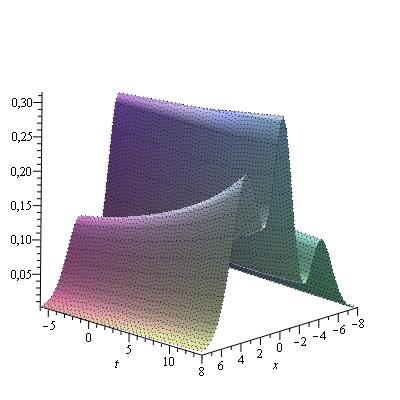}
\caption{A singular (two solitons) wave plotted for $|q(t,x)|^2$ corresponding to (\ref{solex18}) with the parameters $k_1=\frac{4}{10}+\frac{5}{10}i, \ell_1=\frac{4}{10}-\frac{5}{10}i, k_2=-\frac{42}{100}-\frac{5}{10}i, \ell_2=-\frac{42}{100}+\frac{5}{10}i, e^{\alpha_j}=-1+i, e^{\delta_j}=1+i, j=1, 2, k=-1, a=2i$.}
\end{minipage}%
\end{figure}
\end{center}

\subsubsection{Case b. ($S$-symmetric): $(\varepsilon_1,\varepsilon_2)=(1,-1)$, $r=k\bar{q}(t,-x)$}

\noindent This case gives $\bar{a}=-a$, $\ell_i=-\bar{k}_i$, and
$e^{\alpha_i}=ke^{\bar{\delta}_i}$, $i=1, 2$. Consider the following examples.

\noindent\textbf{Example 19.}\, Choose the parameters as $(k_1, \ell_1, k_2, \ell_2)=(\frac{4}{10}+\frac{5}{10}i, -\frac{4}{10}+\frac{5}{10}i, -\frac{42}{100}+\frac{5}{10}i,  \frac{42}{100}+\frac{5}{10}i )$ with $(e^{\alpha_j},e^{\delta_j},k,a)=(1, 1, 1, 2i)$, $j=1, 2$. Then the solution becomes
\begin{equation}\label{solex19}\displaystyle
q(t,x)=\frac{A}{B},
\end{equation}
where
\begin{align*}
A=&e^{(\frac{2}{5}+\frac{1}{2}i)x-(\frac{23}{1600}+\frac{59}{2000}i)t}+e^{(-\frac{21}{50}+\frac{1}{2}i)x+(-\frac{349}{2000}+\frac{15057}{500000}i)t}\\
&+\frac{(-1376739+6892100i)}{17480761}e^{(-\frac{21}{50}+\frac{3}{2}i)x+(-\frac{231}{5000}+\frac{15057}{500000}i)t}\\
&-\frac{(1376739+6892100i)}{17480761}e^{(\frac{2}{5}+\frac{3}{2}i)x-(\frac{1971}{40000}+\frac{59}{2000}i)t},
\end{align*}
and
\begin{align*}
B=&1+e^{ix-\frac{23}{800}t}+\frac{(2047500+10250000i)}{17480761}e^{(\frac{41}{50}+i)x-(\frac{1273}{40000}+\frac{29807}{50000}i)t}+e^{ix-\frac{379}{10000}t}\\
&+\frac{(2047500-10250000i)}{17480761}e^{(-\frac{41}{50}+i)x+(-\frac{1273}{4000}+\frac{29807}{500000}i)t}
+\frac{2825761}{17480761}e^{2ix-\frac{1273}{20000}t}.
\end{align*}
It is a singular solution. The graph of the corresponding function $|q(t,x)|^2$ is given in Figure 19.\\

\noindent\textbf{Example 20.}\, For the parameters $(k_1, \ell_1, k_2, \ell_2)=(\frac{4}{10}+\frac{5}{10}i, -\frac{4}{10}+\frac{5}{10}i, -\frac{42}{100}+\frac{5}{10}i,  \frac{42}{100}+\frac{5}{10}i )$ with $(e^{\alpha_j},e^{\delta_j},k,a)=(-1+i, 1+i, -1, -2i)$, $j=1, 2$, the function $q(t,x)$ becomes
\begin{equation}\label{solex20}\displaystyle
q(t,x)=\frac{A}{B},
\end{equation}
where
\begin{align*}
A=&(1+i)e^{(\frac{2}{5}+\frac{1}{2}i)x+(\frac{23}{1600}+\frac{59}{2000}i)t}+(1+i)e^{(-\frac{21}{50}+\frac{1}{2}i)x+(\frac{349}{20000}-\frac{15057}{500000}i)t}\\
&+\frac{(16537678-11030722i)}{17480761}e^{(-\frac{21}{50}+\frac{3}{2}i)x+(\frac{231}{40000}-\frac{15057}{500000}i)t}\\
&+\frac{(-11030722+16537678i)}{17480761}e^{(\frac{2}{5}+\frac{1}{2}i)x+(\frac{1971}{40000}+\frac{59}{2000}i)t},
\end{align*}
and
\begin{align*}
B=&1-2e^{ix+\frac{23}{800}t}-\frac{(4095000+20500000i)}{17480761}e^{(\frac{41}{50}+i)x+(\frac{1273}{40000}+\frac{29807}{500000}i)t}-2e^{ix+\frac{349}{10000}t}\\
&+\frac{(-4095000+20500000i)}{17480761}e^{(-\frac{41}{50}+i)x+(\frac{1273}{40000}-\frac{29807}{500000}i)t}+\frac{11303044}{17480761}e^{2ix+\frac{1273}{20000}t},
\end{align*}
which is a singular function. The graph of the function $|q(t,x)|^2$ is given in Figure 20.

\newpage
\begin{center}
\begin{figure}[h]
\centering
\begin{minipage}[t]{0.4\linewidth}
\centering
\includegraphics[angle=0,scale=.28]{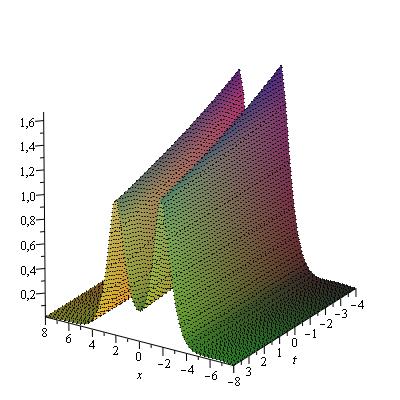}
\caption{A singular (two solitons) wave plotted for $|q(t,x)|^2$ corresponding to (\ref{solex19}) with the parameters $k_1=\frac{4}{10}+\frac{5}{10}i,\ell_1= -\frac{4}{10}+\frac{5}{10}i, k_2=-\frac{42}{100}+\frac{5}{10}i, \ell_2 \frac{42}{100}+\frac{5}{10}i, e^{\alpha_j}=e^{\delta_j}=k=1, j=1, 2, a=2i$.}
\end{minipage}%
\hfill
\begin{minipage}[t]{0.4\linewidth}
\centering
\includegraphics[angle=0,scale=.28]{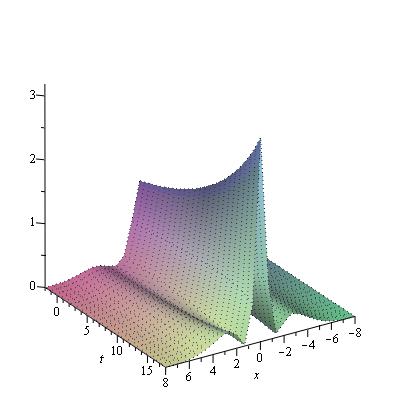}
\caption{A singular (two solitons) wave plotted for $|q(t,x)|^2$ corresponding to (\ref{solex20}) with the parameters $k_1=\frac{4}{10}+\frac{5}{10}i,\ell_1= -\frac{4}{10}+\frac{5}{10}i, k_2=-\frac{42}{100}+\frac{5}{10}i, \ell_2=\frac{42}{100}+\frac{5}{10}i, e^{\alpha_j}=-1+i, e^{\delta_j}=1+i, j=1, 2, k=-1, a=-2i$.}
\end{minipage}%
\end{figure}
\end{center}

\subsubsection{Case c. ($ST$-symmetric): $(\varepsilon_1,\varepsilon_2)=(-1,-1)$, $r=k\bar{q}(-t,-x)$}

\noindent In this case the parameters satisfy $\bar{a}=a$, $\ell_i=-\bar{k}_i$, and
$e^{\alpha_i}=ke^{\bar{\delta}_i}$, $i=1, 2$. Here we give the following examples.

\noindent\textbf{Example 21.}\, Let us choose the parameters $(k_1, \ell_1, k_2, \ell_2)=(\frac{4}{10}+\frac{5}{10}i, -\frac{4}{10}+\frac{5}{10}i, -\frac{42}{100}+\frac{5}{10}i, \frac{42}{100}+\frac{5}{10}i)$ with $(e^{\alpha_j},e^{\delta_j},k,a)=(-1+i,1+i,-1,1)$, $j=1, 2$. Then the solution becomes
\begin{equation}\label{solex21}\displaystyle
q(t,x)=\frac{A}{B}
\end{equation}
where
\begin{align*}
A=&(1+i)e^{(\frac{2}{5}+\frac{1}{2}i)x+(\frac{59}{1000}-\frac{23}{800}i)t}+(1+i)e^{(-\frac{21}{50}+\frac{1}{2}i)x-(\frac{15057}{250000}+\frac{349}{10000}i)t}\\
&+\frac{(16537678-11030722i)}{17480761}e^{(-\frac{21}{50}+\frac{3}{2}i)x-(\frac{15057}{250000}+\frac{231}{2500}i)t}\\
&+\frac{(-11030722+16537678i)}{17480761}e^{(\frac{2}{5}+\frac{3}{2}i)x+(\frac{59}{1000}-\frac{1971}{20000}i)t},
\end{align*}
and
\begin{align*}
B=&1-2e^{ix-\frac{23}{400}it}-\frac{(4095000+20500000i)}{17480761}e^{(\frac{41}{50}+i)x+(\frac{29807}{250000}-\frac{1273}{20000}i)t}-2e^{ix-\frac{349}{5000}it}\\
&+\frac{(-4095000+20500000i)}{17480761}e^{(-\frac{41}{50}+i)x-(\frac{29807}{250000}+\frac{1273}{20000}i)t}+\frac{11303044}{17480761}e^{2ix-\frac{1273}{10000}it}.
\end{align*}
It is a singular function. The graph of the function $|q(t,x)|^2$ is given in Figure 21.\\

\noindent\textbf{Example 22.}\, For the parameters $(k_1, \ell_1, k_2, \ell_2)=(2i, 2i, i, i)$ with $(e^{\alpha_j}, e^{\delta_j}, k, a)=(1, 1, 1, 1)$, $j=1, 2$, we have the non-singular
solution
\begin{equation}\label{solex22}\displaystyle
q(t,x)=\frac{e^{2ix+2it}+e^{ix+\frac{1}{4}it}+\frac{1}{144}e^{5ix+\frac{17}{4}it}+\frac{1}{36}e^{4ix+\frac{5}{2}it}}
{1+\frac{1}{16}e^{4ix+4it}+\frac{2}{9}e^{3ix+\frac{9}{4}it}+\frac{1}{4}e^{2ix+\frac{1}{2}it}+\frac{1}{5184}e^{6ix+\frac{9}{2}it}}.
\end{equation}
The graph of the function $|q(t,x)|^2$ is given in Figure 22.

\begin{center}
\begin{figure}[h]
\centering
\begin{minipage}[t]{0.4\linewidth}
\centering
\includegraphics[angle=0,scale=.24]{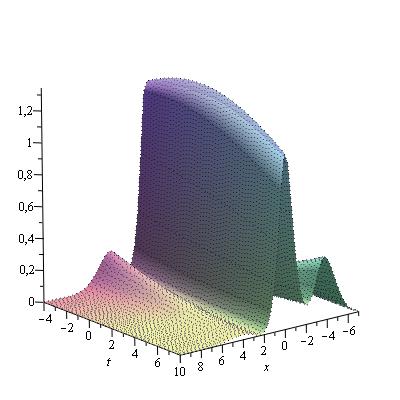}
\caption{A singular (two solitons) wave plotted for $|q(t,x)|^2$ corresponding to (\ref{solex21})
with the parameters $k_1=\frac{4}{10}+\frac{5}{10}i, \ell_1=-\frac{4}{10}+\frac{5}{10}i, k_2=-\frac{42}{100}+\frac{5}{10}i, \ell_2=\frac{42}{100}+\frac{5}{10}i), e^{\alpha_j}=-1+i, e^{\delta_j}=1+i, j=1, 2, k=-1, a=1$.}
\end{minipage}%
\hfill
\begin{minipage}[t]{0.4\linewidth}
\centering
\includegraphics[angle=0,scale=.24]{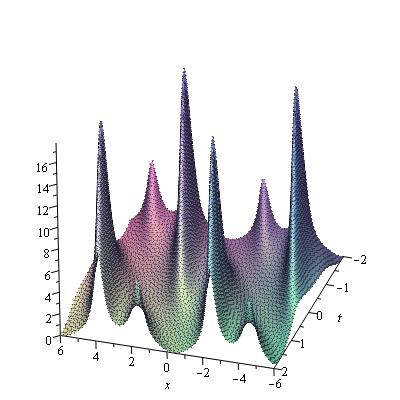}
\caption{A periodical breather-type (two solitons) wave plotted for $|q(t,x)|^2$ corresponding to (\ref{solex22})
with the parameters $k_1=\ell_1=2i, k_2=\ell_2=i, e^{\alpha_j}=e^{\delta_j}=k=a=1$, $j=1, 2$.}
\end{minipage}%
\end{figure}
\end{center}

\subsection{Three-Soliton Solution for the Nonlocal CMKdV Equations:\\
 $(r=k\bar{q}(\varepsilon_1t,\varepsilon_2x))$}

\noindent We first find the conditions on the parameters of
three-soliton solution of the mKdV system to satisfy the equality (\ref{non}) where $r(t,x)$ is given by (\ref{mKdV3solr(t,x)}) and
\begin{equation}
\displaystyle k\bar{q}(\varepsilon_1t,\varepsilon_2x)=k\frac{e^{\bar{\theta}_1}+e^{\bar{\theta}_2}+e^{\bar{\theta}_3}+\sum_{\substack{1\leq i,j,s\leq 3 \\ i<j}} \bar{A}_{ijs}e^{\bar{\theta}_i+\bar{\theta}_j+\bar{\eta}_s}
+\sum_{\substack{1\leq i,j\leq 3 \\ i<j}}\bar{V}_{ij}e^{\bar{\theta}_1+\bar{\theta}_2+\bar{\theta}_3+\bar{\eta}_i+\bar{\eta}_j}}{1+\sum_{1\leq i,j\leq3}e^{\bar{\theta}_i+\bar{\eta}_j+\bar{\alpha}_{ij}}+\sum_{\substack{1\leq i<j\leq 3 \\ 1\leq p<r \leq3}} \bar{M}_{ijpr}e^{\bar{\theta}_i+\bar{\theta}_j+\bar{\eta}_p+\bar{\eta}_r}+\bar{H}e^{\bar{\theta}_1+\bar{\theta}_2+\bar{\theta}_3+\bar{\eta}_1+\bar{\eta}_2+\bar{\eta}_3}},
\end{equation}
where
\begin{align*}\displaystyle
& \bar{\theta}_i=\varepsilon_2\bar{k}_ix-\varepsilon_1\frac{\bar{k}_i^3}{4\bar{a}}t+\bar{\delta}_i, \,i=1, 2, 3\\
& \bar{\eta}_i=\varepsilon_2\bar{\ell}_ix-\varepsilon_1\frac{\bar{\ell}_i^3}{4\bar{a}}t+\bar{\alpha}_i, \,i=1, 2, 3.
\end{align*}
Here we obtain that (\ref{non}) is satisfied by the following conditions:
\begin{equation}\label{threenonlocalmKdVcond}
1)\, \bar{a}=\varepsilon_1\varepsilon_2a, \quad 2)\, \ell_i=\varepsilon_2\bar{k}_i, i=1, 2, 3, \quad 3)\, e^{\alpha_i}=ke^{\bar{\delta}_i}, i=1, 2, 3.
\end{equation}

\noindent Now we will present some examples of three-soliton solutions of the nonlocal reductions of the mKdV system for particular parameters.\\

 \subsubsection{Case a. ($T$-Symmetric): $(\varepsilon_1,\varepsilon_2)=(-1,1)$, $r=k\bar{q}(-t,x)$}

\noindent Here the parameters satisfy $\bar{a}=-a$, $\ell_i=\bar{k}_i$, and
$e^{\alpha_i}=ke^{\bar{\delta}_i}$, $i=1, 2, 3$. We give the following example.

\noindent\textbf{Example 23.}\, For the set of parameters given by $(k_1, l_1, k_2, l_2, k_3, l_3)=(\frac{1}{4},\frac{1}{4},-\frac{1}{2},-\frac{1}{2},-\frac{1}{6},-\frac{1}{6})$
with $(e^{\alpha_j}, e^{\delta_j}, k, a)=(-1+i, 1+i, -1, 10i)$, $j=1, 2, 3$ the solution $q(t,x)$ becomes
\begin{equation}\label{solex23}\displaystyle
q(t,x)=\frac{A}{B},
\end{equation}
where
\begin{align*}
A=&(1+i)[e^{\frac{1}{4}x+\frac{1}{2560}it}+e^{-\frac{1}{2}x-\frac{1}{320}it}+e^{-\frac{1}{6}x-\frac{1}{8640}it}+72e^{-\frac{3}{1280}it}
+18e^{-\frac{3}{4}x-\frac{3}{512}it}\\
&+889e^{-\frac{5}{12}x-\frac{197}{69120}it}+200e^{\frac{1}{3}x+\frac{23}{34560}it}+450e^{-\frac{1}{12}x+\frac{11}{69120}it}
+\frac{1}{2}e^{-\frac{7}{6}x-\frac{11}{1728}it}\\
&+\frac{9}{2}e^{-\frac{5}{6}x-\frac{29}{8640}it}+5400e^{-\frac{2}{3}x-\frac{193}{34560}it}+202500e^{-\frac{1}{3}x-\frac{89}{34560}it}
+\frac{2025}{4}e^{-\frac{13}{12}x-\frac{421}{69120}it}
]
\end{align*}
and
\begin{align*}
B=&1+9e^{-\frac{2}{3}x-\frac{7}{2160}it}+64e^{-\frac{1}{4}x-\frac{7}{2560}it}+2592e^{-\frac{11}{12}x-\frac{413}{69120}it}+1296e^{-\frac{1}{2}x-\frac{7}{1280}it}\\
&+2e^{-x-\frac{1}{160}it}+16200e^{-\frac{1}{6}x-\frac{17}{6912}it}+18e^{-\frac{1}{3}x-\frac{1}{4320}it}+8e^{\frac{1}{2}x+\frac{1}{1280}it}
+\frac{9}{4}e^{-\frac{4}{3}x-\frac{7}{1080}it}\\
&+903474e^{-\frac{5}{6}x-\frac{197}{34560}it}+7200e^{-\frac{7}{12}x-\frac{41}{13824}it}+576e^{\frac{1}{12}x+\frac{19}{69120}it}+90000e^{\frac{1}{6}x+\frac{19}{34560}it}.
\end{align*}
The graph of the corresponding function $|q(t,x)|^2$ is given in Figure 23.
\begin{center}
\begin{figure}[h]
\centering
\begin{minipage}{1\linewidth}
\centering
\includegraphics[angle=0,scale=.22]{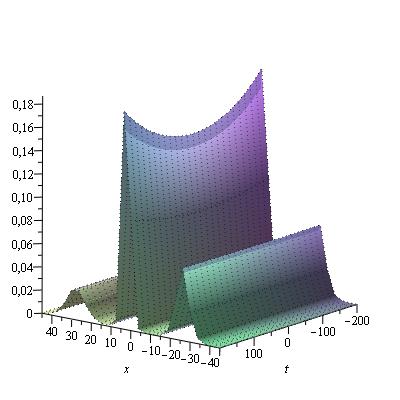}
\caption{A singular (three solitons) wave plotted for $|q(t,x)|^2$ corresponding to (\ref{solex23})
with parameters $k_1=\ell_1=\frac{1}{4}, k_2=\ell_2=-\frac{1}{2}, k_3=\ell_3=-\frac{1}{6}, e^{\alpha_j}=-1+i, e^{\delta_j}=1+i, j=1, 2, 3, k=-1, a=10i$.}
\end{minipage}
\end{figure}
\end{center}

\subsubsection{Case b. ($S$-symmetric): $(\varepsilon_1,\varepsilon_2)=(1,-1)$, $r=k\bar{q}(t,-x)$}

\noindent In this case we have $\bar{a}=-a$, $\ell_i=-\bar{k}_i$, and
$e^{\alpha_i}=ke^{\bar{\delta}_i}$, $i=1, 2, 3$. Consider the following example.

\noindent \textbf{Example 24.}\, Let us take the set of parameters as: $(k_1, l_1, k_2, l_2, k_3, l_3)=(\frac{i}{4},\frac{i}{4},-\frac{i}{2},\\
-\frac{i}{2},-\frac{i}{6},-\frac{i}{6})$
with $(e^{\alpha_j}, e^{\delta_j}, k, a)=(1, 1, 1, \frac{i}{2})$, $j=1, 2, 3$. Then we have the solution $q(t,x)$
\begin{equation}\label{solex24}\displaystyle
q(t,x)=\frac{A}{B},
\end{equation}
where
\begin{align*}
A=&e^{\frac{1}{4}ix+\frac{1}{128}t}+e^{-\frac{1}{2}ix-\frac{1}{16}t}+e^{-\frac{1}{6}ix-\frac{1}{432}t}+36e^{-\frac{3}{64}t}+9e^{-\frac{3}{4}ix-\frac{15}{128}t}\\
&+\frac{889}{2}e^{-\frac{5}{12}ix-\frac{197}{3456}t}+100e^{\frac{1}{3}ix+\frac{23}{1728}t}+225e^{-\frac{1}{12}ix+\frac{11}{3456}t}
+\frac{1}{4}e^{-\frac{7}{6}ix-\frac{55}{432}t}\\
&+\frac{9}{4}e^{-\frac{5}{6}ix-\frac{29}{432}t}+1350e^{-\frac{2}{3}ix-\frac{193}{1728}t}+50625e^{-\frac{1}{3}ix-\frac{89}{1728}t}
+\frac{2025}{16}e^{-\frac{13}{12}ix-\frac{421}{3456}t}
\end{align*}
and
\begin{align*}
B=&1+4e^{\frac{1}{2}ix+\frac{1}{64}t}+32e^{-\frac{1}{4}ix-\frac{7}{128}t}+288e^{\frac{1}{12}ix+\frac{19}{3456}t}+e^{-ix-\frac{1}{8}t}
+\frac{9}{2}e^{-\frac{2}{3}ix-\frac{7}{108}t}\\
&+9e^{-\frac{1}{3}ix-\frac{1}{216}t}+648e^{-\frac{11}{12}ix-\frac{413}{3456}t}+4050e^{-\frac{1}{6}ix-\frac{85}{1728}t}+1800e^{-\frac{7}{12}ix-\frac{205}{3456}t}\\
&+324e^{-\frac{1}{2}ix-\frac{7}{64}t}+22500e^{\frac{1}{6}ix+\frac{19}{1728}t}
+\frac{9}{16}e^{-\frac{4}{3}ix-\frac{7}{54}t}+\frac{451737}{4}e^{-\frac{5}{6}ix-\frac{197}{1728}t}.
\end{align*}
We give the graph of the function $|q(t,x)|^2$ in Figure 24.

\begin{center}
\begin{figure}[h]
\centering
\begin{minipage}{1\linewidth}
\centering
\includegraphics[angle=0,scale=.21]{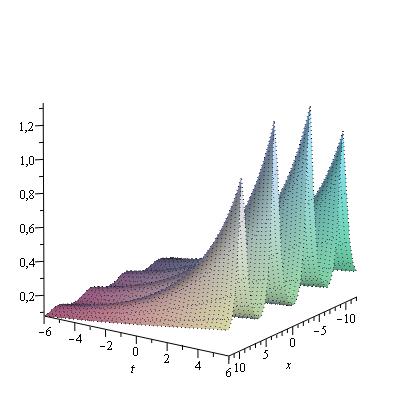}
\caption{A singular periodical (three solitons) wave plotted for $|q(t,x)|^2$ corresponding to (\ref{solex24}) with the parameters
$k_1=\ell_1=\frac{i}{4}, k_2=\ell_2=-\frac{i}{2}, k_3=\ell_3=-\frac{i}{6}, e^{\alpha_j}=e^{\delta_j}=k=1, j=1, 2, 3, a=\frac{i}{2}$.}
\end{minipage}
\end{figure}
\end{center}
\subsubsection{Case c. ($ST$-symmetric): $(\varepsilon_1,\varepsilon_2)=(-1,-1)$, $r=k\bar{q}(-t,-x)$}
\noindent Here we have $\bar{a}=a$, $\ell_i=-\bar{k}_i$, and
$e^{\alpha_i}=ke^{\bar{\delta}_i}$, $i=1, 2, 3$. We give the following example.

\noindent \textbf{Example 25.}\, Consider the following the set of parameters: $(k_1, l_1, k_2, l_2, k_3, l_3)=(\frac{i}{4},\frac{i}{4},\frac{i}{2},\frac{i}{2},\\
-\frac{i}{6},-\frac{i}{6})$
with $(e^{\alpha_j}, e^{\delta_j}, k, a)=(-1+i, 1+i, -1, 2)$, $j=1, 2, 3$. Then the solution $q(t,x)$ becomes
\begin{equation}\label{solex25}\displaystyle
q(t,x)=\frac{A}{B},
\end{equation}
where
\begin{align*}
A=&(1+i)[e^{\frac{1}{4}ix+\frac{1}{512}it}+e^{\frac{1}{2}ix+\frac{1}{64}it}+e^{-\frac{1}{6}ix-\frac{1}{1728}it}-\frac{8}{9}e^{ix+\frac{5}{256}it}
-\frac{2}{9}e^{\frac{5}{4}ix+\frac{17}{512}it}\\
&-\frac{3556}{9}e^{\frac{7}{12}ix+\frac{235}{13824}it}-200e^{\frac{1}{3}ix+\frac{23}{6912}it}-450e^{-\frac{1}{12}ix+\frac{11}{13824}it}
-8e^{\frac{5}{6}ix+\frac{53}{1728}it}\\
&-72e^{\frac{1}{6}ix+\frac{25}{1728}it}+\frac{3200}{243}e^{\frac{4}{3}ix+\frac{239}{6912}it}+40000e^{\frac{2}{3}ix+\frac{127}{6912}it}
+1600e^{\frac{11}{12}ix+\frac{443}{13824}it}]
\end{align*}
and
\begin{align*}
B=&1-18e^{-\frac{1}{3}ix-\frac{1}{864}it}+90000e^{\frac{1}{6}ix+\frac{19}{6912}it}+512e^{\frac{13}{12}ix+\frac{451}{13824}it}
-\frac{64}{9}e^{\frac{3}{4}ix+\frac{9}{512}it}\\
&+576e^{\frac{2}{3}ix+\frac{13}{432}it}-36e^{\frac{1}{3}ix+\frac{13}{864}it}-576e^{\frac{1}{12}ix+\frac{19}{13824}it}-2e^{ix+\frac{1}{32}it}
+800e^{\frac{5}{6}ix+\frac{131}{6912}it}\\
&+12800e^{\frac{5}{12}ix+\frac{227}{13824}it}-8e^{\frac{1}{2}ix+\frac{1}{256}it}
-\frac{47027200}{1323}e^{\frac{7}{6}ix+\frac{235}{6912}it}+\frac{16}{81}e^{\frac{3}{2}ix+\frac{9}{256}it}.
\end{align*}
The graph of the corresponding function $|q(t,x)|^2$ is given in Figure 25.
\begin{center}
\begin{figure}[h]
\centering
\begin{minipage}{1\linewidth}
\centering
\includegraphics[angle=0,scale=.21]{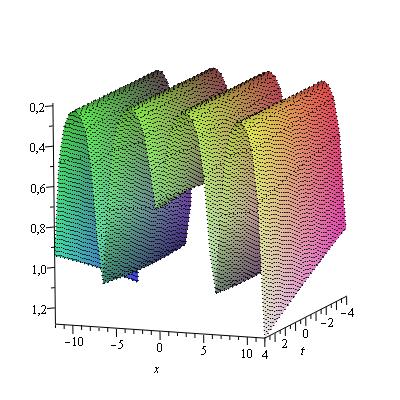}
\caption{A singular (three solitons) wave plotted for $|q(t,x)|^2$ corresponding to (\ref{solex25})
with the parameters $k_1=\ell_1=\frac{i}{4}, k_2=\ell_2=\frac{i}{2}, k_3=l_3=-\frac{i}{6}, e^{\alpha_j}=-1+i
, e^{\delta_j}=1+i, j=1, 2, 3, k=-1, a=2$.}
\end{minipage}\hfill
\end{figure}
\end{center}

\subsection{One-Soliton Solution for the Nonlocal MKdV Equations:\\
 $r=kq(-t,-x)$}

\noindent When we apply the nonlocal reduction $r(t,x)=kq(\varepsilon_1t,\varepsilon_2x)$ to the mKdV system (\ref{mKdV1}) and
(\ref{mKdV2}) we obtain the equation
\begin{equation}\displaystyle
aq_t(t,x)=-\frac{1}{4}q_{xxx}(t,x)+\frac{3}{2}kq(t,x)q(\varepsilon_1t,\varepsilon_2x)q_x(t,x)
\end{equation}
provided that $\varepsilon_1\varepsilon_2=1$ which is possible when $(\varepsilon_1,\varepsilon_2)=(1, 1)$ or
$(\varepsilon_1,\varepsilon_2)=(-1, -1)$. The first case gives a local equation. By the latter one the nonlocal
reduction becomes $r=kq(-t, -x)$ and we get the real
nonlocal $ST$-symmetric equation (\ref{mKdV44}).

\noindent If we consider one-soliton solution (\ref{mKdVsystemonesol}) obtained by the Hirota direct method with the reduction $r=kq(-t,-x)$ through the Type 1 approach,
we get $k_1=-k_2$ which gives trivial solution $q(t,x)=0$. Thus we will use the Type 2 to find nontrivial one-soliton solution
of the nonlocal equation (\ref{mKdV44}).

\noindent We have
\begin{equation}\displaystyle
\frac{e^{\theta_2}}{1+Ae^{\theta_1+\theta_2}}=k\frac{e^{\theta_1^{-}}}{1+Ae^{\theta_1^{-}+\theta_2^{-}}},
\end{equation}
so from the cross multiplication,
\begin{equation}
e^{\theta_2}+Ae^{2\delta_2}e^{\theta_1^{-}}=ke^{\theta_1^{-}}+Ake^{2\delta_1}e^{\theta_2}
\end{equation}
where
\begin{align*}
&\theta_1=k_1x-\frac{k_1^3}{4a}t+\delta_1,\quad \theta_1^{-}=-k_1x+\frac{k_1^3}{4a}t+\delta_1,\\
&\theta_2=k_2x-\frac{k_2^3}{4a}t+\delta_2,\quad \theta_2^{-}=-k_2x+\frac{k_2^3}{4a}t+\delta_2.\quad
\end{align*}
Hence we obtain the conditions
\begin{equation}
i)\, Ake^{2\delta_1}=1,\quad ii)\, Ae^{2\delta_2}=k,
\end{equation}
yielding $e^{\delta_1}=\pm i\frac{(k_1+k_2)}{\sqrt{k}}$ and $e^{\delta_2}=\pm i\sqrt{k}(k_1+k_2)$. Therefore
one-soliton solution of the equation (\ref{mKdV44}) is
\begin{equation}\label{realstsoln}\displaystyle
q(t,x)=\frac{\sigma_1 e^{k_1x-\frac{k_1^3}{4a}t}(k_1+k_2)}{\sqrt{k}(1+\sigma_2e^{(k_1+k_2)x-\frac{(k_1^3+k_2^3)}{4a}t})}, \quad \sigma_j=\pm 1, j=1, 2.
\end{equation}
The solution (\ref{realstsoln}) is non-singular if $\sigma_2=1$.

\noindent Assuming $a=\frac{1}{4}$, $k=-1$, $k_1=-2\eta$, and $k_2=-2\bar{\eta}$, the solution (\ref{realstsoln}) is reduced to the one given by Ablowitz and Musslimani \cite{AbMu3}. In \cite{JZ1}, Ji and
Zhu also considered the equation (\ref{mKdV44}) and found one- and two-soliton solution of this equation by using the Darboux transformation. The solution (\ref{realstsoln}) can be transformed to the one-soliton solution in \cite{JZ1} by taking $a=\frac{1}{4}$, $k=-1$, $k_1=4i\mu_1$ and $k_2=-2v_1+2i\mu_1$. Besides this type of solution they obtained rogue-wave and rational solutions of (\ref{mKdV44}). In \cite{JZ2}, Ji and Zhu obtained the same one-soliton solution through inverse scattering transform. Let us give particular examples.

\noindent \textbf{Example 26.}\, Consider the following set of the parameters $(k_1, k_2, \sigma_1, \sigma_2, k, a)=(\frac{1}{2}, \frac{1}{2}, 1, 1, \\-1, 2)$.
Then the solution becomes
\begin{equation}\label{solex26}\displaystyle
q(t,x)=\frac{e^{\frac{1}{2}x-\frac{1}{64}t}}{1+e^{x-\frac{1}{32}t}}=\frac{1}{2}\mathrm{sech}\Big(\frac{1}{2}x-\frac{1}{64}t\Big).
\end{equation}
It is one-soliton solution. The graph of (\ref{solex26})  is given in Figure 26.

\noindent \textbf{Example 27.}\, Choose $(k_1, k_2, \sigma_1, \sigma_2, k, a)=(\frac{1}{2}, 1, -1, 1, 1, 2)$.
Then the function $|q(t,x)|^2$ corresponding to one-soliton solution becomes
\begin{equation}\label{solex27}\displaystyle
|q(t,x)|^2=\frac{9e^{x+\frac{1}{4}t}}{4(e^{\frac{9}{64}t}+e^{\frac{3}{2}x})^2}
\end{equation}
which is a nonsingular function. It is a soliton-like solution. The graph of (\ref{solex27}) is given in Figure 27.\\

\begin{center}
\begin{figure}[h]
\centering
\begin{minipage}[t]{0.4\linewidth}
\centering
\includegraphics[angle=0,scale=.21]{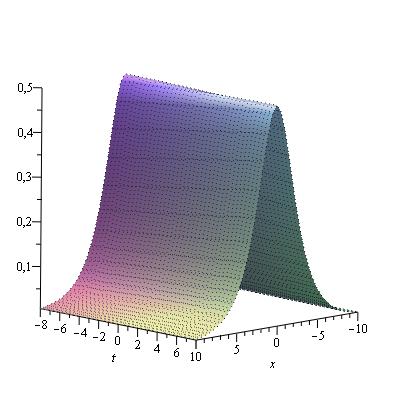}
\caption{One soliton for (\ref{realstsoln}) with the parameters
$k_1=k_2=\frac{1}{2}, \sigma_1=\sigma_2=1, k=-1, a=2.$}
\end{minipage}%
\hfill
\begin{minipage}[t]{0.4\linewidth}
\centering
\includegraphics[angle=0,scale=.21]{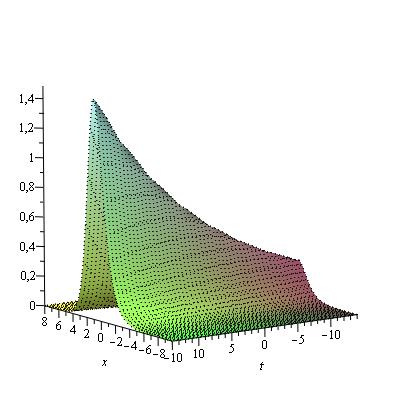}
\caption{A soliton-like wave for $|q(t,x)|^2$ corresponding to (\ref{realstsoln})
with the parameters $k_1=\frac{1}{2}, k_2=1, \sigma_1=-1, \sigma_2=1, k=1, a=2$.}
\end{minipage}%
\end{figure}
\end{center}

\noindent \textbf{Example 28.}\, Take $(k_1, k_2, \sigma_1, \sigma_2, k, a)=(i, 1+\frac{i}{2}, 1, 1, -1, \frac{1}{4})$.
Then we have the one-soliton solution
\begin{equation}\displaystyle
q(t,x)=\frac{(1+\frac{3}{2}i)e^{ix+it}}{1+e^{(1+\frac{3}{2}i)x-(\frac{1}{4}+\frac{3}{8}i)t}}
\end{equation}
so
\begin{equation}\label{solex28}\displaystyle
|q(t,x)|^2=\frac{13e^{\frac{3}{4}t}}{4[(e^{x+\frac{1}{8}t}+e^{\frac{3}{8}t})^2+2e^{x+\frac{1}{2}t}\cos(\frac{3}{2}x-\frac{3}{8}t)]}.
\end{equation}
This is a complexiton solution. The graph of (\ref{solex28}) is given in Figure 28.\\

\noindent \textbf{Example 29.}\, Take $(k_1, k_2, \sigma_1, \sigma_2, k, a)=(\frac{i}{2}, -4+\frac{i}{4}, 1, 1, 1, 10)$.
So one-soliton solution becomes
\begin{equation}\displaystyle
q(t,x)=\frac{-(\frac{3}{4}+4i)e^{\frac{1}{2}ix+\frac{1}{320}it}}{1+e^{(-4+\frac{3}{4}i)x+(\frac{253}{160}-\frac{759}{2560}i)t}}
\end{equation}
so the function $|q(t,x)|^2$ is
\begin{equation}\label{solex29}\displaystyle
|q(t,x)|^2=\frac{265e^{8x}}{16[(e^{4x}+e^{\frac{253}{160}t})^2+2e^{4x+\frac{253}{160}t}\cos(\frac{3}{4}x-\frac{759}{2560}t)]}.
\end{equation}
This is a kink-type solution. The graph of (\ref{solex29}) is given in Figure 29.\\

\begin{center}
\begin{figure}[h]
\centering
\begin{minipage}[t]{0.4\linewidth}
\centering
\includegraphics[angle=0,scale=.21]{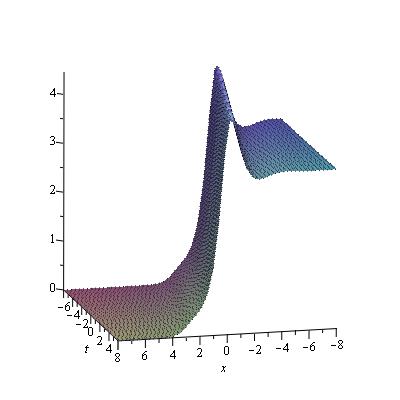}
\caption{A complexiton for $|q(t,x)|^2$ corresponding to (\ref{solex28}) with the parameters
$k_1=i, k_2=1+\frac{i}{2}, \sigma_1=\sigma_2=1, k=-1, a=\frac{1}{4}$.}
\end{minipage}%
\hfill
\begin{minipage}[t]{0.4\linewidth}
\centering
\includegraphics[angle=0,scale=.21]{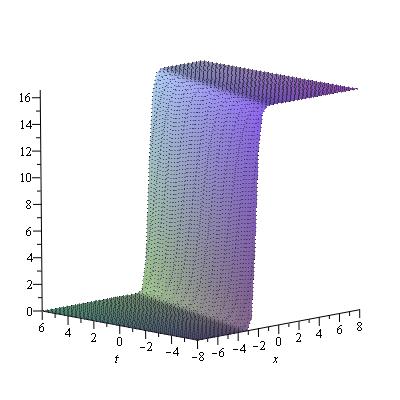}
\caption{A kink-type wave for $|q(t,x)|^2$ corresponding to (\ref{solex29})
with the parameters $k_1=\frac{i}{2}, k_2=-4+\frac{i}{4}, \sigma_1=\sigma_2=k=1, a=10$.}
\end{minipage}%
\end{figure}
\end{center}

\subsection{Two-Soliton Solution for the Nonlocal MKdV Equations:\\
 $r=kq(-t,-x)$}

\noindent Similar to the one-soliton case, if we use two-soliton solution (\ref{mKdVtwosolq(t,x)})-(\ref{mKdVtwosolr(t,x)}) with
the nonlocal reduction $r=kq(-t,-x)$ and apply cross multiplication we obtain the following constraints to be satisfied
by two-soliton solution of (\ref{mKdV44}),
\begin{equation}\displaystyle
e^{\delta_j}=\tau_r i\frac{(k_j+\ell_1)(k_j+\ell_2)}{\sqrt{k}(k_1-k_2)},\quad e^{\alpha_j}=\rho_r i\sqrt{k}\frac{(k_1+\ell_j)(k_2+\ell_j)}{(\ell_1-\ell_2)}, \tau_r=\pm 1, \rho_r=\pm 1, r=1, 2,
\end{equation}
 for $j=1, 2$. Hence two-soliton solution of (\ref{mKdV44}) is
 \begin{equation}\label{twosolq(-t,-x)}\displaystyle
 q(t,x)=\frac{A}{B},
 \end{equation}
 where
 \begin{align}\label{twosolq(-t,-x)A}\displaystyle
 A=&\frac{i}{\sqrt{k}(k_1-k_2)}[\tau_1(k_1+\ell_2)(k_1+\ell_1)e^{k_1x-\frac{k_1^3}{4a}t}+\tau_2(k_2+\ell_2)(k_2+\ell_1)e^{k_2x-\frac{k_2^3}{4a}t}]\nonumber\\
 &+\frac{i\tau_1\tau_2}{\sqrt{k}(\ell_1-\ell_2)}[ (k_1+\ell_2)(k_2+\ell_2)\rho_1e^{(k_1+k_2+\ell_1)x-\frac{(k_1^3+k_2^3+\ell_1^3)}{4a}t}\nonumber\\
 &+(k_1+\ell_1)(k_2+\ell_1)\rho_2e^{(k_1+k_2+\ell_2)x-\frac{(k_1^3+k_2^3+\ell_2^3)}{4a}t}]
 \end{align}
 and
 \begin{align}\label{twosolq(-t,-x)B}\displaystyle
 B=&1+\frac{1}{(k_1-k_2)(\ell_1-\ell_2)}[\tau_1\rho_1(k_1+\ell_2)(k_2+\ell_1)e^{(k_1+\ell_1)x-\frac{(k_1^3+\ell_1^3)}{4a}t}\nonumber\\
 &+\tau_1\rho_2(k_1+\ell_1)(k_2+\ell_2)e^{(k_1+\ell_2)x-\frac{(k_1^3+\ell_2^3)}{4a}t}+\tau_2\rho_1(k_2+\ell_2)(k_1+\ell_1)e^{(k_2+\ell_1)x
 -\frac{(k_2^3+\ell_1^3)}{4a}t}\nonumber\\
 &+\tau_2\rho_2(k_2+\ell_1)(k_1+\ell_2)e^{(k_2+\ell_2)x-\frac{(k_2^3+\ell_2^3)}{4a}t}]+\tau_1\tau_2\rho_1\rho_2
 e^{(k_1+k_2+\ell_1+\ell_2)x-\frac{(k_1^3+k_2^3+\ell_1^3+\ell_2^3)}{4a}t}.
 \end{align}
\noindent Now we consider some particular examples.\\

\noindent \textbf{Example 30.}\, Take the parameters as $(k_1, k_2, \ell_1, \ell_2, k, a)=(1,\frac{1}{2}, 1,\frac{1}{2}, -1, \frac{1}{4})$ with
$\tau_j=\rho_j=1$, $j=1, 2$. The solution becomes
\begin{equation}\label{solex30}\displaystyle
q(t,x)=\frac{6e^{x-t}+3e^{\frac{1}{2}x-\frac{1}{8}t}+3e^{\frac{5}{2}x-\frac{17}{8}t}+6e^{2x-\frac{5}{4}t}}
{1+9e^{2x-2t}+16e^{\frac{3}{2}x-\frac{9}{8}t}+9e^{x-\frac{1}{4}t}+e^{3x-\frac{9}{4}t}}.
\end{equation}
This is a non-singular solution and its graph is given in Figure 30.\\

\noindent \textbf{Example 31.}\, Consider the following set of parameters: $(k_1, k_2, \ell_1, \ell_2, k, a)=(1,0, 1,\frac{1}{2}, -1, \frac{1}{4})$ with
$\tau_j=\rho_j=1$, $j=1, 2$. We have the solution
\begin{equation}\label{solex31}\displaystyle
q(t,x)=\frac{1+3e^{x-t}+\frac{3}{2}e^{2x-2t}+4e^{\frac{3}{2}x-\frac{9}{8}t}}
{1+3e^{2x-2t}+2e^{\frac{3}{2}x-\frac{9}{8}t}+2e^{x-t}+3e^{\frac{1}{2}x-\frac{1}{8}t}+e^{\frac{5}{2}x-\frac{17}{8}t}}.
\end{equation}
The graph of (\ref{solex31}) is given in Figure 31. It represents interaction of soliton and kink-type wave.

\begin{center}
\begin{figure}[h]
\centering
\begin{minipage}[t]{0.4\linewidth}
\centering
\includegraphics[angle=0,scale=.21]{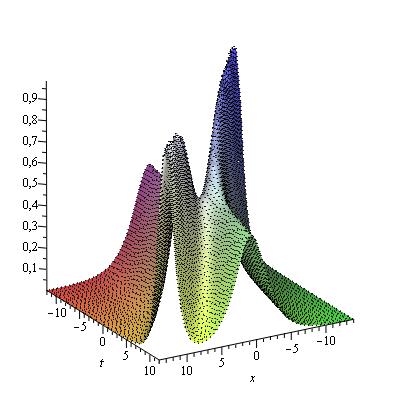}
\caption{Two solitons for (\ref{twosolq(-t,-x)}) with the parameters $k_1=\ell_1=1, k_2=\ell_2=\frac{1}{2}, k=-1, a=\frac{1}{4}, \tau_j=\rho_j=1$, $j=1, 2$.}
\end{minipage}%
\hfill
\begin{minipage}[t]{0.4\linewidth}
\centering
\includegraphics[angle=0,scale=.21]{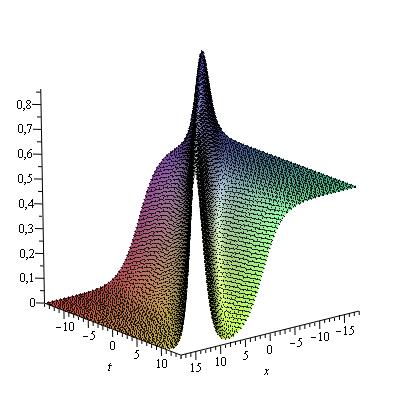}
\caption{Soliton and kink-type wave for (\ref{twosolq(-t,-x)}) with the parameters $k_1=\ell_1=1, k_2=0, \ell_2=\frac{1}{2}, k=-1, a=\frac{1}{4}, \tau_j=\rho_j=1$, $j=1, 2$.}
\end{minipage}%
\end{figure}
\end{center}

\noindent \textbf{Example 32.}\, Choose $(k_1, k_2, \ell_1, \ell_2, k, a)=(1,0, 2,\frac{1}{2}, -1, \frac{1}{4})$ with
$\tau_j=\rho_j=-1$, $j=1, 2$. The two-soliton solution becomes
\begin{equation}\label{solex32}\displaystyle
q(t,x)=-\frac{1+\frac{9}{2}e^{x-t}+\frac{1}{2}e^{3x-9t}+4e^{\frac{3}{2}x-\frac{9}{8}t}}
{1+2e^{3x-9t}+e^{\frac{3}{2}x-\frac{9}{8}t}+e^{2x-8t}+2e^{\frac{1}{2}x-\frac{1}{8}t}+e^{\frac{7}{2}x-\frac{73}{8}t}}
\end{equation}
The graph of the above solution is given in Figure 32. This solution represents interaction of soliton and kink-type wave.

\noindent \textbf{Example 33.}\, Choose $(k_1, k_2, \ell_1, \ell_2, k, a)=(1+\frac{i}{2},-1+\frac{i}{2},1+\frac{i}{2},-1+\frac{i}{2}, -1,\frac{1}{4})$ with
$\tau_j=\rho_j=1$, $j=1, 2$. Then the two-soliton solution becomes
\begin{equation}\label{solex33}\displaystyle
q(t,x)=\frac{e^{(1+\frac{1}{2}i)x-(\frac{1}{4}+\frac{11}{8}i)t}
+e^{(-1+\frac{1}{2}i)x+(\frac{1}{4}-\frac{11}{8}i)t}+e^{(1+\frac{3}{2}i)x-(\frac{1}{4}+\frac{33}{8}i)t}
+e^{(-1+\frac{3}{2}i)x+(\frac{1}{4}-\frac{33}{8}i)t}}
{1-\frac{1}{4}e^{(2+i)x-(\frac{1}{2}+\frac{11}{4}i)t}-\frac{5}{2}e^{ix-\frac{11}{4}it}
-\frac{1}{4}e^{(-2+i)x+(\frac{1}{2}-\frac{11}{4}i)t}+e^{2ix-\frac{11}{2}it}  }
\end{equation}
This solution is a breather-type wave solution. The graph of the corresponding function $|q(t,x)|^2$ is given in Figure 33.

\begin{center}
\begin{figure}[h]
\centering
\begin{minipage}[t]{0.4\linewidth}
\centering
\includegraphics[angle=0,scale=.21]{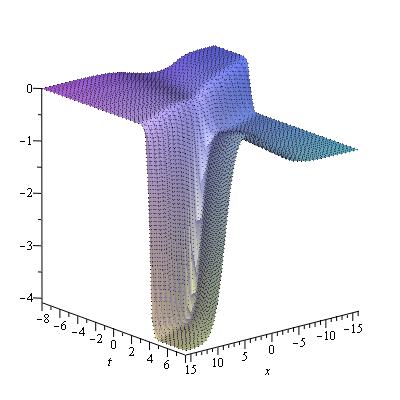}
\caption{Soliton and kink-type wave for (\ref{twosolq(-t,-x)}) with the parameters $k_1=1, k_2=0, \ell_1=2, \ell_2=\frac{1}{2}, k=-1, a=\frac{1}{4}, \tau_j=\rho_j=-1$, $j=1, 2$.}
\end{minipage}%\begin{minipage}[t]{0.24\linewidth}
\hfill
\begin{minipage}[t]{0.4\linewidth}
\centering
\includegraphics[angle=0,scale=.21]{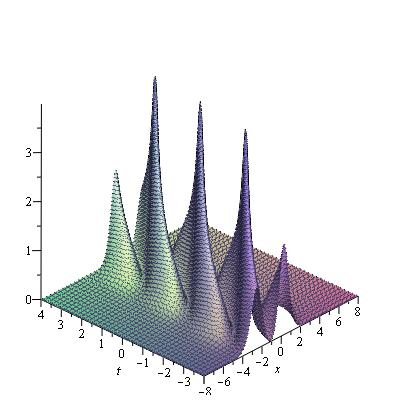}
\caption{A breather-type wave plotted for $|q(t,x)|^2$ corresponding to (\ref{twosolq(-t,-x)}) with the parameters
$k_1=\ell_1=1+\frac{i}{2}, k_2=\ell_2=-1+\frac{i}{2}, k=-1, a=\frac{1}{4}, \tau_j=\rho_j=1$, $j=1, 2$.}
\end{minipage}%
\end{figure}
\end{center}

Note that in \cite{JZ1} and \cite{JZ2}, Ji and
Zhu also obtained two-soliton solution of the equation (\ref{mKdV44}) representing interactions of bright-bright solitons, bright-dark solitons, soliton-kink, and also breather solutions, by using the Darboux transformation and inverse scattering transform, respectively. Actually, if we let $a=\frac{1}{4}$, $k=-1$, $k_1=-2v_2-2i\mu_2$, $k_2=2v_2-2i\mu_2$, $l_1=-2v_1+2i\mu_1$, $l_2=2v_1+2i\mu_1$, and $\rho_j=\tau_j=1$, $j=1, 2$, our solution (\ref{twosolq(-t,-x)}) with (\ref{twosolq(-t,-x)A}) and (\ref{twosolq(-t,-x)B}) can be transformed to the two-soliton solution given in \cite{JZ1}. Similarly, assuming $a=\frac{1}{4}$, $k=-1$, and using the notations for the parameters in \cite{JZ2} as $k_1=-2i\bar{k}_1$, $k_2=-2i\bar{k}_2$, $l_1=2ik_1$, $l_2=2ik_2$, and $\tau_j=\bar{\sigma}_j$, $\rho_j=\sigma_j$, $j=1,2$, our two-soliton solution turns to be the solution given in \cite{JZ2}.

By using the same approach we can also find three-soliton solution of the nonlocal equation (\ref{mKdV44}).

\section{Concluding Remarks}

In this work, we have studied local and nonlocal reductions of the mKdV system of nonlinear equations. There are two local and also two nonlocal reductions of these equations. Nonlocal reductions provide us  time ($T$)-, space ($S$)-, and space-time ($ST$)-reversal symmetric nonlocal cmKdV equations and space-time ($ST$)-reversal symmetric nonlocal mKdV equation. Each reduced equations are integrable by construction. It means that they have Lax pairs and recursion operators. Starting from any one of the nonlocal cmKdV equations we can generate infinitely many other nonlocal higher order cmKdV equations. Here in this work we mainly focus on the soliton solutions of the nonlocal cmKdV and nonlocal mKdV equations. When we use the Hirota method to obtain soliton solutions of these equations we observed that there are two different types.  We present one-soliton solutions of the nonlocal equations of all types but we present only one type of two- and three- soliton solutions.  We also plot the graphs of these solutions for particular values of the parameters of the solutions.

From the study of NLS and mKdV systems we observed that they both have local and nonlocal reductions. Moreover in both of these systems there corresponds at least one nonlocal reduction to a local reduction. Both systems have $r(t,x)=k \bar{q}(t,x)$ as local reduction and the corresponding nonlocal reductions are $r(t,x)=k \bar{q}(\epsilon_{1} t,\epsilon_{2}x)$ where $k$ is real constant and $ (\epsilon_{1}, \epsilon_{2})=(1,-1),(-1,1), (-1,-1)$. From these reductions we obtain local and nonlocal NLS equations and local and nonlocal complex mKdV equations. The mKdV system has additional local and nonlocal reductions. Local reduction $r(t,x)=k q(t,x)$, $k$ is real constant, and its corresponding nonlocal reduction $r(t,x)=k q(-t,-x)$ give the nonlocal mKdV equation. From all these experiences we conclude with a conjecture: If a system of equations admits a local reduction then there exists at least one corresponding nonlocal reduction of the same system.

\section{Appendix}
In this section we present the coefficients explicitly given in Section 2.3.
\begin{align*}
&S_{12}=3M_{2312}(k_1^2\ell_1-k_1k_2^2-k_1\ell_1^2+k_2^2\ell_1+k_2\ell_1^2-k_1k_3^2+k_3^2\ell_1+k_3\ell_1^2+k_2k_3^2+k_1^2\ell_2-k_1\ell_2^2+k_2^2\ell_2\\
&+k_2\ell_2^2+k_3^2\ell_2+k_3\ell_2^2-2k_1k_2\ell_2-2k_1k_3\ell_1-2k_1k_2\ell_1-2k_1k_3\ell_2+2k_2k_3\ell_1+2k_2k_3\ell_2+2k_3\ell_1\ell_2\\
&+k_1^2k_2+k_1^2k_3+k_2^2k_3+2k_2\ell_1\ell_2-2k_1\ell_1\ell_2+\ell_1^2\ell_2+\ell_1\ell_2^2-2k_1k_2k_3)\\
&+3M_{1312}(k_1^2\ell_1+k_1k_2^2+k_1\ell_1^2+k_2^2\ell_1-k_2\ell_1^2+k_1k_3^2+k_3^2\ell_1+k_3\ell_1^2-k_2k_3^2+k_1^2\ell_2+k_1\ell_2^2+k_2^2\ell_2\\
&-k_2\ell_2^2+k_3^2\ell_2+k_3\ell_2^2-2k_1k_2\ell_2+2k_1k_3\ell_2-2k_1k_2\ell_1-2k_2k_3\ell_2-2k_2k_3\ell_1+2k_1k_3\ell_1+2k_3\ell_1\ell_2\\
&-k_1^2k_2+k_1^2k_3+k_2^2k_3-2k_2\ell_1\ell_2+\ell_1^2\ell_2+\ell_1\ell_2^2+2k_1\ell_1\ell_2-2k_1k_2k_3)\\
&+3M_{1212}(k_1^2\ell_1+k_1k_2^2+k_1\ell_1^2+k_2^2\ell_1+k_2\ell_1^2+k_1k_3^2+k_3^2\ell_1-k_3\ell_1^2+k_2k_3^2+k_1^2\ell_2+k_1\ell_2^2+k_2^2\ell_2\\
&+k_2\ell_2^2+k_3^2\ell_2-k_3\ell_2^2+2k_1k_2\ell_2-2k_1k_3\ell_2+2k_1k_2\ell_1-2k_2k_3\ell_2-2k_2k_3\ell_1-2k_1k_3\ell_1-2k_3\ell_1\ell_2\\
&+k_1^2k_2-k_1^2k_3-k_2^2k_3+2k_2\ell_1\ell_2+\ell_1^2\ell_2+\ell_1\ell_2^2+2k_1\ell_1\ell_2-2k_1k_2k_3)\\
&+\frac{3A_{122}}{(k_3+\ell_1)^2}(-k_1^2\ell_1+k_1k_2^2+k_1\ell_1^2-k_2^2\ell_1+k_2\ell_1^2+k_1k_3^2-k_3^2\ell_1-k_3\ell_1^2+k_2k_3^2+k_1^2\ell_2+k_1\ell_2^2+k_2^2\ell_2\\
&+k_2\ell_2^2+k_3^2\ell_2-k_3\ell_2^2+2k_1k_2\ell_2+2k_1k_3\ell_1-2k_1k_2\ell_1-2k_1k_3\ell_2+2k_2k_3\ell_1-2k_2k_3\ell_2+2k_3\ell_1\ell_2\\
&+k_1^2k_2-k_1^2k_3-k_2^2k_3-2k_2\ell_1\ell_2-2k_1\ell_1\ell_2+\ell_1^2\ell_2-\ell_1\ell_2^2-2k_1k_2k_3)\\
&+\frac{3A_{132}}{(k_2+\ell_1)^2}(-k_1^2\ell_1+k_1k_2^2+k_1\ell_1^2-k_2^2\ell_1-k_2\ell_1^2+k_1k_3^2-k_3^2\ell_1+k_3\ell_1^2-k_2k_3^2+k_1^2\ell_2+k_1\ell_2^2+k_2^2\ell_2\\
&-k_2\ell_2^2+k_3^2\ell_2+k_3\ell_2^2-2k_1k_2\ell_2-2k_1k_3\ell_1+2k_1k_2\ell_1+2k_1k_3\ell_2+2k_2k_3\ell_1-2k_2k_3\ell_2-2k_3\ell_1\ell_2\\
&-k_1^2k_2+k_1^2k_3+k_2^2k_3+2k_2\ell_1\ell_2-2k_1\ell_1\ell_2+\ell_1^2\ell_2-\ell_1\ell_2^2-2k_1k_2k_3)\\
&+\frac{3A_{131}}{(k_2+\ell_2)^2}(k_1^2\ell_1+k_1k_2^2+k_1\ell_1^2+k_2^2\ell_1-k_2\ell_1^2+k_1k_3^2+k_3^2\ell_1+k_3\ell_1^2-k_2k_3^2-k_1^2\ell_2+k_1\ell_2^2-k_2^2\ell_2\\
&-k_2\ell_2^2-k_3^2\ell_2+k_3\ell_2^2+2k_1k_2\ell_2+2k_1k_3\ell_1-2k_1k_2\ell_1-2k_1k_3\ell_2-2k_2k_3\ell_1+2k_2k_3\ell_2-2k_3\ell_1\ell_2\\
&-k_1^2k_2+k_1^2k_3+k_2^2k_3+2k_2\ell_1\ell_2-2k_1\ell_1\ell_2-\ell_1^2\ell_2+\ell_1\ell_2^2-2k_1k_2k_3)\\
&+\frac{3A_{121}}{(k_3+\ell_2)^2}(k_1^2\ell_1+k_1k_2^2+k_1\ell_1^2+k_2^2\ell_1+k_2\ell_1^2+k_1k_3^2+k_3^2\ell_1-k_3\ell_1^2+k_2k_3^2-k_1^2\ell_2+k_1\ell_2^2-k_2^2\ell_2\\
&+k_2\ell_2^2-k_3^2\ell_2-k_3\ell_2^2-2k_1k_2\ell_2-2k_1k_3\ell_1+2k_1k_2\ell_1+2k_1k_3\ell_2-2k_2k_3\ell_1+2k_2k_3\ell_2+2k_3\ell_1\ell_2\\
&+k_1^2k_2-k_1^2k_3-k_2^2k_3-2k_2\ell_1\ell_2-2k_1\ell_1\ell_2-\ell_1^2\ell_2+\ell_1\ell_2^2-2k_1k_2k_3)\\
&+\frac{3A_{231}}{(k_1+\ell_2)^2}(k_1^2\ell_1-k_1k_2^2-k_1\ell_1^2+k_2^2\ell_1+k_2\ell_1^2-k_1k_3^2+k_3^2\ell_1+k_3\ell_1^2+k_2k_3^2-k_1^2\ell_2-k_1\ell_2^2-k_2^2\ell_2\\
&+k_2\ell_2^2-k_3^2\ell_2+k_3\ell_2^2+2k_1k_2\ell_2-2k_1k_3\ell_1-2k_1k_2\ell_1+2k_1k_3\ell_2+2k_2k_3\ell_1-2k_2k_3\ell_2-2k_3\ell_1\ell_2\\
&+k_1^2k_2+k_1^2k_3+k_2^2k_3-2k_2\ell_1\ell_2+2k_1\ell_1\ell_2-\ell_1^2\ell_2+\ell_1\ell_2^2-2k_1k_2k_3)\\
&+\frac{3A_{232}}{(k_1+\ell_1)^2}(-k_1^2\ell_1-k_1k_2^2-k_1\ell_1^2-k_2^2\ell_1+k_2\ell_1^2-k_1k_3^2-k_3^2\ell_1+k_3\ell_1^2+k_2k_3^2+k_1^2\ell_2-k_1\ell_2^2+k_2^2\ell_2\\
&+k_2\ell_2^2+k_3^2\ell_2+k_3\ell_2^2-2k_1k_2\ell_2+2k_1k_3\ell_1+2k_1k_2\ell_1-2k_1k_3\ell_2-2k_2k_3\ell_1+2k_2k_3\ell_2-2k_3\ell_1\ell_2\\
&+k_1^2k_2+k_1^2k_3+k_2^2k_3-2k_2\ell_1\ell_2+2k_1\ell_1\ell_2+\ell_1^2\ell_2-\ell_1\ell_2^2-2k_1k_2k_3),
\end{align*}
\begin{align*}
&S_{13}=3M_{1313}(k_1^2\ell_1+k_1k_2^2+k_1\ell_1^2+k_2^2\ell_1-k_2\ell_1^2+k_1k_3^2+k_3^2\ell_1+k_3\ell_1^2-k_2k_3^2+k_1^2\ell_3+k_1\ell_3^2+k_2^2\ell_3\\
&-k_2\ell_3^2+k_3^2\ell_3+k_3\ell_3^2-2k_1k_2\ell_3-2k_1k_2\ell_1-2k_2k_3\ell_3-2k_2k_3\ell_1+2k_1k_3\ell_1+2k_1k_3\ell_3+2k_1\ell_1\ell_3\\
&-2k_2\ell_1\ell_3+2k_3\ell_1\ell_3-k_1^2k_2+k_1^2k_3+k_2^2k_3+\ell_1^2\ell_3+\ell_1\ell_3^2-2k_1k_2k_3)\\
&+3M_{1213}(k_1^2\ell_1+k_1k_2^2+k_1\ell_1^2+k_2^2\ell_1+k_2\ell_1^2+k_1k_3^2+k_3^2\ell_1-k_3\ell_1^2+k_2k_3^2+k_1^2\ell_3+k_1\ell_3^2+k_2^2\ell_3\\
&+k_2\ell_3^2+k_3^2\ell_3-k_3\ell_3^2+2k_1k_2\ell_3+2k_1k_2\ell_1-2k_2k_3\ell_3-2k_2k_3\ell_1-2k_1k_3\ell_1-2k_1k_3\ell_3+2k_1\ell_1\ell_3\\
&+2k_2\ell_1\ell_3-2k_3\ell_1\ell_3+k_1^2k_2-k_1^2k_3-k_2^2k_3+\ell_1^2\ell_3+\ell_1\ell_3^2-2k_1k_2k_3)\\
&+3M_{2313}(k_1^2\ell_1-k_1k_2^2-k_1\ell_1^2+k_2^2\ell_1+k_2\ell_1^2-k_1k_3^2+k_3^2\ell_1+k_3\ell_1^2+k_2k_3^2+k_1^2\ell_3-k_1\ell_3^2+k_2^2\ell_3\\
&+k_2\ell_3^2+k_3^2\ell_3+k_3\ell_3^2-2k_1k_2\ell_3-2k_1k_2\ell_1+2k_2k_3\ell_3+2k_2k_3\ell_1-2k_1k_3\ell_1-2k_1k_3\ell_3-2k_1\ell_1\ell_3\\
&+2k_2\ell_1\ell_3+2k_3\ell_1\ell_3+k_1^2k_2+k_1^2k_3+k_2^2k_3+\ell_1^2\ell_3+\ell_1\ell_3^2-2k_1k_2k_3)\\
&+\frac{3A_{123}}{(k_3+\ell_1)^2}(-k_1^2\ell_1+k_1k_2^2+k_1\ell_1^2-k_2^2\ell_1+k_2\ell_1^2+k_1k_3^2-k_3^2\ell_1-k_3\ell_1^2+k_2k_3^2+k_1^2\ell_3+k_1\ell_3^2+k_2^2\ell_3\\
&+k_2\ell_3^2+k_3^2\ell_3-k_3\ell_3^2+2k_1k_2\ell_3-2k_1k_2\ell_1-2k_2k_3\ell_3+2k_2k_3\ell_1+2k_1k_3\ell_1-2k_1k_3\ell_3-2k_1\ell_1\ell_3\\
&-2k_2\ell_1\ell_3+2k_3\ell_1\ell_3+k_1^2k_2-k_1^2k_3-k_2^2k_3+\ell_1^2\ell_3-\ell_1\ell_3^2-2k_1k_2k_3)\\
&+\frac{3A_{131}}{(k_2+\ell_3)^2}(k_1^2\ell_1+k_1k_2^2+k_1\ell_1^2+k_2^2\ell_1-k_2\ell_1^2+k_1k_3^2+k_3^2\ell_1+k_3\ell_1^2-k_2k_3^2-k_1^2\ell_3+k_1\ell_3^2-k_2^2\ell_3\\
&-k_2\ell_3^2-k_3^2\ell_3+k_3\ell_3^2+2k_1k_2\ell_3-2k_1k_2\ell_1+2k_2k_3\ell_3-2k_2k_3\ell_1+2k_1k_3\ell_1-2k_1k_3\ell_3-2k_1\ell_1\ell_3\\
&+2k_2\ell_1\ell_3-2k_3\ell_1\ell_3-k_1^2k_2+k_1^2k_3+k_2^2k_3-\ell_1^2\ell_3+\ell_1\ell_3^2-2k_1k_2k_3)\\
&+\frac{3A_{121}}{(k_3+\ell_3)^2}(k_1^2\ell_1+k_1k_2^2+k_1\ell_1^2+k_2^2\ell_1+k_2\ell_1^2+k_1k_3^2+k_3^2\ell_1-k_3\ell_1^2+k_2k_3^2-k_1^2\ell_3+k_1\ell_3^2-k_2^2\ell_3\\
&+k_2\ell_3^2-k_3^2\ell_3-k_3\ell_3^2-2k_1k_2\ell_3+2k_1k_2\ell_1+2k_2k_3\ell_3-2k_2k_3\ell_1-2k_1k_3\ell_1+2k_1k_3\ell_3-2k_1\ell_1\ell_3\\
&-2k_2\ell_1\ell_3+2k_3\ell_1\ell_3+k_1^2k_2-k_1^2k_3-k_2^2k_3-\ell_1^2\ell_3+\ell_1\ell_3^2-2k_1k_2k_3)\\
&+\frac{3A_{133}}{(k_2+\ell_1)^2}(-k_1^2\ell_1+k_1k_2^2+k_1\ell_1^2-k_2^2\ell_1-k_2\ell_1^2+k_1k_3^2-k_3^2\ell_1+k_3\ell_1^2-k_2k_3^2+k_1^2\ell_3+k_1\ell_3^2+k_2^2\ell_3\\
&-k_2\ell_3^2+k_3^2\ell_3+k_3\ell_3^2-2k_1k_2\ell_3+2k_1k_2\ell_1-2k_2k_3\ell_3+2k_2k_3\ell_1-2k_1k_3\ell_1+2k_1k_3\ell_3-2k_1\ell_1\ell_3\\
&+2k_2\ell_1\ell_3-2k_3\ell_1\ell_3-k_1^2k_2+k_1^2k_3+k_2^2k_3+\ell_1^2\ell_3-\ell_1\ell_3^2-2k_1k_2k_3)\\
&+\frac{3A_{231}}{(k_1+\ell_3)^2}(k_1^2\ell_1-k_1k_2^2-k_1\ell_1^2+k_2^2\ell_1+k_2\ell_1^2-k_1k_3^2+k_3^2\ell_1+k_3\ell_1^2+k_2k_3^2-k_1^2\ell_3-k_1\ell_3^2-k_2^2\ell_3\\
&+k_2\ell_3^2-k_3^2\ell_3+k_3\ell_3^2+2k_1k_2\ell_3-2k_1k_2\ell_1-2k_2k_3\ell_3+2k_2k_3\ell_1-2k_1k_3\ell_1+2k_1k_3\ell_3+2k_1\ell_1\ell_3\\
&-2k_2\ell_1\ell_3-2k_3\ell_1\ell_3+k_1^2k_2+k_1^2k_3+k_2^2k_3-\ell_1^2\ell_3+\ell_1\ell_3^2-2k_1k_2k_3)\\
&+\frac{3A_{233}}{(k_1+\ell_1)^2}(-k_1^2\ell_1-k_1k_2^2-k_1\ell_1^2-k_2^2\ell_1+k_2\ell_1^2-k_1k_3^2-k_3^2\ell_1+k_3\ell_1^2+k_2k_3^2+k_1^2\ell_3-k_1\ell_3^2+k_2^2\ell_3\\
&+k_2\ell_3^2+k_3^2\ell_3+k_3\ell_3^2-2k_1k_2\ell_3+2k_1k_2\ell_1+2k_2k_3\ell_3-2k_2k_3\ell_1+2k_1k_3\ell_1-2k_1k_3\ell_3+2k_1\ell_1\ell_3\\
&-2k_2\ell_1\ell_3-2k_3\ell_1\ell_3+k_1^2k_2+k_1^2k_3+k_2^2k_3+\ell_1^2\ell_3-\ell_1\ell_3^2-2k_1k_2k_3),
\end{align*}
\begin{align*}
&S_{23}=3M_{1223}(k_1k_2^2+k_1k_3^2+k_2k_3^2+k_1^2\ell_2+k_1\ell_2^2+k_2^2\ell_2+k_2\ell_2^2+k_3^2\ell_2-k_3\ell_2^2+k_1^2\ell_3+k_1\ell_3^2+k_2^2\ell_3\\
&+k_2\ell_3^2+k_3^2\ell_3-k_3\ell_3^2-2k_1k_3\ell_2+2k_1k_2\ell_2-2k_1k_3\ell_3+2k_1k_2\ell_3-2k_2k_3\ell_2-2k_2k_3\ell_3+2k_1\ell_2\ell_3\\
&+2k_2\ell_2\ell_3-2k_3\ell_2\ell_3+k_1^2k_2-k_1^2k_3-k_2^2k_3+\ell_2^2\ell_3+\ell_2\ell_3^2-2k_1k_2k_3)\\
&+3M_{1323}(k_1k_2^2+k_1k_3^2-k_2k_3^2+k_1^2\ell_2+k_1\ell_2^2+k_2^2\ell_2-k_2\ell_2^2+k_3^2\ell_2+k_3\ell_2^2+k_1^2\ell_3+k_1\ell_3^2+k_2^2\ell_3\\
&-k_2\ell_3^2+k_3^2\ell_3+k_3\ell_3^2+2k_1k_3\ell_2-2k_1k_2\ell_2+2k_1k_3\ell_3-2k_1k_2\ell_3-2k_2k_3\ell_2-2k_2k_3\ell_3+2k_1\ell_2\ell_3\\
&-2k_2\ell_2\ell_3+2k_3\ell_2\ell_3-k_1^2k_2+k_1^2k_3+k_2^2k_3+\ell_2^2\ell_3+\ell_2\ell_3^2-2k_1k_2k_3)\\
&+3M_{2323}(-k_1k_2^2-k_1k_3^2+k_2k_3^2+k_1^2\ell_2-k_1\ell_2^2+k_2^2\ell_2+k_2\ell_2^2+k_3^2\ell_2+k_3\ell_2^2+k_1^2\ell_3-k_1\ell_3^2+k_2^2\ell_3\\
&+k_2\ell_3^2+k_3^2\ell_3+k_3\ell_3^2-2k_1k_3\ell_2-2k_1k_2\ell_2-2k_1k_3\ell_3-2k_1k_2\ell_3+2k_2k_3\ell_2+2k_2k_3\ell_3-2k_1\ell_2\ell_3\\
&+2k_2\ell_2\ell_3+2k_3\ell_2\ell_3+k_1^2k_2+k_1^2k_3+k_2^2k_3+\ell_2^2\ell_3+\ell_2\ell_3^2-2k_1k_2k_3)\\
&+\frac{3A_{122}}{(k_3+\ell_3)^2}(k_1k_2^2+k_1k_3^2+k_2k_3^2+k_1^2\ell_2+k_1\ell_2^2+k_2^2\ell_2+k_2\ell_2^2+k_3^2\ell_2-k_3\ell_2^2-k_1^2\ell_3+k_1\ell_3^2-k_2^2\ell_3\\
&+k_2\ell_3^2-k_3^2\ell_3-k_3\ell_3^2-2k_1k_3\ell_2+2k_1k_2\ell_2+2k_1k_3\ell_3-2k_1k_2\ell_3-2k_2k_3\ell_2+2k_2k_3\ell_3-2k_1\ell_2\ell_3\\
&-2k_2\ell_2\ell_3+2k_3\ell_2\ell_3+k_1^2k_2-k_1^2k_3-k_2^2k_3-\ell_2^2\ell_3+\ell_2\ell_3^2-2k_1k_2k_3)\\
&+\frac{3A_{123}}{(k_3+\ell_2)^2}(k_1k_2^2+k_1k_3^2+k_2k_3^2-k_1^2\ell_2+k_1\ell_2^2-k_2^2\ell_2+k_2\ell_2^2-k_3^2\ell_2-k_3\ell_2^2+k_1^2\ell_3+k_1\ell_3^2+k_2^2\ell_3\\
&+k_2\ell_3^2+k_3^2\ell_3-k_3\ell_3^2+2k_1k_3\ell_2-2k_1k_2\ell_2-2k_1k_3\ell_3+2k_1k_2\ell_3+2k_2k_3\ell_2-2k_2k_3\ell_3-2k_1\ell_2\ell_3\\
&-2k_2\ell_2\ell_3+2k_3\ell_2\ell_3+k_1^2k_2-k_1^2k_3-k_2^2k_3+\ell_2^2\ell_3-\ell_2\ell_3^2-2k_1k_2k_3)\\
&+\frac{3A_{132}}{(k_2+\ell_3)^2}(k_1k_2^2+k_1k_3^2-k_2k_3^2+k_1^2\ell_2+k_1\ell_2^2+k_2^2\ell_2-k_2\ell_2^2+k_3^2\ell_2+k_3\ell_2^2-k_1^2\ell_3+k_1\ell_3^2-k_2^2\ell_3\\
&-k_2\ell_3^2-k_3^2\ell_3+k_3\ell_3^2+2k_1k_3\ell_2-2k_1k_2\ell_2-2k_1k_3\ell_3+2k_1k_2\ell_3-2k_2k_3\ell_2+2k_2k_3\ell_3-2k_1\ell_2\ell_3\\
&+2k_2\ell_2\ell_3-2k_3\ell_2\ell_3-k_1^2k_2+k_1^2k_3+k_2^2k_3-\ell_2^2\ell_3+\ell_2\ell_3^2-2k_1k_2k_3)\\
&+\frac{3A_{133}}{(k_2+\ell_2)^2}(k_1k_2^2+k_1k_3^2-k_2k_3^2-k_1^2\ell_2+k_1\ell_2^2-k_2^2\ell_2-k_2\ell_2^2-k_3^2\ell_2+k_3\ell_2^2+k_1^2\ell_3+k_1\ell_3^2+k_2^2\ell_3\\
&-k_2\ell_3^2+k_3^2\ell_3+k_3\ell_3^2-2k_1k_3\ell_2+2k_1k_2\ell_2+2k_1k_3\ell_3-2k_1k_2\ell_3+2k_2k_3\ell_2-2k_2k_3\ell_3-2k_1\ell_2\ell_3\\
&+2k_2\ell_2\ell_3-2k_3\ell_2\ell_3-k_1^2k_2+k_1^2k_3+k_2^2k_3+\ell_2^2\ell_3-\ell_2\ell_3^2-2k_1k_2k_3)\\
&+\frac{3A_{232}}{(k_1+\ell_3)^2}(-k_1k_2^2-k_1k_3^2+k_2k_3^2+k_1^2\ell_2-k_1\ell_2^2+k_2^2\ell_2+k_2\ell_2^2+k_3^2\ell_2+k_3\ell_2^2-k_1^2\ell_3-k_1\ell_3^2-k_2^2\ell_3\\
&+k_2\ell_3^2-k_3^2\ell_3+k_3\ell_3^2-2k_1k_3\ell_2-2k_1k_2\ell_2+2k_1k_3\ell_3+2k_1k_2\ell_3+2k_2k_3\ell_2-2k_2k_3\ell_3+2k_1\ell_2\ell_3\\
&-2k_2\ell_2\ell_3-2k_3\ell_2\ell_3+k_1^2k_2+k_1^2k_3+k_2^2k_3-\ell_2^2\ell_3+\ell_2\ell_3^2-2k_1k_2k_3)\\
&+\frac{3A_{233}}{(k_1+\ell_2)^2}(-k_1k_2^2-k_1k_3^2+k_2k_3^2-k_1^2\ell_2-k_1\ell_2^2-k_2^2\ell_2+k_2\ell_2^2-k_3^2\ell_2+k_3\ell_2^2+k_1^2\ell_3-k_1\ell_3^2+k_2^2\ell_3\\
&+k_2\ell_3^2+k_3^2\ell_3+k_3\ell_3^2+2k_1k_3\ell_2+2k_1k_2\ell_2-2k_1k_3\ell_3-2k_1k_2\ell_3-2k_2k_3\ell_2+2k_2k_3\ell_3+2k_1\ell_2\ell_3\\
&-2k_2\ell_2\ell_3-2k_3\ell_2\ell_3+k_1^2k_2+k_1^2k_3+k_2^2k_3+\ell_2^2\ell_3-\ell_2\ell_3^2-2k_1k_2k_3),
\end{align*}
\begin{align*}
&Q_{12}=3M_{1213}(k_1k_2^2-k_1^2\ell_2+k_1\ell_2^2-k_2^2\ell_2+k_2\ell_2^2+k_1^2\ell_3+k_1\ell_3^2+k_2^2\ell_3+k_2\ell_3^2-2k_1k_2\ell_2+2k_1k_2\ell_3\\
&-2k_1\ell_2\ell_3-2k_2\ell_2\ell_3+k_1^2k_2+\ell_2^2\ell_3-\ell_2\ell_3^2+2k_1\ell_1k_2+\ell_1^2k_1+\ell_1\ell_2^2+\ell_1k_1^2+\ell_1^2k_2+\ell_1k_2^2-2\ell_1\ell_2\ell_3\\
&-2\ell_1\ell_2k_1-\ell_1^2\ell_2-2\ell_1\ell_2k_2+2\ell_1\ell_3k_1+2\ell_1\ell_3k_2+\ell_1\ell_3^2+\ell_1^2\ell_3)\\
&+3M_{1223}(k_1k_2^2+k_1^2\ell_2+k_1\ell_2^2+k_2^2\ell_2+k_2\ell_2^2+k_1^2\ell_3+k_1\ell_3^2+k_2^2\ell_3+k_2\ell_3^2+2k_1k_2\ell_2+2k_1k_2\ell_3\\
&+2k_1\ell_2\ell_3+2k_2\ell_2\ell_3+k_1^2k_2+\ell_2^2\ell_3-\ell_2\ell_3^2-2k_1\ell_1k_2+\ell_1^2k_1-\ell_1\ell_2^2-\ell_1k_1^2+\ell_1^2k_2-\ell_1k_2^2-2\ell_1\ell_2\ell_3\\
&-2\ell_1\ell_2k_1+\ell_1^2\ell_2-2\ell_1\ell_2k_2-2\ell_1\ell_3k_1-2\ell_1\ell_3k_2-\ell_1\ell_3^2+\ell_1^2\ell_3)\\
&+3M_{1212}(k_1k_2^2+k_1^2\ell_2+k_1\ell_2^2+k_2^2\ell_2+k_2\ell_2^2-k_1^2\ell_3+k_1\ell_3^2-k_2^2\ell_3+k_2\ell_3^2+2k_1k_2\ell_2-2k_1k_2\ell_3\\
&-2k_1\ell_2\ell_3-2k_2\ell_2\ell_3+k_1^2k_2-\ell_2^2\ell_3+\ell_2\ell_3^2+2k_1\ell_1k_2+\ell_1^2k_1+\ell_1\ell_2^2+\ell_1k_1^2+\ell_1^2k_2+\ell_1k_2^2-2\ell_1\ell_2\ell_3\\
&+2\ell_1\ell_2k_1+\ell_1^2\ell_2+2\ell_1\ell_2k_2-2\ell_1\ell_3k_1-2\ell_1\ell_3k_2+\ell_1\ell_3^2-\ell_1^2\ell_3)\\
&+\frac{3B_{121}}{(k_2+\ell_3)^2}(k_1k_2^2+k_1^2\ell_2+k_1\ell_2^2+k_2^2\ell_2-k_2\ell_2^2-k_1^2\ell_3+k_1\ell_3^2-k_2^2\ell_3-k_2\ell_3^2-2k_1k_2\ell_2+2k_1k_2\ell_3\\
&-2k_1\ell_2\ell_3+2k_2\ell_2\ell_3-k_1^2k_2-\ell_2^2\ell_3+\ell_2\ell_3^2-2k_1\ell_1k_2+\ell_1^2k_1+\ell_1\ell_2^2+\ell_1k_1^2-\ell_1^2k_2+\ell_1k_2^2-2\ell_1\ell_2\ell_3\\
&+2\ell_1\ell_2k_1+\ell_1^2\ell_2-2\ell_1\ell_2k_2-2\ell_1\ell_3k_1+2\ell_1\ell_3k_2+\ell_1\ell_3^2+\ell_1^2\ell_3)\\
&+\frac{3B_{122}}{(k_1+\ell_3)^2}(-k_1k_2^2+k_1^2\ell_2-k_1\ell_2^2+k_2^2\ell_2+k_2\ell_2^2-k_1^2\ell_3-k_1\ell_3^2-k_2^2\ell_3+k_2\ell_3^2-2k_1k_2\ell_2+2k_1k_2\ell_3\\
&+2k_1\ell_2\ell_3-2k_2\ell_2\ell_3+k_1^2k_2-\ell_2^2\ell_3+\ell_2\ell_3^2-2k_1\ell_1k_2-\ell_1^2k_1+\ell_1\ell_2^2+\ell_1k_1^2+\ell_1^2k_2+\ell_1k_2^2-2\ell_1\ell_2\ell_3\\
&-2\ell_1\ell_2k_1+\ell_1^2\ell_2+2\ell_1\ell_2k_2+2\ell_1\ell_3k_1-2\ell_1\ell_3k_2+\ell_1\ell_3^2-\ell_1^2\ell_3)\\
&+\frac{3B_{132}}{(k_1+\ell_2)^2}(-k_1k_2^2-k_1^2\ell_2-k_1\ell_2^2-k_2^2\ell_2+k_2\ell_2^2+k_1^2\ell_3-k_1\ell_3^2+k_2^2\ell_3+k_2\ell_3^2+2k_1k_2\ell_2-2k_1k_2\ell_3\\
&+2k_1\ell_2\ell_3-2k_2\ell_2\ell_3+k_1^2k_2+\ell_2^2\ell_3-\ell_2\ell_3^2-2k_1\ell_1k_2-\ell_1^2k_1+\ell_1\ell_2^2+\ell_1k_1^2+\ell_1^2k_2+\ell_1k_2^2-2\ell_1\ell_2\ell_3\\
&+2\ell_1\ell_2k_1-\ell_1^2\ell_2-2\ell_1\ell_2k_2-2\ell_1\ell_3k_1+2\ell_1\ell_3k_2+\ell_1\ell_3^2+\ell_1^2\ell_3)\\
&+\frac{3B_{131}}{(k_2+\ell_2)^2}(k_1k_2^2-k_1^2\ell_2+k_1\ell_2^2-k_2^2\ell_2-k_2\ell_2^2+k_1^2\ell_3+k_1\ell_3^2+k_2^2\ell_3+k_2\ell_3^2+2k_1k_2\ell_2-2k_1k_2\ell_3\\
&-2k_1\ell_2\ell_3+2k_2\ell_2\ell_3-k_1^2k_2+\ell_2^2\ell_3-\ell_2\ell_3^2-2k_1\ell_1k_2+\ell_1^2k_1+\ell_1\ell_2^2+\ell_1k_1^2-\ell_1^2k_2+\ell_1k_2^2-2\ell_1\ell_2\ell_3\\
&-2\ell_1\ell_2k_1-\ell_1^2\ell_2+2\ell_1\ell_2k_2+2\ell_1\ell_3k_1-2\ell_1\ell_3k_2+\ell_1\ell_3^2+\ell_1^2\ell_3)\\
&+\frac{3B_{231}}{(k_2+\ell_1)^2}(k_1k_2^2+k_1^2\ell_2+k_1\ell_2^2+k_2^2\ell_2-k_2\ell_2^2+k_1^2\ell_3+k_1\ell_3^2+k_2^2\ell_3-k_2\ell_3^2-2k_1k_2\ell_2-2k_1k_2\ell_3\\
&+2k_1\ell_2\ell_3-2k_2\ell_2\ell_3-k_1^2k_2+\ell_2^2\ell_3+\ell_2\ell_3^2+2k_1\ell_1k_2+\ell_1^2k_1-\ell_1\ell_2^2-\ell_1k_1^2-\ell_1^2k_2-\ell_1k_2^2-2\ell_1\ell_2\ell_3\\
&-2\ell_1\ell_2k_1+\ell_1^2\ell_2+2\ell_1\ell_2k_2-2\ell_1\ell_3k_1+2\ell_1\ell_3k_2-\ell_1\ell_3^2+\ell_1^2\ell_3)\\
&+\frac{3B_{232}}{(k_1+\ell_1)^2}(k_1k_2^2-k_1^2\ell_2+k_1\ell_2^2-k_2^2\ell_2+k_2\ell_2^2+k_1^2\ell_3+k_1\ell_3^2+k_2^2\ell_3+k_2\ell_3^2-2k_1k_2\ell_2+2k_1k_2\ell_3\\
&-2k_1\ell_2\ell_3-2k_2\ell_2\ell_3+k_1^2k_2+\ell_2^2\ell_3-\ell_2\ell_3^2+2k_1\ell_1k_2+\ell_1^2k_1+\ell_1\ell_2^2+\ell_1k_1^2+\ell_1^2k_2+\ell_1k_2^2-2\ell_1\ell_2\ell_3\\
&-2\ell_1\ell_2k_1-\ell_1^2\ell_2-2\ell_1\ell_2k_2+2\ell_1\ell_3k_1+2\ell_1\ell_3k_2+\ell_1\ell_3^2+\ell_1^2\ell_3),
\end{align*}
\begin{align*}
&Q_{13}=3M_{1313}(k_1k_3^2-k_1^2\ell_2+k_1\ell_2^2-k_3^2\ell_2+k_3\ell_2^2+k_1^2\ell_3+k_1\ell_3^2+k_3^2\ell_3+k_3\ell_3^2-2k_1k_3\ell_2+2k_1k_3\ell_3\\
&-2k_1\ell_2\ell_3-2k_3\ell_2\ell_3+k_1^2k_3+2k_1\ell_1k_3+\ell_2^2\ell_3-\ell_2\ell_3^2+\ell_1^2k_1+\ell_1\ell_2^2+\ell_1k_1^2+\ell_1^2k_3+\ell_1k_3^2\\
&-2\ell_1\ell_2\ell_3-2\ell_1\ell_2k_1-\ell_1^2\ell_2-2\ell_1\ell_2k_3+2\ell_1\ell_3k_1+2\ell_1\ell_3k_3+\ell_1\ell_3^2+\ell_1^2\ell_3)\\
&+3M_{1312}(k_1k_3^2+k_1^2\ell_2+k_1\ell_2^2+k_3^2\ell_2+k_3\ell_2^2-k_1^2\ell_3+k_1\ell_3^2-k_3^2\ell_3+k_3\ell_3^2+2k_1k_3\ell_2-2k_1k_3\ell_3\\
&-2k_1\ell_2\ell_3-2k_3\ell_2\ell_3+k_1^2k_3+2k_1\ell_1k_3-\ell_2^2\ell_3+\ell_2\ell_3^2+\ell_1^2k_1+\ell_1\ell_2^2+\ell_1k_1^2+\ell_1^2k_3+\ell_1k_3^2\\
&-2\ell_1\ell_2\ell_3+2\ell_1\ell_2k_1+\ell_1^2\ell_2+2\ell_1\ell_2k_3-2\ell_1\ell_3k_1-2\ell_1\ell_3k_3+\ell_1\ell_3^2-\ell_1^2\ell_3)\\
&+3M_{1323}(k_1k_3^2+k_1^2\ell_2+k_1\ell_2^2+k_3^2\ell_2+k_3\ell_2^2+k_1^2\ell_3+k_1\ell_3^2+k_3^2\ell_3+k_3\ell_3^2+2k_1k_3\ell_2+2k_1k_3\ell_3\\
&+2k_1\ell_2\ell_3+2k_3\ell_2\ell_3+k_1^2k_3-2k_1\ell_1k_3+\ell_2^2\ell_3+\ell_2\ell_3^2+\ell_1^2k_1-\ell_1\ell_2^2-\ell_1k_1^2+\ell_1^2k_3-\ell_1k_3^2\\
&-2\ell_1\ell_2\ell_3-2\ell_1\ell_2k_1+\ell_1^2\ell_2-2\ell_1\ell_2k_3-2\ell_1\ell_3k_1-2\ell_1\ell_3k_3-\ell_1\ell_3^2+\ell_1^2\ell_3)\\
&+\frac{3B_{123}}{(k_1+\ell_3)^2}(-k_1k_3^2+k_1^2\ell_2-k_1\ell_2^2+k_3^2\ell_2+k_3\ell_2^2-k_1^2\ell_3-k_1\ell_3^2-k_3^2\ell_3+k_3\ell_3^2-2k_1k_3\ell_2+2k_1k_3\ell_3\\
&+2k_1\ell_2\ell_3-2k_3\ell_2\ell_3+k_1^2k_3-2k_1\ell_1k_3-\ell_2^2\ell_3+\ell_2\ell_3^2-\ell_1^2k_1+\ell_1\ell_2^2+\ell_1k_1^2+\ell_1^2k_3+\ell_1k_3^2\\
&-2\ell_1\ell_2\ell_3-2\ell_1\ell_2k_1+\ell_1^2\ell_2+2\ell_1\ell_2k_3+2\ell_1\ell_3k_1-2\ell_1\ell_3k_3+\ell_1\ell_3^2-\ell_1^2\ell_3)\\
&+\frac{3B_{133}}{(k_1+\ell_2)^2}(-k_1k_3^2-k_1^2\ell_2-k_1\ell_2^2-k_3^2\ell_2+k_3\ell_2^2+k_1^2\ell_3-k_1\ell_3^2+k_3^2\ell_3+k_3\ell_3^2+2k_1k_3\ell_2-2k_1k_3\ell_3\\
&+2k_1\ell_2\ell_3-2k_3\ell_2\ell_3+k_1^2k_3-2k_1\ell_1k_3+\ell_2^2\ell_3-\ell_2\ell_3^2-\ell_1^2k_1+\ell_1\ell_2^2+\ell_1k_1^2+\ell_1^2k_3+\ell_1k_3^2\\
&-2\ell_1\ell_2\ell_3+2\ell_1\ell_2k_1-\ell_1^2\ell_2-2\ell_1\ell_2k_3-2\ell_1\ell_3k_1+2\ell_1\ell_3k_3+\ell_1\ell_3^2+\ell_1^2\ell_3)\\
&+\frac{3B_{121}}{(k_3+\ell_3)^2}(k_1k_3^2+k_1^2\ell_2+k_1\ell_2^2+k_3^2\ell_2-k_3\ell_2^2-k_1^2\ell_3+k_1\ell_3^2-k_3^2\ell_3-k_3\ell_3^2-2k_1k_3\ell_2+2k_1k_3\ell_3\\
&-2k_1\ell_2\ell_3+2k_3\ell_2\ell_3-k_1^2k_3-2k_1\ell_1k_3-\ell_2^2\ell_3+\ell_2\ell_3^2+\ell_1^2k_1+\ell_1\ell_2^2+\ell_1k_1^2-\ell_1^2k_3+\ell_1k_3^2\\
&-2\ell_1\ell_2\ell_3+2\ell_1\ell_2k_1+\ell_1^2\ell_2-2\ell_1\ell_2k_3-2\ell_1\ell_3k_1+2\ell_1\ell_3k_3+\ell_1\ell_3^2-\ell_1^2\ell_3)\\
&+\frac{3B_{231}}{(k_3+\ell_1)^2}(k_1k_3^2-k_1^2\ell_2+k_1\ell_2^2-k_3^2\ell_2+k_3\ell_2^2+k_1^2\ell_3+k_1\ell_3^2+k_3^2\ell_3+k_3\ell_3^2-2k_1k_3\ell_2+2k_1k_3\ell_3\\
&-2k_1\ell_2\ell_3-2k_3\ell_2\ell_3+k_1^2k_3+2k_1\ell_1k_3+\ell_2^2\ell_3-\ell_2\ell_3^2+\ell_1^2k_1+\ell_1\ell_2^2+\ell_1k_1^2+\ell_1^2k_3+\ell_1k_3^2\\
&-2\ell_1\ell_2\ell_3-2\ell_1\ell_2k_1-\ell_1^2\ell_2-2\ell_1\ell_2k_3+2\ell_1\ell_3k_1+2\ell_1\ell_3k_3+\ell_1\ell_3^2+\ell_1^2\ell_3)\\
&+\frac{3B_{233}}{(k_1+\ell_1)^2}(-k_1k_3^2+k_1^2\ell_2-k_1\ell_2^2+k_3^2\ell_2+k_3\ell_2^2+k_1^2\ell_3-k_1\ell_3^2+k_3^2\ell_3+k_3\ell_3^2-2k_1k_3\ell_2-2k_1k_3\ell_3\\
&-2k_1\ell_2\ell_3+2k_3\ell_2\ell_3+k_1^2k_3+2k_1\ell_1k_3+\ell_2^2\ell_3+\ell_2\ell_3^2-\ell_1^2k_1-\ell_1\ell_2^2-\ell_1k_1^2+\ell_1^2k_3-\ell_1k_3^2\\
&-2\ell_1\ell_2\ell_3+2\ell_1\ell_2k_1+\ell_1^2\ell_2-2\ell_1\ell_2k_3+2\ell_1\ell_3k_1-2\ell_1\ell_3k_3-\ell_1\ell_3^2+\ell_1^2\ell_3)\\
&+\frac{3B_{131}}{(k_3+\ell_2)^2}(k_1k_3^2-k_1^2\ell_2+k_1\ell_2^2-k_3^2\ell_2-k_3\ell_2^2+k_1^2\ell_3+k_1\ell_3^2+k_3^2\ell_3-k_3\ell_3^2+2k_1k_3\ell_2-2k_1k_3\ell_3\\
&-2k_1\ell_2\ell_3+2k_3\ell_2\ell_3-k_1^2k_3-2k_1\ell_1k_3+\ell_2^2\ell_3-\ell_2\ell_3^2+\ell_1^2k_1+\ell_1\ell_2^2+\ell_1k_1^2-\ell_1^2k_3+\ell_1k_3^2\\
&-2\ell_1\ell_2\ell_3-2\ell_1\ell_2k_1-\ell_1^2\ell_2+2\ell_1\ell_2k_3+2\ell_1\ell_3k_1-2\ell_1\ell_3k_3+\ell_1\ell_3^2+\ell_1^2\ell_3),
\end{align*}
\begin{align*}
&Q_{23}=3M_{2312}(k_2k_3^2+k_2^2\ell_2+k_2\ell_2^2+k_3^2\ell_2+k_3\ell_2^2-k_2^2\ell_3+k_2\ell_3^2-k_3^2\ell_3+k_3\ell_3^2+2k_2k_3\ell_2-2k_2k_3\ell_3\\
&-2k_2\ell_2\ell_3-2k_3\ell_2\ell_3+k_2^2k_3-\ell_2^2\ell_3+\ell_2\ell_3^2+\ell_1\ell_2^2+\ell_1^2k_2+\ell_1k_2^2+\ell_1^2k_3+\ell_1k_3^2-2\ell_1\ell_2\ell_3+\ell_1^2\ell_2\\
&+2k_2\ell_1k_3+2\ell_1\ell_2k_2+2\ell_1\ell_2k_3-2\ell_1\ell_3k_2-2\ell_1\ell_3k_3+\ell_1\ell_3^2-\ell_1^2\ell_3)\\
&+3M_{2313}(k_2k_3^2-k_2^2\ell_2+k_2\ell_2^2-k_3^2\ell_2+k_3\ell_2^2+k_2^2\ell_3+k_2\ell_3^2+k_3^2\ell_3+k_3\ell_3^2-2k_2k_3\ell_2+2k_2k_3\ell_3\\
&-2k_2\ell_2\ell_3-2k_3\ell_2\ell_3+k_2^2k_3+\ell_2^2\ell_3-\ell_2\ell_3^2+\ell_1\ell_2^2+\ell_1^2k_2+\ell_1k_2^2+\ell_1^2k_3+\ell_1k_3^2-2\ell_1\ell_2\ell_3-\ell_1^2\ell_2\\
&+2k_2\ell_1k_3-2\ell_1\ell_2k_2-2\ell_1\ell_2k_3+2\ell_1\ell_3k_2+2\ell_1\ell_3k_3+\ell_1\ell_3^2+\ell_1^2\ell_3)\\
&+3M_{2323}(k_2k_3^2+k_2^2\ell_2+k_2\ell_2^2+k_3^2\ell_2+k_3\ell_2^2+k_2^2\ell_3+k_2\ell_3^2+k_3^2\ell_3+k_3\ell_3^2+2k_2k_3\ell_2+2k_2k_3\ell_3\\
&+2k_2\ell_2\ell_3+2k_3\ell_2\ell_3+k_2^2k_3+\ell_2^2\ell_3+\ell_2\ell_3^2-\ell_1\ell_2^2+\ell_1^2k_2-\ell_1k_2^2+\ell_1^2k_3-\ell_1k_3^2-2\ell_1\ell_2\ell_3+\ell_1^2\ell_2\\
&-2k_2\ell_1k_3-2\ell_1\ell_2k_2-2\ell_1\ell_2k_3-2\ell_1\ell_3k_2-2\ell_1\ell_3k_3-\ell_1\ell_3^2+\ell_1^2\ell_3)\\
&+\frac{3B_{122}}{(k_3+\ell_3)^2}(k_2k_3^2+k_2^2\ell_2+k_2\ell_2^2+k_3^2\ell_2-k_3\ell_2^2-k_2^2\ell_3+k_2\ell_3^2-k_3^2\ell_3-k_3\ell_3^2-2k_2k_3\ell_2+2k_2k_3\ell_3\\
&-2k_2\ell_2\ell_3+2k_3\ell_2\ell_3-k_2^2k_3-\ell_2^2\ell_3+\ell_2\ell_3^2+\ell_1\ell_2^2+\ell_1^2k_2+\ell_1k_2^2-\ell_1^2k_3+\ell_1k_3^2-2\ell_1\ell_2\ell_3+\ell_1^2\ell_2\\
&-2k_2\ell_1k_3+2\ell_1\ell_2k_2-z2\ell_1\ell_2k_3-2\ell_1\ell_3k_2+2\ell_1\ell_3k_3+\ell_1\ell_3^2-\ell_1^2\ell_3)\\
&+\frac{3B_{123}}{(k_2+\ell_3)^2}(-k_2k_3^2+k_2^2\ell_2-k_2\ell_2^2+k_3^2\ell_2+k_3\ell_2^2-k_2^2\ell_3-k_2\ell_3^2-k_3^2\ell_3+k_3\ell_3^2-2k_2k_3\ell_2+2k_2k_3\ell_3\\
&+2k_2\ell_2\ell_3-2k_3\ell_2\ell_3+k_2^2k_3-\ell_2^2\ell_3+\ell_2\ell_3^2+\ell_1\ell_2^2-\ell_1^2k_2+\ell_1k_2^2+\ell_1^2k_3+\ell_1k_3^2-2\ell_1\ell_2\ell_3+\ell_1^2\ell_2\\
&-2k_2\ell_1k_3-2\ell_1\ell_2k_2+2\ell_1\ell_2k_3+2\ell_1\ell_3k_2-2\ell_1\ell_3k_3+\ell_1\ell_3^2-\ell_1^2\ell_3)\\
&+\frac{3B_{132}}{(k_3+\ell_2)^2}(k_2k_3^2-k_2^2\ell_2+k_2\ell_2^2-k_3^2\ell_2-k_3\ell_2^2+k_2^2\ell_3+k_2\ell_3^2+k_3^2\ell_3-k_3\ell_3^2+2k_2k_3\ell_2-2k_2k_3\ell_3\\
&-2k_2\ell_2\ell_3+2k_3\ell_2\ell_3-k_2^2k_3+\ell_2^2\ell_3-\ell_2\ell_3^2+\ell_1\ell_2^2+\ell_1^2k_2+\ell_1k_2^2-\ell_1^2k_3+\ell_1k_3^2-2\ell_1\ell_2\ell_3-\ell_1^2\ell_2\\
&-2k_2\ell_1k_3-2\ell_1\ell_2k_2+2\ell_1\ell_2k_3+2\ell_1\ell_3k_2-2\ell_1\ell_3k_3+\ell_1\ell_3^2+\ell_1^2\ell_3)\\
&+\frac{3B_{133}}{(k_2+\ell_2)^2}(-k_2k_3^2-k_2^2\ell_2-k_2\ell_2^2-k_3^2\ell_2+k_3\ell_2^2+k_2^2\ell_3-k_2\ell_3^2+k_3^2\ell_3+k_3\ell_3^2+2k_2k_3\ell_2-2k_2k_3\ell_3\\
&+2k_2\ell_2\ell_3-2k_3\ell_2\ell_3+k_2^2k_3+\ell_2^2\ell_3-\ell_2\ell_3^2+\ell_1\ell_2^2-\ell_1^2k_2+\ell_1k_2^2+\ell_1^2k_3+\ell_1k_3^2-2\ell_1\ell_2\ell_3-\ell_1^2\ell_2\\
&-2k_2\ell_1k_3+2\ell_1\ell_2k_2-2\ell_1\ell_2k_3-2\ell_1\ell_3k_2+2\ell_1\ell_3k_3+\ell_1\ell_3^2+\ell_1^2\ell_3)\\
&+\frac{3B_{232}}{(k_3+\ell_1)^2}(k_2k_3^2+k_2^2\ell_2+k_2\ell_2^2+k_3^2\ell_2-k_3\ell_2^2+k_2^2\ell_3+k_2\ell_3^2+k_3^2\ell_3-k_3\ell_3^2-2k_2k_3\ell_2-2k_2k_3\ell_3\\
&+2k_2\ell_2\ell_3-2k_3\ell_2\ell_3-k_2^2k_3+\ell_2^2\ell_3+\ell_2\ell_3^2-\ell_1\ell_2^2+\ell_1^2k_2-\ell_1k_2^2-\ell_1^2k_3-\ell_1k_3^2-2\ell_1\ell_2\ell_3+\ell_1^2\ell_2\\
&+2k_2\ell_1k_3-2\ell_1\ell_2k_2+2\ell_1\ell_2k_3-2\ell_1\ell_3k_2+2\ell_1\ell_3k_3-\ell_1\ell_3^2+\ell_1^2\ell_3)\\
&+\frac{3B_{233}}{(k_2+\ell_1)^2}(-k_2k_3^2+k_2^2\ell_2-k_2\ell_2^2+k_3^2\ell_2+k_3\ell_2^2+k_2^2\ell_3-k_2\ell_3^2+k_3^2\ell_3+k_3\ell_3^2-2k_2k_3\ell_2-2k_2k_3\ell_3\\
&-2k_2\ell_2\ell_3+2k_3\ell_2\ell_3+k_2^2k_3+\ell_2^2\ell_3+\ell_2\ell_3^2-\ell_1\ell_2^2-\ell_1^2k_2-\ell_1k_2^2+\ell_1^2k_3-\ell_1k_3^2-2\ell_1\ell_2\ell_3+\ell_1^2\ell_2\\
&+2k_2\ell_1k_3+2\ell_1\ell_2k_2-2\ell_1\ell_2k_3+2\ell_1\ell_3k_2-2\ell_1\ell_3k_3-\ell_1\ell_3^2+\ell_1^2\ell_3).
\end{align*}
\begin{align*}\displaystyle
&H=\Big[\frac{M_{2323}}{(k_1+\ell_1)^2}(k_1-k_2-k_3+\ell_1-\ell_2-\ell_3)^2+\frac{M_{2312}}{(k_1+\ell_3)^2}(k_1-k_2-k_3-\ell_1-\ell_2+\ell_3)^2
\\&+\frac{M_{1313}}{(k_2+\ell_2)^2}(-k_1+k_2-k_3-\ell_1+\ell_2-\ell_3)^2+\frac{M_{2313}}{(k_1+\ell_2)^2}(k_1-k_2-k_3-\ell_1+\ell_2-\ell_3)^2
\\&+\frac{M_{1212}}{(k_3+\ell_3)^2}(-k_1-k_2+k_3-\ell_1-\ell_2+\ell_3)^2+\frac{M_{1323}}{(k_2+\ell_1)^2}(-k_1+k_2-k_3+\ell_1-\ell_2-\ell_3)^2
\\&+\frac{M_{1312}}{(k_2+\ell_3)^2}(-k_1+k_2-k_3-\ell_1-\ell_2+\ell_3)^2+\frac{M_{1223}}{(k_3+\ell_1)^2}(-k_1-k_2+k_3+\ell_1-\ell_2-\ell_3)^2
\\&+\frac{M_{1213}}{(k_3+\ell_2)^2}(-k_1-k_2+k_3-\ell_1+\ell_2-\ell_3)^2-V_{12}-V_{13}-V_{23}-W_{12}-W_{13}-W_{23}
\\&
-B_{231}A_{231}-B_{233}A_{121}-B_{132}A_{132}-B_{121}A_{233}-B_{122}A_{133}-B_{133}A_{122}-B_{123}A_{123}
\\&-B_{131}A_{232}-B_{232}A_{131}\Big]\Big/(k_1+k_2+k_3+\ell_1+\ell_2+\ell_3)^2.
\end{align*}

\section{Acknowledgment}
  This work is partially supported by the Scientific
and Technological Research Council of Turkey (T\"{U}B\.{I}TAK).\\


\section{References}

\begin{thebibliography}{99}

\bibitem{AKNS} Ablowitz M.J, Kaup D.J., Newell A.C., and Segur H., {\it The inverse scattering transform-Fourier analysis for nonlinear problems}, Stud.
Appl. Math. {\bf 53}, Issue: 4, 249--315 (1974).

\bibitem{AbMu1} Ablowitz M.J. and Musslimani Z.H., {\it Integrable nonlocal nonlinear Schr\"{o}dinger equation}, Phys. Rev. Lett. {\bf 110}, 064105 (2013).

\bibitem{AbMu2} Ablowitz M.J. and Musslimani Z.H., {\it Inverse scattering transform for the integrable nonlocal nonlinear Schr\"{o}dinger equation}, Nonlinearity {\bf 29}, 915--946 (2016).

\bibitem{AbMu3} Ablowitz M.J. and Musslimani Z.H., {\it Integrable nonlocal nonlinear equations}, Stud. Appl. Math. {\bf 139}, Issue: 1, 7--59 (2016).

\bibitem{ma}  Ma L.Y., Shen S.F., and Zhu Z.N., {\it Integrable nonlocal complex mKdV equation: soliton solution and gauge equivalence}, {(\tt
    arXiv:1612.06723 [nlin.SI])}.

\bibitem{JZ1} Ji J.L. and Zhu Z.N., {\it On a nonlocal modified Korteweg-de Vries equation: Integrability, Darboux transformation and soliton solutions},
Commun. Non. Sci. Numer. Simulat. {\bf 42}, 699--708 (2017).

\bibitem{JZ2} Ji J.L. and Zhu Z.N., {\it Soliton solutions of an integrable nonlocal modified Korteweg-de Vries equation through
inverse scattering transform}, J. Math. An. and App. {\bf 453}, 973--984 (2017). {(\tt arXiv:1603.03994 [nlin.SI])}.

\bibitem{Yang} Yang B. and Yang J., {\it Transformations between nonlocal and local integrable equations}, {(\tt arXiv:1705.00332v1 [nlin.PS])}.

\bibitem{fok} Fokas A.S., {\it Integrable multidimensional versions of the nonlocal Schr\"{o}dinger equation}, Nonlinearity {\bf 29}, 319--324 (2016).

\bibitem{sak} Sakkaravarthi K. and Kanna T., {\it Bright solitons in coherently coupled nonlinear Schr\"{o}dinger equations with alternate
signs of nonlinearities}, J. Math.Phys. {\bf 54}, 013701 (2013).

\bibitem{gerd} Gerdjikov V.S. and Saxena A., {\it Complete integrability of nonlocal nonlinear Schr\"{o}dinger equation}, J. Math. Phys. {\bf 58}, Issue:1, 013502 (2017). {(\tt arXiv:1510.00480[nlin.SI])}.

\bibitem{sin} Sinha D. and Ghosh P.K., {\it Integrable nonlocal vector nonlinear Schr\"{o}dinger equation with self-induced parity-time symmetric potential}, Phys. Lett. {\bf A 381}, 124-128 (2017).

\bibitem{gerd1}  Gerdjikov V.S., Grahovski D.G., and Ivanov R.I., {\it On the integrable wave interactions and Lax pairs on symmetric spaces}, {(\tt arXiv:1607.06940[nlin.SI])}, in the special issue on Mathematical modelling and physical dynamics of solitary waves: From continuum mechanics to field theory, Eds. Ivan C. Christov, M.D. Todorov, S. Yoshida. (Wave Motion, http://dx.doi.org/10.1016/j.wavemoti.2016.07.012)

\bibitem{gerd2} Gerdjikov V.S., Grahovski D.G., and Ivanov R.I., {\it On the N-wave equations with PT symmetry}, Theor. and Math. Phys. {\bf 188}, No.3, 1305--1321 (2016).

\bibitem{gerd3} Gerdjikov V.S., {\it On nonlocal models of Kulish-Sklyanin type and generalized Fourier transforms}, Stud. Comp. Int. {\bf 681}, 37--52 (2017). {(\tt arXiv:1703.03705[nlin.SI])}.

\bibitem{Sax} Khare A. and Saxena A., {\it Periodic and hyperbolic soliton solutions of a number of nonlocal nonlinear
equations}, J. Math. Phys. {\bf 56},  032104 (2015).

\bibitem{li}  Li M. and Xu T., {\it Dark and antidark soliton interactions in the nonlocal nonlinear Schr\"{o}dinger equation
with the self-induced parity-time-symmetric potential},  Phys. Rev. E {\bf91}, 033202 (2015).

\bibitem{huang} Huang X. and King L., {\it Soliton solutions for the nonlocal nonlinear Schr\"{o}dinger equation}, Eur. Phys. J. Plus {\bf 131}, 148 (2016).

\bibitem{Wen} Wen X.Y., Yan Z., and Yang Y., {\it Dynamics of higher-order rational solitons for the nonlocal nonlinear Schr\"{o}dinger
equation with the self-induced parity-time-symmetric potential},  Chaos {\bf 26}, 063123 (2015).

\bibitem{gur2} G{\" u}rses M., {\it Nonlocal Fordy-Kulish equations on symmetric spaces}, Phys. Lett. A {\bf 381}, 1791-1794 (2017). {(\tt arXiv:1702.03371[nlin.SI])}.

\bibitem{gur3} G{\" u}rses M., {\it Nonlocal super integrable equations}, {(\tt arXiv:1704.01273[nlin.SI])}.

\bibitem{Vincent} Caudrelier V., {\it Interplay between the inverse scattering method and the unified transform with an application}, {(\tt arXiv:1704.05306[math-ph])}.

\bibitem{GurPek}  G{\" u}rses M. and Pekcan A., {\it Nonlocal nonlinear Schr\"{o}dinger equations and their soliton solutions}, {(\tt arXiv:1707.07610 [nlin.SI])}.
\end{thebibliography}
\end{document}